\documentclass[12pt]{article} 
\usepackage[sectionbib]{natbib}
\usepackage{array,epsfig,fancyheadings,rotating}
\usepackage[]{hyperref}  

\usepackage{amsmath}
\usepackage{amssymb}
\usepackage{amsfonts}
\usepackage{multirow}
\usepackage{amsthm}

\usepackage{boxedminipage}
\usepackage{natbib}
\usepackage{multirow}
\usepackage{pdflscape}
\usepackage{booktabs}
\usepackage{array}
\usepackage{rotating}
\usepackage{graphicx}
\usepackage{subfigure}
\usepackage{epsfig}
\usepackage{enumerate}
\usepackage{textcomp}
\usepackage{amsmath, amsfonts,amssymb}
\usepackage{dsfont}
\usepackage{color}
\usepackage{fullpage}
\usepackage{setspace}
\usepackage{algorithm}
\usepackage{algpseudocode}
\usepackage{epstopdf}

\usepackage[usenames,dvipsnames,svgnames,table]{xcolor}
\usepackage{hyperref}

\newtheorem{theorem}{Theorem}
\newtheorem{lemma}{Lemma}
\newtheorem{corollary}{Corollary}
\newtheorem{proposition}{Proposition}
\theoremstyle{definition}
\newtheorem{definition}{Definition}
\newtheorem{example}{Example}
\newtheorem{remark}{Remark}

\usepackage{titlesec}

\titleformat*{\section}{\large\bfseries}
\titleformat*{\subsection}{\large\bfseries}
\titleformat*{\subsubsection}{\bfseries}

\newcommand{\ignore}[1]{}{}

\newcommand{\beq}{\begin{equation}}
\newcommand{\eeq}{\end{equation}}
\newcommand{\beas}{\begin{eqnarray*}}
\newcommand{\eeas}{\end{eqnarray*}}
\newcommand{\bea}{\begin{eqnarray}}
\newcommand{\eea}{\end{eqnarray}}
\newcommand{\bei}{\begin{itemize}}
\newcommand{\eei}{\end{itemize}}
\newcommand{\ben}{\begin{enumerate}}
\newcommand{\een}{\end{enumerate}}
\newcommand{\bet}{\begin{theorem}}
\newcommand{\eet}{\end{theorem}}
\newcommand{\bel}{\begin{lemma}}
\newcommand{\eel}{\end{lemma}}
\newcommand{\bep}{\begin{proposition}}
\newcommand{\eep}{\end{proposition}}
\newcommand{\bed}{\begin{definition}}
\newcommand{\eed}{\end{definition}}
\newcommand{\bec}{\begin{corollary}}
\newcommand{\eec}{\end{corollary}}
\newcommand{\bex}{\begin{example}}
\newcommand{\eex}{\end{example}}

\newcommand{\ep}{\epsilon}

\newcommand{\argmin}{\mathop{\rm arg\min}}

\def\0{\mathbf{0}}

\def\E{\mathbb{E}}
\def\a{\mathbf{a}}
\def\bb{\mathbf{b}}

\def\x{\mathbf{x}}
\def\bx{\mathbf{x}}
\def\by{\mathbf{y}}
\def\Ga{\boldsymbol{\Gamma}}

\def\bG{\boldsymbol{\Gamma}}
\def\brho{\boldsymbol{\rho}}

\def\R{\mathbb{R}}

\def\D{\mathbf{D}}
\def\A{\mathbf{A}}
\def\z{\mathbf{z}}

\def\be{\boldsymbol{\beta}}

\def\X{\mathbf{X}}

\def\bX{\mathbf{X}}
\def\I{\mathbf{I}}
\def\Y{\mathbf{Y}}
\def\Z{\mathbf{Z}}
\def\S{\boldsymbol{\Sigma}}

\def\ps{\boldsymbol{\Psi}}

\def\O{\boldsymbol{\Omega}}
\def\bal{\boldsymbol{\alpha}}

\def\bO{\boldsymbol{\Omega}}
\def\m{\boldsymbol{\mu}}

\def\be{\boldsymbol{\beta}}

\def\va{\boldsymbol{\varepsilon}}
\def\bgamma{\boldsymbol{\gamma}}

\def\m{\boldsymbol{\mu}}
\def\bmu{\boldsymbol{\mu}}

\def\pr{\mathbb{P}} 
\def\ep{\mathbb{E}} 
\def\Cov{\mathrm{Cov}} 
\def\Var{\mathrm{Var}} 

\def\tr{\mathrm{tr}}
\def\supp{{\mathrm{supp}}}
\def\diag{{\mathrm{diag}}}

\def\nuu{\boldsymbol{\nu}}

\let\hat\widehat
\let\tilde\widetilde

\usepackage{xr}
\usepackage{fancyhdr}
\renewcommand\headrulewidth{0pt}
\begin{document}

\markright{ \hbox{\footnotesize\rm Statistica Sinica
}\hfill\\[-13pt]
\hbox{\footnotesize\rm
}\hfill }

\renewcommand{\thefootnote}{}
$\ $\par
$\ $\par


\fontsize{12}{14pt plus.8pt minus .6pt}\selectfont \vspace{0.8pc}
\centerline{\Large\bf Graph Estimation for Matrix-variate Gaussian Data}
\vspace{.4cm}
\centerline{Xi Chen ~ and ~ Weidong Liu} \vspace{.4cm} \centerline{\it
Department of Information, Operations \& Management Sciences, Stern School of Business}
\centerline{\it  New York University}
\centerline{\it
and ~ Department of Mathematics,  Institute of Natural Sciences and MOE-LSC}
\centerline{\it  Shanghai Jiao Tong University}
\vspace{.55cm} \fontsize{9}{11.5pt plus.8pt minus
.6pt}\selectfont

\begin{quotation}
\noindent {\it Abstract: }
Matrix-variate Gaussian graphical models (GGM) have been widely used for modeling matrix-variate data. Since the support of sparse precision matrix represents the conditional independence graph among matrix entries,  conducting support recovery yields valuable information. A commonly used approach is the penalized log-likelihood method. However, due to the complicated structure of precision matrices in the form of Kronecker product, the log-likelihood is non-convex, which presents challenges for both computation and theoretical analysis.
In this paper, we propose an alternative approach by formulating the support recovery problem into a multiple testing problem. A new test statistic is developed and based on that, we use the popular Benjamini and Hochberg's procedure to control false discovery rate (FDR) asymptotically. Our method involves only convex optimization, making it computationally attractive. Theoretically, our method allows very weak conditions, i.e., even when the sample size is finite and the dimensions go to infinity, the asymptotic normality of the test statistics and FDR control can still be guaranteed. We further provide the power analysis result. The finite sample performance of the proposed method is illustrated by both simulated and real data analysis.\par

\vspace{9pt}
\par
\end{quotation}\par

\def\thefigure{\arabic{figure}}
\def\thetable{\arabic{table}}

\renewcommand{\theequation}{\thesection.\arabic{equation}}

\fontsize{12}{14pt plus.8pt minus .6pt}\selectfont

\section{Introduction}

In the era of big data, matrix-variate observations are becoming prevalent in a wide range of domains, such as biomedical imaging, genomics, financial markets, spatio-temporal environmental data analysis, and more. A typical example is the gene expression data in genomics, in which each observation contains expression levels of $p$ genes on $q$ microarrays of the same subject  (see, e.g., \cite{Efron09,YinLi12}). Another example of such data is the multi-channel electroencephalography (EEG) data for brain imaging studies (see, e.g., \cite{Bijma05}), in which each measurement can be expressed as a matrix with rows corresponding to $p$ different channels and columns to $q$ time points. \cite{LengTang2012} provided more interesting examples of matrix-variate data such as the volatility of equality option. Due to the prevalence of matrix-variate observations (especially high-dimensional observations), it is important for us to understand the structural information encoded in these observations.

To study matrix-variate data where each observation $\bX$ is a $p \times q$ matrix, it is commonly assumed that $\bX$ follows a matrix-variate Gaussian distribution, e.g., \cite{Efron09,AllenTibshirani2010,LengTang2012,YinLi12,Zhou2014}. The matrix-variate Gaussian distribution is a generalization of the familiar multivariate normal distribution for vector-variate data. In particular, let $\text{vec}(\bX) \in \mathbb{R}^{pq \times 1}$ be the vectorization of matrix $\bX$ obtained by stacking the columns of $\bX$ on top of each other. We say that $\bX$ follows a matrix-variate Gaussian distribution $ \bX \sim N(\bmu, \S \otimes \ps) $ with mean matrix $\bmu \in \mathbb{R}^{p \times q}$, row covariance matrix $\S\in \mathbb{R}^{p \times p}$ and column covariance matrix $\ps \in \R^{q\times q}$ if and only  if $\mathrm{vec}(\bX') \sim N(\mathrm{vec}(\bmu'), \S \otimes \ps)$, where $\bX'$ denotes the transpose of $\bX$ and  $\otimes$ is the Kronecker product.

\pagestyle{fancy}
\renewcommand\headrulewidth{0pt}
\lhead{}
\chead{GRAPH ESTIMATION FOR MATRIX-VARIATE GGM}
\rhead{\thepage}
The readers might refer to  \cite{Dawid1981} and \cite{GuptaNagar1999} for more properties of matrix-variate Gaussian distribution.  Similar to the vector-variate Gaussian graphical models (GGMs) in which the conditional independence graph is encoded in the support of the precision matrix, one can analogously define matrix-variate Gaussian graphical models (MGGM) (a.k.a. Gaussian bigraphical models).  Let us denote \emph{conditional independence graph} by the undirected graph $G=(V,E)$, where $V=\{V_{ij}\}_{1 \leq i \leq p, 1 \leq j \leq q}$ contains $p \times q$ nodes and each node corresponds to an entry in  the random matrix $\mathbf{X}$. We can regard the edge set $E$ as a $pq \times pq$ matrix where  there is no edge between $X_{ij}$ and $X_{kl}$ if and only if $X_{ij}$ and $X_{kl}$ are conditionally independent given the rest of the entries. The goal of the \emph{graph estimation} is to estimate the edge set  $E$, which unveils important structural information on the conditional independence relationship.

The estimation of the conditional independence graph is equivalent to the estimation of \emph{the support of the precision matrix}. In particular, let  $\O=\S^{-1}=(\omega_{ik})_{p\times p}$, $\Ga=\ps^{-1}=(\gamma_{jl})_{q\times q}$. The precision matrix of the MGGM  is a $pq \times pq$ matrix $\O \otimes \Ga = (\S \otimes \ps)^{-1}$, where $(\O \otimes \Ga)_{q(i-1)+j,\; q(k-1)+l}= \omega_{ik}\cdot \gamma_{jl}$. The conditional independence among entries of $\bX$ can be presented by the support of $\O \otimes \Ga$ (denoted by $\text{supp}(\O \otimes \Ga)$), which is equivalent to $\text{supp}(\O) \otimes \text{supp}(\Ga)$. To see this, we recall  the well-known fact that $X_{ij}$ and $X_{kl}$ are conditionally independent given the rest of  the entries, if and only if $\varrho_{ij, kl}=0$, where $ \varrho_{ij, kl}$ is the partial correlation between $X_{ij}$ and $X_{kl}$:
 \begin{equation}\label{eq:partial}
 \varrho_{ij, kl} = - \frac{\omega_{ik}}{\sqrt{\omega_{ii}\omega_{kk}}} \cdot \frac{\gamma_{jl}}{\sqrt{\omega_{jj}\omega_{ll}}}.
\end{equation}
In other words, $X_{ij}$ and $X_{kl}$ are conditionally independent  if and only if there is at least one zero in $\omega_{ik}$ or $\gamma_{jl}$.
Therefore, to estimate the conditional independence graph, one only needs to estimate $\text{supp}(\O)$ and $\text{supp}(\Ga)$. Their Kronecker product $\text{supp}(\O) \otimes \text{supp}(\Ga)$ gives the edge set $E$.  It is worthwhile to note that for a given matrix-variate Gaussian distribution, multiplying a constant to $\bO$ and dividing $\bG$ by the same constant will lead to the same distribution.   The existing literature usually assumes that $\omega_{11}=1$ to make the model identifiable (see, e.g.,  \cite{LengTang2012}). However, if we are interested in  support recovery rather than values of $\omega_{ik}$ or $\gamma_{jl}$, then there is no identifiability issue.



%

Due to the complicated structure in the precision matrices of MGGMs,  research on matrix-variate GGMs (MGGMs) is scarce compared to the large body of literature on vector-variate GGMs. The vector-variate GGM with random vector observations can be viewed as a special case of MGGM with $p=1$ or $q=1$ and readers may refer to, e.g., \cite{MeinshausenBuhlmannP2006,YuanLin2007,RothmanBickelLevinaZhu2008,dAspremont08,
FriedmanHastieTibshirani2008,Yuan2010,RavikumarWainwrightRaskuttiYu2011,CaiLiuLuo2011,
LiuHanYuanLaffertyWasserman2012,XueZou2012,Liu2013,Zhu:14,FanY:16,Ren:16} for the recent development in vector-variate GGMs.  Among these works, our work is closely related to \cite{Liu2013}, which conducts graph estimation via false discovery rate (FDR) control for vector-variate GGMs. However, due to the complicated structure of MGGMs, the proposed test statistics are fundamentally different from the ones in \cite{Liu2013} and the theoretical analysis is much more challenging. The details on the comparisons to \cite{Liu2013} are deferred to Section \ref{sec:disc}.

For estimating sparse precision matrices of  matrix-variate Gaussian data, one popular approach is based on the penalized likelihood method. However, since the precision matrices are in the form of a Kronecker product, the negative log-likelihood function is no longer convex, which makes both computation and theoretical analysis significantly more challenging than in the case of classical vector-variate GGMs. A few recent works  \citep{AllenTibshirani2010,LengTang2012,YinLi12,Kalaitzis13Bi,Tsiligkaridis13,Ying13,HuangChen2015} have focused on  developing various penalized likelihood approaches for estimating MGGMs or extensions of MGGMs (e.g.,  multiple MGGMs in \cite{HuangChen2015} and semiparametric extension in \cite{Ying13}).  In particular, \cite{LengTang2012} provided pioneering theoretical guarantees on the estimated precision matrices, e.g., rate of convergence under the Frobenius norm and sparsistency. One limitation is that the theoretical results are stated in terms that there \emph{exists} a local minimizer that enjoys good properties. In practice, it might be difficult to determine whether the obtained local minimizer from an optimization solver is a desired local minimizer. 
In addition, most convergence results require certain conditions on the sample size $n$ and dimensionality $p$ and $q$, e.g., $p$ and $q$ cannot differ too much from each other and $n$ should go to infinity at a certain rate. On the other hand, we will show later that $n\rightarrow \infty$ is not necessary for the control of false discovery rate (FDR) in support recovery for MGGMs. \cite{Zhou2014} developed new penalized methods for estimating $\S \otimes \ps$ and $\O \otimes \Ga$ and established the convergence rate under both the spectral norm and the Frobenius norm. However, the main focus of \cite{Zhou2014} is not the support recovery of $\O \otimes \Ga$, and our goal of accurate FDR control cannot be achieved by the method in \cite{Zhou2014} nor other penalized optimization approaches.



The main goal of this work is to infer the support of the precision matrix for an MGGM in the high-dimensional setting, which fully characterizes the conditional independence relationship.
Our method differs from the common approaches that turn the problem into a joint optimization over $\O$ and $\Ga$ as penalized likelihood methods.  Instead, we utilize the large-scale testing framework and formulate the problem into multiple testing problems for $\O$ and $\Ga$:
\begin{equation}\label{eq:test_Omega}
  H^{\O}_{0ij}: \omega_{ij}=0 \quad \text{vs} \quad  H^{\O}_{0ij}: \omega_{ij} \neq 0,  \quad  1 \leq i < j \leq p
\end{equation}
and
\begin{equation}\label{eq:test_Gamma}
  H^{\Ga}_{0ij}: \gamma_{ij}=0 \quad \text{vs} \quad   H^{\Ga}_{0ij}: \gamma_{ij} \neq 0, \quad 1 \leq i < j \leq q.
\end{equation}
By conducting the multiple testing for  \eqref{eq:test_Omega} and \eqref{eq:test_Gamma}, we obtain the estimates for the support of $\O$ and $\Ga$, denoted by $\widehat{\mathrm{supp}(\O)}$ and $\widehat{\mathrm{supp}(\Ga)}$, respectively. Then, the support of $\O \otimes \Ga$  can be naturally estimated by $\widehat{\mathrm{supp}(\O)} \otimes \widehat{\mathrm{supp}(\Ga)}$. Instead of aiming for perfect support recovery, which will require strong conditions, our goal is to asymptotically control the false discovery rate (FDR). The FDR, originally introduced for multiple testing \citep{Benjamini95}, has  been considered one of the most important criterion for evaluating the quality of estimated networks (e.g., in the application of genetic data analysis \citep{SchaferStrimmer2005,MaGongBohnert2007}). Please refer to \eqref{eq:FDP_FDR_1} and \eqref{eq:FDP_FDR_2} in Section \ref{sec:FDR_theory} for the definition of FDR in our graph estimation problems. 

Although conducting variable selection via multiple testing is not a new idea, how to implement such a high-level idea for MGGMs with a complicated covariance structure requires several innovations in the
methodology development.  In particular,  to conduct the multiple testing in \eqref{eq:test_Omega} and \eqref{eq:test_Gamma}, it is critical to construct a test statistic
with the explicit asymptotic null distribution for each edge. To this end, we propose a new approach that fully utilizes the correlation structures among rows and columns of $\mathbf{X}$. In particular, suppose that there are $n$ \emph{i.i.d.} $p\times q$ matrix-variate samples $\mathbf{X}^{(1)}, \ldots, \mathbf{X}^{(n)}$.  To conduct the testing in \eqref{eq:test_Gamma} and estimate the support of  $q \times q$ matrix $\Ga$, we treat  each row of $\mathbf{X}^{(k)}$ as a $q$-dimensional sample. In such a way, we construct $n\cdot p$ \emph{correlated} vector-variate samples for the testing problem in  \eqref{eq:test_Gamma}, where the correlation among these ``row samples" is characterized by the covariance matrix $\S$.  One important advantage of this approach is that it only requires number of row samples $np \rightarrow \infty$  to control FDR asymptotically and thus allows the finite sample size $n$ even when $p$ and $q$ go to infinity.  On the other hand, the correlation structure among row samples also presents a significant challenge to the development of the FDR control approach, and  most existing inference techniques for vector-variate GGMs heavily rely on the independence assumption (see, e.g., \cite{Liu2013, vandegeer2014, RenSunZhangZhou2015}). To address this challenge, we summarize the effect of correlation among ``row samples" into a simple quantity depending on $\S$, and based on that, introduce a variance correction technique into the construction of the test statistics (see Section \ref{sec:test_stat}).  It is noteworthy that the testing of the $\text{supp}(\O)$ in \eqref{eq:test_Omega} can be performed in a completely symmetric way with $nq$ correlated ``column samples" from the data.

More specifically, the high-level description of the proposed large-scale testing approach is as follows.  Given the ``row samples'' from the data, the first step is to construct an asymptotically normal test statistic for each $H_{0ij}^{\Ga}$ in \eqref{eq:test_Gamma}.
 \ignore{Let us recall the property of partial correlation coefficient $\rho^{\bG}_{ij\cdot}=-\frac{\gamma_{ij}}{\sqrt{\gamma_{ii}}\sqrt{\gamma_{jj}}}$ and notice that $\gamma_{ij}=0$ is equivalent to $\rho^{\bG}_{ij\cdot}=0$. To test whether $\rho^{\bG}_{ij\cdot}=0$,} We utilize a fundamental result from multivariate analysis  which relates the partial correlation coefficient to the correlation coefficient of residuals from linear regression. To compute the  sample version of the correlation coefficient of the residuals,  we first construct an initial estimator for the regression coefficients.  With the initial estimator in place, we can directly show that the sample correlation coefficient of the residuals is asymptotically normal under the null. We further apply the aforementioned variance correction technique and obtain the final test statistic for each $H_{0ij}^{\Ga}$. Combining the developed test statistics with the Benjamini and Hochberg approach \citep{Benjamini95},  we show that the resulting procedure asymptotically controls the FDR for both $\widehat{\mathrm{\supp}(\Ga)}$ and $\widehat{\mathrm{\supp}(\O)}$ (and thus $\widehat{\mathrm{\supp}(\Ga)} \otimes \widehat{\mathrm{\supp}(\O)}$) under some sparsity conditions of $\Ga$ and $\O$.


 The proposed method described above is the first to investigate FDR control in MGGMs. This work greatly generalizes the method on FDR control for \emph{i.i.d.} vector-variate GGMs in \cite{Liu2013} and improves over optimization-based approaches. The main contribution and difference between our results and the existing ones for vector-variate GGMs (e.g., \cite{Liu2013}) are summarized as follows,
\begin{enumerate}
  \item  We propose a novel \emph{test statistic} in Section \ref{sec:test_stat}. By introducing a new construction of the initial regression coefficients (i.e.,  setting a particular element in each initial Lasso estimator to zero), our testing approach no longer requires a complicated bias-correlation step as in Eq. (6) in \cite{Liu2013}. Furthermore, the  limiting null distribution of the sample covariance coefficient between residuals (see $\hat{r}_{ij}$ in \eqref{eq:def_eps}) can be easily obtained.  In fact, this idea can be used to provide simpler testing procedure
      for ordinary vector-variate high-dimensional graphical models.

  \item   Instead of relying on the \emph{i.i.d.} assumption in GGM literature, we propose to extract $np$ \emph{correlated} vector-variate ``rows samples" (as well $nq$ correlated ``column samples") from  matrix-variate observations. By utilizing correlation structure among rows and columns,  our approach allows for finite sample size, which is a very attractive property from both theoretical and practical perspectives. More specifically, even in the case that $n$ is a constant and $p \rightarrow \infty$ and $q \rightarrow \infty$, our method still guarantees the asymptotic normality of  the test statistics and FDR control. This is fundamentally different from the case of vector-variate GGMs, which always requires $n \rightarrow \infty$ for the support recovery. %
      Therefore, this work greatly generalizes \cite{Liu2013}, which only deals with \emph{i.i.d.} vector-variate  Gaussian samples, to correlated data.

      In this paper, we developed new techniques and theoretical analysis to address the challenges arising from correlated samples. For example, the proposed \emph{variance correlation technique} can be used as a general technique for high-dimensional inference with correlated samples.  Moreover, the initial estimator is now based on the Lasso with correlated samples. To this end, we establish the  consistency result for the Lasso with correlated samples, which itself is of independent interest for high-dimensional linear regression.


  \item  Theoretically, by utilizing the  Kronecker product structure of the covariance matrix of $\textbf{X}$, the proposed method  allows the partial correlation between $X_{ij}$ and $X_{kl}$ (i.e.,  $\varrho_{ij, kl}$ in \eqref{eq:partial}) to be of the order of $\frac{1}{n-1}\sqrt{\frac{\log p \log q}{pq}}$ so that the corresponding edge can be detected (please see the power analysis in Section \ref{sec:power} and Theorem \ref{th4} for details.) This is essentially different from any vector-variate GGM estimator (e.g., the one from \cite{Liu2013}) that requires the partial correlation to be at least $C \frac{1}{\sqrt{n}}$. 
\end{enumerate}

Moreover, in terms of support recovery and computation cost, the proposed method  has several advantages as compared to  popular penalized likelihood approaches:
\begin{enumerate}
  \item As we mentioned before, the FDR is a basic performance measure of support recovery of MGGMs. The proposed method provides an accurate control of FDR (see Theorem \ref{th3}); however, the existing optimization-based approaches remain unclear about how to choose the tuning parameter to control a desired FDR while keeping nontrivial statistical power. In fact, existing support recovery results only guarantee that the estimated positions of zeros are supersets of the positions of true zeros in $\O$ and $\Ga$ with probability tending to one when $n, p, q$ all go to infinity at certain rates. (see, e.g., \cite{LengTang2012, YinLi12}).
  \item In terms of computation, as compared to  existing penalized likelihood methods whose objective functions are non-convex, our approach is completely based on convex optimization and thus computationally more attractive. In particular, the main computational cost of our approach is the construction of initial estimates for $p+q$  regression coefficient vectors, which will directly lead to our test statistics. The corresponding computational problems are completely independent allowing an efficient parallel implementation.
  \item Theoretically, our approach allows a wider range of $p$ and $q$. In particular, our result on FDR control holds for $(p,q)$ such that $q^{r_2} \leq p \leq q^{r_1}$ for some $0< r_2 \leq r_1$, while in comparison, \cite{LengTang2012} require that $p\log(p) =o(nq)$ and $q \log (q) =  o(np)$. As one can see, if $n$ is a constant or $n \rightarrow \infty$ but $n=o(\min(\log p, \log q))$, such a condition will not hold.
\end{enumerate}


\ignore {Before we conclude the introduction, we briefly review some related works for vector-variate GGMs. When $p=1$ or $q=1$, MGGMs reduce to vector-variate GGMs, which have been extensively studied in the high-dimensional setting over the past ten years. The major research problems in high-dimensional GGMs include the estimation and support recovery of precision matrices, which have been investigated by \cite{MeinshausenBuhlmannP2006,YuanLin2007,RothmanBickelLevinaZhu2008,Yuan2010,dAspremont08,FriedmanHastieTibshirani2008,RavikumarWainwrightRaskuttiYu2011,CaiLiuLuo2011}, etc.
\cite{LiuHanYuanLaffertyWasserman2012} and \cite{XueZou2012} further extended GGMs to a more general class of multivariate distributions, i.e., semiparametric Gaussian copula graphical models. Recently, \cite{Liu2013,RenSunZhangZhou2015,Jankova15} developed various approaches to address statistical inference problem in high-dimensional GGMs and established asymptotic normality for elements of the precision matrix. }

\section{Notations and Organization of the paper}

We introduce some necessary notations. Let $\bX^{(k)}=(X^{(k)}_{ij})_{p\times q}$ for $k=1, \ldots, n \;$ be the $n$ \emph{i.i.d.} matrix-variate observations from $N_{p,q}(\bmu, \S \otimes \ps)$ and let  $\bar{\bX}=\frac{1}{n} \sum_{k=1}^n \bX^{(k)}$. Put $\S=(\sigma_{ij})$ and $\ps=(\psi_{ij})$. For any vector $\x=(x_{1},\ldots,x_{p})^{'}$, let $\x_{-i}$ denote $p-1$ dimensional vector by removing $x_{i}$ from $\x$. Let $\langle \bx, \by \rangle$ be the inner product of two vectors $\bx$ and $\by$. For any $p\times q$ matrix $\A$, let $\A_{i, \cdot}$ denote the $i$-th row of $\A$ (or $\A_i$ when it is clear from the context)  and $\A_{\cdot,j}$  denote the $j$-th column of $\A$ (or $\A_j$ when it is clear from the context). Further, let $\A_{i,-j}$ denote the $i$-th row of $\A$ with its $j$-th entry being removed and $\A_{-i,j}$ denotes the $j$-th column of $\A$ with its $i$-th entry being removed. $\A_{-i,-j}$ denote a $(p-1)\times (q-1)$ matrix by removing the $i$-th row and $j$-th column of $\A$. Define $[n] = \{1,\ldots, n\}$, $[p] = \{1,\ldots, p\}$ and $[q]=  \{1,\ldots, q\}$.
For a $p$-dimensional vector $\bx$, let $|\bx|_0=\sum_{j=1}^p I(x_j \neq 0)$,  $|\bx|_1=\sum_{j=1}^p |x_j|$ and $|\bx|_2=\sqrt{\sum_{j=1}^p x_j^2}$ be the $\ell_0$, $\ell_1$ and Euclidean norm of $\bx$, respectively. For a matrix $\A=(a_{ij}) \in \R^{p \times q}$, let $\|\A\|_{\text{F}}=\sqrt{\sum_{i\in[p], j \in [q]} a_{ij}^2}$ be the Frobenius norm of $\A$,  $|\A|_{\infty} =\max_{i \in [p], j \in [q]} |a_{ij}|$ be the element-wise $\ell_\infty$-norm of $\A$ and $\|\A\|_2= \sup_{|\x|_2 \leq 1} |\A\x|_2$ be  the spectral  norm of  $\A$. For a square matrix $\A$, let $\text{tr}(\A)$ denote the trace of $\A$. 
For a given set $\mathcal{H}$, let $Card(\mathcal{H})$ be the cardinality of $\mathcal{H}$. Throughout the paper, we use $\mathbf{I}_p$ to denote the $p \times p$ identity matrix, and  use
$C$, $c$, etc. to denote generic constants whose values might change from place to place.

The rest of the paper is organized as follows. In Section \ref{sec:method}, we introduce our test statistics and then describe the FDR control procedure for MGGM estimation. Theoretical results on asymptotic normality of our test statistic and FDR control are given in Section \ref{sec:theory}. Simulations and real data analysis are given in Section \ref{sec:exp}. In Section \ref{sec:disc}, we provide further discussions on the proposed method and also point out some interesting future work. All of the technical proofs and some additional experimental results are relegated to the supplementary material.


\section{Methodology}
\label{sec:method}

Recall the definition of false discover proportion (FDP) as the proportion of false discoveries among total rejections. Note that, if $\widehat{\text{supp}(\O)}$ and $\widehat{\text{supp}(\Ga)}$ are the estimators of supp$(\O)$ and supp$(\Ga)$, respectively, under the control of FDP at level $\alpha  \in [0,1] $, it is clear that the FDP of $\widehat{\text{supp}(\O)}\otimes\widehat{\text{supp}(\Ga)}$ as an estimator of supp($\O\otimes \Ga$) will be controlled at some level $\alpha{'}$. Here, the level $\alpha'$ (explicitly given later in \eqref{eq:FDP_est}) is a monotonically increasing function in $\alpha$. Therefore, we reduce our task to design an estimator of supp$(\Ga)$ under the FDP level $\alpha$ and the estimator of supp$(\O)$ can be constructed in a completely symmetric way. We propose to estimate supp$(\Ga)$ by implementing the following large-scale multiple tests:
\begin{eqnarray}\label{eq:multi_test}
H_{0ij}: \gamma_{ij}=0\quad\mbox{vs.}\quad H_{1ij}: \gamma_{ij}\neq 0,\quad 1\leq i<j\leq q.
\end{eqnarray}

\subsection{Construction of test statistics }
\label{sec:test_stat}

In this section, we propose our test statistic for each $H_{0ij}$ in \eqref{eq:multi_test} constructed from $n$ \emph{i.i.d.} $p\times q$ matrix-variate samples $\mathbf{X}^{(1)}, \ldots, \mathbf{X}^{(n)}$ with the population distribution $\mathbf{X} \sim N(\bmu, \S \otimes \ps)$. Let us denote the partial correlation matrix associated with $\Ga$ by
$\brho^{\bG}=\left(\rho^{\bG}_{ij\cdot}\right)_{q \times q}$,
where each  $\rho^{\bG}_{ij\cdot}=-\frac{\gamma_{ij}}{\sqrt{\gamma_{ii}\gamma_{jj}}}$ is the partial correlation coefficient between $X_{li}$ and $X_{lj}$ for any $1 \leq l \leq p$.
\ignore{Note that each of the null hypothesis $H_{0ij}$ in \eqref{eq:multi_test} can also be stated as $H_{0ij}: \rho^{\bG}_{ij\cdot}=0$.} The following well-known result relates the partial correlation coefficient $\rho^{\bG}_{ij\cdot}$ to regression problems. In particular, for $1 \leq i< j \leq q$ and any $1 \leq l \leq p$, let us define the population regression coefficients:
\begin{align}\label{eq:reg_model}
(\alpha_i, \be_i ) & = \argmin_{a \in \mathbb{R}, \bb \in \mathbb{R}^{q-1}}   \E (X_{li} - a - \bX_{l, -i } \bb)^2
\;,  (\alpha_j, \be_j) & = \argmin_{a \in \mathbb{R}, \bb \in \mathbb{R}^{q-1}, }  \E (X_{lj} - a- \bX_{l, -j } \bb)^2. \end{align}
The standard linear regression result shows that
\begin{equation}\label{eq:popu_be}
  \be_i= -\gamma^{-1}_{ii}\Ga_{-i,i} \; ;  \qquad  \be_j= -\gamma^{-1}_{jj}\Ga_{-j,j}.
\end{equation}
For such defined $\be_i$ and $\be_j$, the corresponding residuals $\epsilon_{li}$ and $\epsilon_{lj}$ take the following form,
\begin{align}\label{eq:eps}
  \varepsilon_{li}=X_{li} -  \alpha_i - \bX_{l, -i } \be_i \; ;  \quad \varepsilon_{lj}=X_{lj} -  \alpha_j - \bX_{l, -j } \be_j.
\end{align}
It is known that $\E(\varepsilon_{li})=\E(\varepsilon_{lj})=0$. Moreover, the correlation between $\varepsilon_{li}$ and $\varepsilon_{lj}$ is $\text{Corr}(\varepsilon_{li},\varepsilon_{lj})=-\rho^{\bG}_{ij\cdot}= \frac{\gamma_{ij}}{\sqrt{\gamma_{ii}\gamma_{jj}}}$. To see this, let $\bgamma_i$ be the $i$-th column of $\Ga$ for $1 \leq i \leq q$. By \eqref{eq:popu_be} and \eqref{eq:eps}, we can equivalently write $\epsilon_{li}=-\alpha_{i}+\gamma_{ii}^{-1}\bX_{l,\cdot} \bgamma_{i}$ and $\epsilon_{lj}=-\alpha_{j}+\gamma_{jj}^{-1}\bX_{l, \cdot} \bgamma_{j}$. Since the covariance of $\bX_{l, \cdot}$, $\text{Cov}(\bX_{l, \cdot})$, is $\sigma_{ll} \ps = \sigma_{ll} \Ga^{-1}$, we have
\begin{equation}\label{eq:cov_i_j}
\mathrm{Cov}(\epsilon_{li},\epsilon_{lj})=\gamma_{ii}^{-1}\gamma_{jj}^{-1} \bgamma_{i}^T\text{Cov}(\bX_{l, \cdot})\bgamma_{j} = \sigma_{ll}\gamma_{ii}^{-1}\gamma_{jj}^{-1} \left(\bgamma_i^T \Ga^{-1}\right) \bgamma_j =\sigma_{ll}\gamma_{ii}^{-1}\gamma_{jj}^{-1}  \gamma_{ij},
\end{equation}
where the last equality is because $\bgamma_{i}^T\bG^{-1}=\mathbf{e}_{i}^T$ and  $\mathbf{e}_{i}$ denotes the $i$-th canonical vector. Similarly, we obtain that $\mathrm{Var}(\epsilon_{li})=\sigma_{ll}\gamma_{ii}^{-1}$ and $\mathrm{Var}(\epsilon_{lj})=\sigma_{ll}\gamma_{jj}^{-1}$, which together with \eqref{eq:cov_i_j} imply that $\text{Corr}(\varepsilon_{li},\varepsilon_{lj})= \frac{\gamma_{ij}}{\sqrt{\gamma_{ii}\gamma_{jj}}}$.
%
Therefore, testing whether $\gamma_{ij}=0$ (or $\rho^{\bG}_{ij\cdot}=0$) is equivalent to testing whether $\text{Corr}(\varepsilon_{li},\varepsilon_{lj})=0$. We will build our test statistics based on this key equivalence relationship.


To implement the aforementioned idea, one needs to construct an initial estimator for each $\be_j$  so that the distribution of sample correlation coefficient of residuals can be deduced easily. To construct the initial estimator, we first
let $\hat{\be}_{j}=(\hat{\beta}_{1,j},\ldots,\hat{\beta}_{q-1,j})^{'}$ be any estimator for $\be_{j}$, which satisfies
\begin{eqnarray}\label{bt1}
\max_{1\leq j\leq q}|\hat{\be}_{j}-\be_{j}|_{1}=O_{\pr}(a_{n1}),\quad\mbox{and}\quad \max_{1\leq j\leq q}|\hat{\be}_{j}-\be_{j}|_{2}=O_{\pr}(a_{n2}),
\end{eqnarray}
where $a_{n1} \rightarrow 0$ and $a_{n2} \rightarrow 0$ at some rate that will be specified later.
The Lasso \citep{Tibshirani:96}, Dantzig selector \citep{CandesETao2007}, or other sparse regression approaches can be adopted provided that \eqref{bt1} is satisfied (see Section \ref{sec:init} for details). Under the null $H_{0ij}$: $\gamma_{ij}=0$, according to \eqref{eq:popu_be},  the $i$-th element in  $\be_j$, which corresponds to the covariate $X_{li}$ in \eqref{eq:eps}, is zero;  and the $(j-1)$-th element in $\be_i$ which corresponds to the covariate $X_{lj}$ in \eqref{eq:eps}, is also zero. Hence, for each pair $i<j$, we construct the initial estimators  $\hat{\be}_{j,\backslash i}=(\hat{\be}_{1,j},\ldots,\hat{\be}_{i-1,j},0,\hat{\be}_{i+1,j},\ldots,\hat{\be}_{q-1,j})^{'}$ and $\hat{\be}_{i,\backslash  j}=(\hat{\be}_{1,i},\ldots,\hat{\be}_{j-2,i},0,\hat{\be}_{j,i},\ldots,\hat{\be}_{q-1,i})^{'}$  for $\be_{j}$ and $\be_{i}$, respectively.
For notation briefness, we let $\hat{\be}_{j,\backslash j}=\hat{\be}_{j}$ so that the ``sample residual" $\hat{\varepsilon}^{(k)}_{ljj}$ introduced in the below  is also well-defined (see \eqref{eq:def_eps}). 

 Given the initial estimators under the null, we construct the ``sample residuals" by treating each row $l \in [p]$ of a matrix-variate observation $\mathbf{X}^{(k)}$ for $k\in [n]$ as a ``row sample", i.e., $\mathbf{X}^{(k)}_{l,\cdot}=(X^{(k)}_{l1}, \ldots X^{(k)}_{lq})$.
The ``sample residuals" corresponding to  $\epsilon_{li}$ and $\epsilon_{lj}$ in \eqref{eq:eps}  are defined as follows:
\begin{eqnarray}\label{eq:def_eps}
\hat{\varepsilon}^{(k)}_{lij}=X^{(k)}_{li}-\bar{X}_{li}-(\textbf{X}^{(k)}_{l,-i}-\bar{\textbf{X}}_{l,-i})\hat{\be}_{i,\setminus j}, \quad
\hat{\varepsilon}^{(k)}_{lji}=X^{(k)}_{lj}-\bar{X}_{lj}-(\textbf{X}^{(k)}_{l,-j}-\bar{\textbf{X}}_{l,-j})\hat{\be}_{j,\setminus i},
\end{eqnarray}
where $\bar{X}_{li}=\frac{1}{n}\sum_{k=1}^{n}X^{(k)}_{li}$  and $\bar{\textbf{X}}_{l,-i}=\frac{1}{n}\sum_{k=1}^{n}\textbf{X}^{(k)}_{l,-i}$. 
Further let $\hat{r}_{ij}$ be the sample covariance coefficient between constructed residuals,
\begin{eqnarray}\label{eq:r}
\hat{r}_{ij}=  \frac{1}{(n-1)p}\sum_{k=1}^{n}\sum_{l=1}^{p} \hat{\varepsilon}^{(k)}_{lij}\hat{\varepsilon}^{(k)}_{lji};
\end{eqnarray}
and $\frac{\hat{r}_{ij}}{\sqrt{\hat{r}_{ii}\hat{r}_{jj}}}$ be the corresponding sample correlation coefficient of residuals.

The proposed construction of the sample correlation of residuals has twofold benefits. First, under the null $H_{0ij}$, incorporating the fact that $\gamma_{ij}=0$ into the regression coefficients enables us to derive the asymptotic null distribution  of the sample correlation coefficients. In particular, later in Proposition \ref{prop}, we show that under the null,
\begin{eqnarray}\label{eq:asy_normality}
\sqrt{\frac{(n-1)p}{A_{p}}}\frac{\hat{r}_{ij}}{\sqrt{\hat{r}_{ii}\hat{r}_{jj}}}\Rightarrow N(0,1),
\end{eqnarray}
where
\begin{equation}\label{eq:def_A_p}
  A_{p}=\frac{p\|\S\|^{2}_{\text{F}}}{(\text{tr}(\S))^{2}}.
\end{equation}
Here, $A_p$ is the asymptotic variance of $\sqrt{(n-1)p} \frac{\hat{r}_{ij}}{\sqrt{\hat{r}_{ii}\hat{r}_{jj}}}$ under the null.
It is noteworthy that the term $A_p$ is critical since it plays the role of \emph{variance correction} when treating rows of matrix-variate data as correlated samples. \ignore{In fact, when rows of $\mathbf{X}$ are \emph{i.i.d.} and thus $\S$ is a diagonal matrix with the same value, we have $A_p=1$.}
The second merit of the proposed construction is that, although the sample correlation coefficients are constructed under the null, we can show that $\frac{\hat{r}_{ij}}{\sqrt{\hat{r}_{ii}\hat{r}_{jj}}}$ converges to $(1-\gamma_{ij}\psi_{ij})\rho^{\Ga}_{ij\cdot}$ in probability as $np \rightarrow \infty$  under both the  null and alternative (see Proposition \ref{propv}). This result indicates that the test statistic based on $\frac{\hat{r}_{ij}}{\sqrt{\hat{r}_{ii}\hat{r}_{jj}}}$ can properly reject the null when the magnitude of partial correlation coefficient $|\rho^{\Ga}_{ij\cdot}|$ is away from zero.  It is worth noting that for many widely studied covariance structures of $\ps$ and $\Ga=\ps^{-1}$, the quantity $\psi_{ij}\gamma_{ij}$ is often non-positive, which makes the signal strength $(1-\gamma_{ij}\psi_{ij})\rho^{\Ga}_{ij\cdot}$ even larger than the partial correlation coefficient and thus leads to good statistical power.
For example, for two variables which are directly positively/negatively correlated, they will often be  positively/negatively conditional correlated.


It is worthwhile to note that the variance correction quantity $A_p \in \mathbb{R}$ in \eqref{eq:asy_normality} is unknown as it involves $\S$. In the next subsection, we will propose a ratio consistent estimator $\hat{A}_{p}$ such that $\hat{A}_{p}/A_p \rightarrow 1$ in probability as $nq \rightarrow \infty$. Using the estimator $\hat{A}_p$, we construct the final test statistic for $H_{0ij}$:
\begin{eqnarray}\label{t1}
\hat{T}_{ij}=\sqrt{\frac{(n-1)p}{\hat{A}_{p}}}\frac{\hat{r}_{ij}}{\sqrt{\hat{r}_{ii}\hat{r}_{jj}}},
\end{eqnarray}
which is asymptotically normal given \eqref{eq:asy_normality} and the ratio consistency of $\hat{A}_p$.

\subsubsection{Estimator for $A_p$}
We propose an estimator $\hat{A}_{p}$ of $A_p$ based on a thresholding estimator of $\S$. We first construct an initial estimator of $\S$ based on $nq$ ``column samples", where each column of $\mathbf{X}^{(k)}$ for $k \in [n]$ is treated as a $p$-dimensional sample. In particular, let $\bar{\textbf{X}}=\frac{1}{n}\sum_{k=1}^{n}\textbf{X}^{(k)}$,  each centered column  sample $\Y_{kj}=\X_{\cdot, j}^{(k)}-\bar{\X}_{\cdot, j} \in \mathbb{R}^{p \times 1}$ for $k \in [n]$ and $j \in [q]$. Moreover, let us define $\hat{\S}=(\hat{\sigma}_{ij})_{p\times p}:=\frac{1}{(n-1)q}\sum_{k=1}^{n} \sum_{j=1}^q (\Y_{kj})(\Y_{kj})'$. In a more succinct notation, let $\textbf{Y}=[\textbf{X}^{(1)}-\bar{\textbf{X}},\ldots,\textbf{X}^{(n)}-\bar{\textbf{X}}]\in \mathbb{R}^{p\times(nq)}  $and  $\hat{\S} = \frac{1}{(n-1)q}\textbf{Y}\textbf{Y}'$. Then, we  threshold the elements of $\hat{\S}$ as follows:
\begin{equation}\label{eq:thresh_sigma}
\hat{\sigma}_{ij,\lambda}=\hat{\sigma}_{ij}I\left\{|\hat{\sigma}_{ij}|\geq \lambda\sqrt{\frac{\log \max(p,nq)}{nq}}\right\} \quad \text{for} \; i \neq j;
\end{equation}
and $\hat{\sigma}_{ii,\lambda}=\hat{\sigma}_{ii}$ for $i \in [p]$. Set $\hat{\S}_{\lambda}=(\hat{\sigma}_{ij,\lambda})_{p\times p}$ and define the plug-in estimator of $A_p$:
\begin{equation}\label{eq:hat_A_p}
\hat{A}_{p}=\frac{p\|\hat{\S}_{\lambda}\|^{2}_{\text{F}}}{(\text{tr}(\hat{\S}_{\lambda}))^{2}}.
\end{equation}
In Proposition \ref{prop2} in Section \ref{sec:theory}, we show that $\hat{A}_{p}/A_p \rightarrow 1$ in probability as $nq \rightarrow \infty$ for a properly chosen $\lambda$ (a data-driven approach for the choice of $\lambda$ will be discussed later in Section \ref{sec:init}). Therefore, by \eqref{eq:asy_normality}, we have the desired asymptotic normality of $\hat{T}_{ij}$ under the null: $\hat{T}_{ij}\Rightarrow N(0,1)$ as $np\rightarrow\infty$.

We further note that when the columns of $\textbf{X}$ are \emph{i.i.d.} (i.e., $\ps=\mathbf{I}_q)$,  consistency under the spectral norm  $\|\hat{\S}_{\lambda}-\S\|_2$ has been established if the sparsity condition of $\S$ satisfies the row sparsity level $s_{0}(p)=O(\sqrt{nq/\log p})$   (see, e.g., \cite{bickel08regularized,Cai2011Adaptive} and references therein). We  do not need such a strong consistency result of $\hat{\S}_{\lambda}$ in the spectral norm to establish the consistency of $\hat{A}_p$. In fact, since $A_p$ only involves $\|\S\|_{\text{F}}$ rather than $\S$ itself, the sparsity condition on $s_{0}(p)$ is no longer necessary; see Proposition \ref{prop2} and its proof for more details.
It is also noteworthy that some other estimators of $\|\S\|_{\text{F}}$ have been proposed (e.g,. by \cite{Chen10Two}). However, those approaches heavily rely on the \emph{i.i.d.} assumption on the  columns of $\textbf{X}$, which is not valid in our setting. 


\subsection{Initial estimators of regression coefficients}
\label{sec:init}

In the construction of the test statistic $\hat{T}_{ij}$, we need the estimate $\hat{\be}_{j}$ that satisfies the condition in \eqref{bt1}. Here, we choose to construct  $\hat{\be}_{j}$ using Lasso for the presentation simplicity, however other approaches such as the Dantzig selector can also be used.  In particular,
let $\textbf{Z}=\left[ (\textbf{X}^{(1)})'-\bar{\textbf{X}}^{'},\ldots,(\textbf{X}^{(n)})'-\bar{\textbf{X}}^{'}\right]\in \R^{q\times(np)}$ be $np$ $q$-dimensional samples extracted from the data, and $\hat{\ps}=\frac{1}{(n-1)p}\textbf{Z}\textbf{Z}^{'}=:(\hat{\psi}_{ij})_{q\times q}$. For $1\leq j\leq q$, define the  scaling/normalizing vector $\D_{j}=\diag(\hat{\ps}_{-j,-j}) \in \mathbb{R}^{(q-1) \times (q-1)}$. The coefficients $\be_{j}$ can be estimated by Lasso as follows:
\begin{eqnarray}\label{so2-2}
\hat{\be}_{j}(\delta)=\D_{j}^{-1/2}\hat{\bal}_{j}(\delta),
\end{eqnarray}
where
\begin{eqnarray}\label{eq:lasso_alpha}
\hat{\bal}_{j}(\delta)=\argmin_{\bal\in \mathbb{R}^{q-1}} \Big{\{} \frac{1}{2np}\sum_{k=1}^{n}\sum_{l=1}^{p}(X^{(k)}_{lj}-\bar{X}_{lj}-(\textbf{X}^{(k)}_{l,-j}-\bar{\textbf{X}}_{l,-j})\D_{j}^{-1/2}\bal)^{2}+\theta_{nj}(\delta)|\bal|_{1}\Big{\}}
\end{eqnarray}
and
\begin{eqnarray}\label{a10}
\theta_{nj}(\delta)=\delta\sqrt{\frac{\hat{\psi}_{jj}\log \max(q,np)}{np}}.
\end{eqnarray}
In the Lasso estimate in \eqref{eq:lasso_alpha}, the $np$ covariates-response pairs $\left(\textbf{X}^{(k)}_{l,-j}-\bar{\textbf{X}}_{l,-j} \;, \;  X^{(k)}_{lj}-\bar{X}_{lj} \right)$ for $k \in [n]$ and $j\in [p]$ are not \emph{i.i.d.} and thus the standard consistency result of Lasso cannot be applied here.  By exploring the correlation structure among rows of $\bX$, we managed to derive the rate of convergence of the Lasso estimator in the $\ell_1$ and the $\ell_2$ norms. This result (see Proposition \ref{pro3-3} and its proof) might be of independent interest for dealing with high-dimensional correlated data. For the choice of tuning parameters,  our theoretical results will hold for any large enough constants $\lambda$ in \eqref{eq:thresh_sigma} for estimating $\hat{A}_p$ (see Proposition \ref{prop2}) and $\delta>0$ in \eqref{a10} for $\hat{\be}_{j}(\delta)$ (see Proposition \ref{pro3-3}). In our experiment, we will adopt a data-driven parameter-tuning strategy from \cite{Liu2013}.


\subsection{FDR control procedure}
\label{sec:FDR_control}
Given the constructed test statistic $\hat{T}_{ij}$, we can carry out $(q^2-q)/2$ tests in \eqref{eq:multi_test} simultaneously using the popular Benjamini and Hochberg (BH) method \citep{Benjamini95}. Let the $p$-values $p_{ij}=2-2\Phi(|\hat{T}_{ij}|)$ for $1\leq i<j\leq q$. We sort these $m=(q^{2}-q)/2$ $p$-values such that $p_{(1)}<\ldots<p_{(m)}$. For a given $0<\alpha<1$, define
\begin{eqnarray*}
\hat{k}=\max\{0\leq k\leq m: p_{(k)}\leq \alpha k/m\}.
\end{eqnarray*}
\ignore{the decision rule  that rejects $H_{0ij}$ when $|\hat{T}_{ij}| \geq t$ ,  it is ideal to select the smallest threshold level $t$ to detect as many nonzero $\gamma_{ij}$ as possible while controlling the FDR/FDP. Recall the definition of FDP and FDR for the multiple testing of $\text{supp}(\Ga)$:
\begin{equation}\label{eq:FDP_Ga}
  \text{FDP}_1(t)=\frac{\sum_{(i,j) \in \mathcal{H}_0}I\{|\hat{T}_{ij}|\geq t \} }{ \max\{\sum_{1\leq i <j\leq q}I\{|\hat{T}_{ij}|\geq t\},1\}};  \quad \text{FDR}_1(t)=\E\left[\text{FDP}_1(t)\right],
\end{equation}
where $\mathcal{H}_0:=\{(i,j): \gamma_{ij}=0, 1 \leq i < j \leq q\}$. A natural choice of the threshold $t$ will be $\hat{t}_{\mathrm{orc}} = \inf\left\{t >0: \text{FDP}_1(t) \leq \alpha \right\}$, where the infimum is taken in order to reject as many hypotheses as possible. However, the oracle $\hat{t}_{\mathrm{orc}}$ cannot be computed since $\mathcal{H}_0$ is unknown in FDP$_1(t)$. In the spirit of the BH method, we use $\mathbb{P}(|N(0,1)| \geq t) (q^2-q)/2 =(1-\Phi(t)) (q^2-q)$ to replace the unknown term $\sum_{(i,j) \in \mathcal{H}_0}I\{|\hat{T}_{ij}|\geq t \}$ in FDP$_1(t)$. Here, $\Phi(t)$ denotes the CDF for standard normal distribution.  When $\Ga$ is sparse, the number of true null hypotheses $|\mathcal{H}_0|$ is close to $(q^2-q)/2$ and thus such an approximation is reasonable. In summary, we describe the procedure to conduct the multiple testing in \eqref{eq:test_Gamma} and estimate $\text{supp}(\Ga)$ as follows. }\vspace{-5mm}
\begin{center}
\begin{boxedminipage}{1.0\textwidth}
For $1\leq i\neq j\leq q$, we reject $H_{0ij}$ if $p_{ij}\leq p_{(\hat{k})}$ and the estimated support of $\Ga$ is
\begin{equation}\label{eq:supp_Ga}
\widehat{\text{supp}(\Ga)}=\{(i,j): p_{ij}\leq p_{(\hat{k})}, 1 \leq i \neq j \leq q\}\cup\{(i,i): 1\leq i\leq q\}.
\end{equation}
Note that we set $\hat{T}_{ji}=\hat{T}_{ij}$ for $1 \leq i<j \leq q$ in \eqref{eq:supp_Ga}.
\end{boxedminipage}
\end{center}

Note that the original results from \cite{Benjamini95} cannot be directly applied to obtain the guarantee of FDR control since the test statistics (and thus the $p$-values) are correlated with each other. By utilizing some proof techniques developed by \cite{Liu2013}, we manage to prove that this procedure controls the FDR/FDP asymptotically in $\widehat{\text{supp}(\Ga)}$ (see Section \ref{sec:FDR_theory} for details).

{\bf Estimation of supp($\O$)}. The estimation of $\mathrm{supp}(\O)$ can be done with  a completely symmetric procedure.  In particular, we only need to consider the transpose of each matrix-variate observation $\bX^{(k)}$, i.e., $(\textbf{X}^{(1)})',\ldots, (\textbf{X}^{(n)})'$ and change some necessary notations (e.g., $p$ to $q$ and $q$ to $p$).

{\bf Estimation of supp($\O\otimes \Ga$)}. Let $\widehat{\text{supp}(\O)}$ and $\widehat{\text{supp}(\Ga)}$ be the estimators of supp$(\O)$ and supp$(\Ga)$, respectively, under the control of FDR at level $\alpha$.
The support of $\O\otimes \Ga$ can then be estimated by $\widehat{\text{supp}(\O)}\otimes\widehat{\text{supp}(\Ga)}$. In Theorem \ref{th3}, we will show that the FDR/FDP of this estimator is controlled at level
\begin{equation}\label{eq:alpha_prime}
\alpha'=\frac{\alpha ((2-\alpha)ab+aq+bp)}{\max(ab+aq+pb,1)}
\end{equation}
asymptotically, where $a$ and $b$ are the  numbers of  total discoveries in $\widehat{\text{supp}(\O)}$ and $\widehat{\text{supp}(\Ga)}$, respectively, excluding the diagonal entries. We also note that when the FDR/FDP level $\alpha'$  for the joint support estimation is given, to determine the FDR level $\alpha$ for the estimation of supp$(\O)$ and supp$(\Ga)$, one can try a sequence of $\alpha$'s from a small value to a large value. For each candidate $\alpha$, we obtain the value $a$ and $b$ by estimating supp$(\O)$ and supp$(\Ga)$ and plug the obtained values into \eqref{eq:alpha_prime}. Finally, we will choose the value $\alpha$ for the separate estimation that leads to the closest value to the pre-given $\alpha'$.

\begin{remark}
\label{remark_decorr}
%

One natural approach that for solving our problem is the de-correlation method. More precisely, if $\S$ is known, the data matrix can be transformed as $\S^{-1/2}\textbf{X}\sim N(\S^{-1/2}\m,\textbf{I}_{p\times p}\otimes\ps)$, based on which the method from \cite{Liu2013} can be applied. Therefore, a natural two-stage approach is to first obtain an consistent estimator of  $\S^{-1/2}$ (e.g., \cite{CaiLiuLuo2011}) and then apply the FDR control approach of \cite{Liu2013} to $\hat{\S}^{-1/2} \mathbf{X}^{(i)}$ for $1 \leq i \leq n$.  However, this ``de-correlation'' approach is not applicable under our problem setup. In fact, to ensure  the estimation error between $\hat{\S}^{-1/2}$ and $\S^{-1/2}$ is negligible for the FDR control, by some elementary calculations, we can find that we need the condition
\begin{eqnarray}\label{rem_a2}
\frac{nq}{(ps^{2}(p))^{1/(1-\tau)}}\rightarrow\infty
 \end{eqnarray}
 to replace $\S^{-1/2}$ by $\hat{\S}^{-1/2}$. (This condition means that we need a large sample number $nq$ to estimate $\S^{-1/2}$ accurately.) Similarly, to replace $\ps^{-1/2}$ by  $\hat{\ps}^{-1/2}$, we need the condition
 \begin{eqnarray}\label{rem_a22}
\frac{np}{(qs^{2}(q))^{1/(1-\tau)}}\rightarrow\infty.
 \end{eqnarray}
Hence, to get $\widehat{\textrm{supp}(\S^{-1})}\otimes\widehat{\textrm{supp}(\ps^{-1})}$, we need (\ref{rem_a2}) and (\ref{rem_a22})  simultaneously. However, when $n$ is fixed or small, (\ref{rem_a2}) and (\ref{rem_a22}) are contrary. Hence, it is impossible to do the  de-correlation for  rows and columns of $\X$ simultaneously.
\end{remark}

\section{Theoretical Results}
\label{sec:theory}

In this section, we provide the properties of the developed test statistic in \eqref{t1}, the guarantee of FDR control, power analysis and convergence rate of the initial estimator. All the proofs are relegated to the supplement.

Let $\lambda_{\min}(\S)=\lambda^{(1)}_{1}\leq\ldots\leq\lambda^{(1)}_{p}=\lambda_{\max}(\S)$ be eigenvalues of $\S$ and  $\lambda_{\min}(\ps)=\lambda^{(2)}_{1}\leq\ldots\leq\lambda^{(2)}_{q}=\lambda_{\max}(\ps)$ be eigenvalues of $\ps$. We make the following typical assumption on eigenvalues throughout this section:

\vspace{5pt}

\noindent{\bf (C1).} We assume that $c^{-1}\leq\lambda^{(1)}_{1}\leq\ldots\leq\lambda^{(1)}_{p}\leq c$ and $c^{-1}\leq\lambda^{(2)}_{1}\leq\ldots\leq\lambda^{(2)}_{q}\leq c$ for some constant $c>0$.

\vspace{5pt}

The condition (C1) is a standard eigenvalue assumption in high-dimensional covariance estimation literature (see the survey \cite{CaiRenZhou:16} and  references therein). This assumption is natural for many important classes of covariance matrices, e.g., bandable, Toeplitz, and sparse covariance matrices.
It is worthwhile to note that the assumption (C1) implies that $1/c' \leq A_{p}\leq c'$ (see $A_p$ in \eqref{eq:def_A_p}) for some constant $c'>0$.  We first provide some key results on the properties of the test statistic  and the estimator $\hat{A}_p$ of $A_p$ in the next subsection.

\subsection{Asymptotic normality and convergence results of the proposed test statistics}

The first result gives the asymptotic normality for the test statistic $\sqrt{\frac{(n-1)p}{A_{p}}}\frac{\hat{r}_{ij}}{\sqrt{\hat{r}_{ii}\hat{r}_{jj}}}$ in \eqref{eq:asy_normality} under the null.

\begin{proposition}\label{prop}
Assume that, as $np\rightarrow\infty$, $\log\max(q,np)=o(np)$, and the  estimator $\hat{\be}_j$ for $j \in [q]$ satisfies \eqref{bt1} with
\begin{eqnarray}\label{c1}
a_{n1}=o(1/\sqrt{\log \max(q,np)})\mbox{\quad and\quad }  a_{n2}=o((np)^{-1/4}).
\end{eqnarray}
Under the null $H_{0ij}: \gamma_{ij}=0$, we have, as $np\rightarrow\infty$,
\begin{eqnarray*}
\sqrt{\frac{(n-1)p}{A_{p}}}\frac{\hat{r}_{ij}}{\sqrt{\hat{r}_{ii}\hat{r}_{jj}}}\Rightarrow N(0,1)
\end{eqnarray*}
in distribution, where $\hat{r}_{ii}$ and $A_{p}$ are defined in \eqref{eq:r}  and \eqref{eq:def_A_p}, respectively.
\end{proposition}

 The next proposition shows that under alternatives,  $\frac{\hat{r}_{ij}}{\sqrt{\hat{r}_{ii}\hat{r}_{jj}}}$ will converge to a nonzero number, which indicates that our test statistics will lead to a non-trivial power.   Recall that  $\rho^{\Ga}_{ij\cdot}$ is the  partial correlation coefficient between $X_{li}$ and $X_{lj}$ (for any $1 \leq l \leq p$).

\begin{proposition}\label{propv}
Suppose that conditions in Proposition \ref{prop} hold. We have, for $1\leq i <  j\leq q$,
\begin{eqnarray*}
\frac{\hat{r}_{ij}}{\sqrt{\hat{r}_{ii}\hat{r}_{jj}}}-(1-\gamma_{ij}\psi_{ij})\rho^{\Ga}_{ij\cdot}\rightarrow 0
\end{eqnarray*}
in probability as $np\rightarrow\infty$.
\end{proposition}

Also note that the condition in \eqref{c1} will be established later in Proposition \ref{pro3-3}. It is interesting to see that in Propositions \ref{prop} and \ref{propv}, we only require $np\rightarrow\infty$, which means that
the sample size $n$ can be a constant. This is a significant difference between the estimation of MGGMs and
that of vector-variate GGMs. In the latter problem, to establish the asymptotic consistency or normality, the sample size is usually required to go to infinity in the existing literature (see, e.g.,  \cite{RothmanBickelLevinaZhu2008,Lam2009Sparsistency,Liu2013,RenSunZhangZhou2015}).

We next establish the convergence rate for the estimator of $A_{p}$. To this end, we need an additional condition on $\S$. \vspace{1mm}

\noindent\textbf{(C2).} For some $0<\tau<2$, assume that $\sum_{j=1}^{p}|\sigma_{ij}|^{\tau}\leq Cs(p)$ with $s(p)=\frac{1}{(\log q)^2}\Big{(}\sqrt{\frac{nq}{\log \max(p,nq)}}\Big{)}^{(2-\tau)\wedge 1}$ uniformly in $1\leq i\leq p$.\vspace{3mm}

Note that when $0 < \tau <1$, this assumption becomes the typical weak sparsity assumption in high-dimensional covariance estimation \citep{HighDimBook:11}.


\begin{proposition}\label{prop2} Let $\tilde{\lambda}=\lambda\sqrt{\frac{\log \max(p,nq)}{nq}}$ with $\lambda$ being sufficiently large.
Suppose that (C2) holds. We have $\hat{A}_{p}/A_{p}=1+O_{\pr}(\tilde{\lambda}^{(2-\tau)\wedge 1}s(p))$ as $nq\rightarrow\infty$.
\end{proposition}

Combining Propositions \ref{prop} and \ref{prop2}, we have the asymptotic normality under the null for our final test statistic $\hat{T}_{ij}=\sqrt{\frac{(n-1)p}{\hat{A}_{p}}}\frac{\hat{r}_{ij}}{\sqrt{\hat{r}_{ii}\hat{r}_{jj}}}$ in \eqref{t1}.

\ignore{By Lemma \ref{le3} in Appendix, we have $b_{ij}-\tr(\S)/p=o_{\pr}(1)$, where $b_{ij}$ is defined in \eqref{eq:bnij}. Therefore, Propositions \ref{prop} and \ref{prop2} also imply that, to detect $\gamma_{ij}\neq 0$ correctly,
$\gamma_{ij}$ should satisfy $\sqrt{np}|\gamma_{ij}|\rightarrow\infty$. Note that, in the estimation of vector-variate GGMs, it requires $\sqrt{n}|\gamma_{ij}|\rightarrow\infty$ to ensure $\gamma_{ij}\neq 0$ can be detected with high probability; see \cite{RenSunZhangZhou2015} for the minimax optimal rate for the estimation of single entry in precision matrix in vector-variate GGMs.}

\subsection{Guarantees on FDP/FDR control}
\label{sec:FDR_theory}
We next show that the FDP and FDR of $\widehat{\supp(\O)}\otimes\widehat{\supp(\Ga)}$ can be controlled asymptotically.  To this end, we discuss  the FDP and FDR of the estimation of supp$(\Ga)$ and supp$(\O)$ separately. For the estimation of supp$(\Ga)$, recall the definition of FDP and FDR:
\begin{eqnarray}\label{eq:FDP_FDR_1}
\mbox{FDP}_{1}=\frac{\sum_{(i,j)\in\mathcal{H}_{0}}I\{(i,j)\in \widehat{\supp(\Ga)}\}}{\max (  \sum_{1\leq i<j\leq q}I\{(i,j)\in \widehat{\supp(\Ga)}\},1) }, \quad
\mbox{FDR}_{1} =  \E (\mbox{FDP}_{1}),
\end{eqnarray}
where $\mathcal{H}_{0}=\{(i,j):\gamma_{ij}=0,\quad 1\leq i<j\leq q\}$. Let $\mathcal{H}_{1}=\{(i,j):\gamma_{ij}\neq 0,\quad 1\leq i<j\leq q\}$. Further define $\varpi_{0}=Card(\mathcal{H}_{0})$ as the total number of true nulls, $\varpi_{1 }=card(\mathcal{H}_1)$ as the number of true alternatives, and $\varpi=(q^{2}-q)/2$ as the total number of hypotheses.  For a constant $\gamma>0$ and $1\leq i\leq q$, define $\mathcal{A}_{i}(\gamma)=\{j: 1\leq j\leq q,~j\neq i,~|\gamma_{ij}|\geq (\log q)^{-2-\gamma}\}.$
Theorem \ref{th1} shows that our procedure controls FDP and FDR at a given level $\alpha$ asymptotically.

\begin{theorem}\label{th1} Let the dimension $(p,q)$ satisfy $q\leq (np)^{r}$ for some $r>0$. Suppose that
\begin{eqnarray}\label{a5}
Card\Big{\{}(i,j): 1\leq i< j\leq q,~~|(1-\gamma_{ij}\psi_{ij})\rho^{\Ga}_{ij\cdot}|\geq 4\sqrt{\frac{A_{p}\log q}{(n-1)p}}\Big{\}}\geq \sqrt{\log\log q},
\end{eqnarray}
where  $A_p$ is defined in  \eqref{eq:def_A_p}.
Assume that $\varpi_{1} \leq c \varpi$ for some $c<1$ and   $\{\hat{\be}_{i}\}_{i \in [q]}$ satisfy \eqref{bt1} with
\begin{eqnarray}\label{cd2}
a_{n1}=o(1/\log \max(q,np)) \mbox{\quad and\quad } a_{n2}=o((np\log q)^{-1/4}).
\end{eqnarray}
Under (C1), (C2), and $\max_{1\leq i\leq q}Card(\mathcal{A}_{i}(\gamma))=O(q^{\vartheta})$ for some $\vartheta<1/2$ and $\gamma>0$, we have
$\lim_{np, q\rightarrow\infty}\frac{\mathrm{FDR}_{1}}{\alpha \varpi_{0}/\varpi}=1$ and  $\frac{\mathrm{FDP}_{1}}{\alpha \varpi_{0}/\varpi}\rightarrow 1$ in probability as $np, q\rightarrow\infty$.
\end{theorem}

 Condition (\ref{a5}) requires the number of true alternatives is at least $\sqrt{\log \log q}$. This condition is very mild and in fact, is a nearly necessary condition for the FDP control. Proposition 2.1 in \cite{liu2014} shows that,
in large-scale multiple testing problems, if the number of true alternatives is fixed, then it is impossible for the Benjamini and Hochberg method \citep{Benjamini95} to control the FDP with probability tending to one at any desired level. Note that (\ref{a5}) is only slightly stronger than the condition that  the number of true alternatives goes to infinity.  The condition on $Card(\mathcal{A}_{i}(\gamma))$ is a sparsity condition for $\Ga$. This condition is also quite weak.
For the estimation of vector-variate GGMs, the existing literature often requires the row sparsity level of precision matrix to be less than $O(\sqrt{n})$.  Note that when the dimension $q$ is much larger than $n$, our condition on $Card(\mathcal{A}_{i}(\gamma))$ in Theorem \ref{th1} is clearly much weaker.  It should be noted that in Theorem \ref{th1}, the sample size $n$ can be a fixed  constant as long as the dimension $p, q\rightarrow\infty$.

As in Theorem \ref{th1}, we have the similar FDP and FDR control result for the estimation of $\supp(\O)$.
Let $\mathcal{H}^{'}_{0}=\{(i,j):\omega_{ij}=0,\quad 1\leq i<j\leq p\}$ and $\mathcal{H}^{'}_{1}=\{(i,j):\omega_{ij}\neq 0,\quad 1\leq i<j\leq p\}$. Further, let $\kappa_{0}=Card(\mathcal{H}^{'}_{0})$, $\kappa_{1}=Card(\mathcal{H}^{'}_{1})$  and $\kappa=(p^{2}-p)/2$.  Recall the definition of FDP and FDR of the estimation of supp$(\O)$:
\begin{eqnarray}\label{eq:FDP_FDR_2}
\mbox{FDP}_{2}=\frac{\sum_{(i,j)\in\mathcal{H}^{'}_{0}}I\{(i,j)\in \widehat{\supp(\O)}\}}{\max ( \sum_{1\leq i<j\leq q}I\{(i,j)\in \widehat{\supp(\O)}\},1) },\quad\mbox{FDR}_{2} = \ep (\mbox{FDP}_{2}).
\end{eqnarray}
For a constant $\gamma>0$ and $1\leq i\leq p$, define $\mathcal{B}_{i}(\gamma)=\{j: 1\leq j\leq p,~j\neq i,~|\omega_{ij}|\geq (\log p)^{-2-\gamma}\}.$
Let $B_{q}=q\|\ps\|^{2}_{\text{F}}/(\text{tr}(\ps))^{2}$ and the partial correlation associated with $\O$ be $\rho^{\bO}_{ij\cdot} = -  \frac{\omega_{ij}}{\sqrt{\omega_{jj}\omega_{jj}}}$ for $1 \leq i < j \leq p$. As (C2), we assume the following condition on $\ps=(\psi_{ij})_{p \times  p}$, \vspace{1mm}

\noindent\textbf{(C3).} For some $0<\tau<2$, assume that $\sum_{j=1}^{q}|\psi_{ij}|^{\tau}\leq Cs(q)$ with $s(q)=\frac{1}{(\log p)^2}\Big{(}\sqrt{\frac{np}{\log \max(q,np)}}\Big{)}^{(2-\tau)\wedge 1}$ uniformly in $1\leq i\leq q$.\vspace{1mm}

\begin{theorem}\label{th2}  Let the dimension $(p,q)$ satisfy $p\leq (nq)^{r}$ for some $r>0$. Suppose that
\begin{eqnarray}\label{a6}
Card\Big{\{}(i,j): 1\leq i< j\leq p,~~|(1-\omega_{ij}\sigma_{ij})\rho^{\O}_{ij\cdot}|\geq 4\sqrt{\frac{B_{q}\log p}{(n-1)q}}\Big{\}}\geq \sqrt{\log\log p}.
\end{eqnarray}
Assume that $\kappa_{1} \leq c\kappa$ for some $c<1$ and  $\{\hat{\be}_{i}\}_{i \in [p]}$ satisfies \eqref{bt1} with
\begin{eqnarray}\label{cd22}
a_{n1}=o(1/\log \max(p,nq)) \mbox{\quad and\quad } a_{n2}=o((nq\log p)^{-1/4}).
\end{eqnarray}
Under (C1), (C3), and $\max_{1\leq i\leq p}Card(\mathcal{B}_{i}(\gamma))=O(p^{\vartheta})$ for some $\vartheta<1/2$ and $\gamma>0$, we have
$\lim_{nq, p\rightarrow\infty}\frac{\mathrm{FDR}_{2}}{\alpha \kappa_{0}/\kappa}=1$ and  $\frac{\mathrm{FDP}_{2}}{\alpha \kappa_{0}/\kappa}\rightarrow 1$ in probability as $nq, p\rightarrow\infty$.
\end{theorem}

By Theorems \ref{th1} and \ref{th2}, we can obtain the FDP and FDR result of the estimator $\widehat{\supp(\O)}\otimes \widehat{\supp(\Ga)}$. In particular, let $a_{0}$ and $a$ be the number of false discoveries and total discoveries in $\widehat{\supp(\O)}$, excluding the diagonal entries. Similarly, let
$b_{0}$ and $b$ be the number of false discoveries and total discoveries in $\widehat{\supp(\Ga)}$, excluding the diagonal entries. It is then easy to calculate that the number of false discoveries in $\widehat{\supp(\O)}\otimes \widehat{\supp(\Ga)}$ is $a_{0}(q+b)+(a-a_{0})b_{0}+pb_{0}$, and the number of total discoveries in $\widehat{\supp(\O)}\otimes \widehat{\supp(\Ga)}$  is $pb+a(q+b)$ (excluding the diagonal entries). We have the following formulas for FDP and FDR of $\widehat{\supp(\O)}\otimes \widehat{\supp(\Ga)}$:
\begin{eqnarray}\label{eq:FDP_all}
\text{FDP}=\frac{a_{0}(q+b)+(a-a_{0})b_{0}+pb_{0}}{\max(pb+a(q+b),1)},\quad \text{FDR}=\ep(\text{FDP}).
\end{eqnarray}
The true FDP in \eqref{eq:FDP_all} cannot be computed in practice since the number of false discoveries $a_0$ and $b_0$ are unknown. One straightforward estimator for FDP is to replace the unknown quantities $a_0$ and $b_0$ with $\alpha a$ and $\alpha b$, respectively, which leads to the following FDP estimator $\alpha'$:
\begin{equation}\label{eq:FDP_est}
 \alpha'=\frac{\alpha ((2-\alpha)ab+aq+bp)}{\max(ab+aq+bp,1)}.
\end{equation}
Note that the values of $a$ and $b$ in \eqref{eq:FDP_est} are known, which represent  the number of total discoveries in   $\widehat{\supp(\O)}$ and $\widehat{\supp(\Ga)}$, respectively.
The FDP estimator $\alpha'$ takes the value in $[0,1]$ and is monotonically increasing as a function of $\alpha$. In the next theorem, we show that $\mathrm{FDP}/\alpha' \rightarrow 1$ in probability as $p, q\rightarrow\infty$.

\begin{theorem}\label{th3} Under the conditions of Theorems \ref{th1} and \ref{th2} with the sparsity condition $\omega_{1}=o(\omega)$ and $\kappa_{1}=o(\kappa)$, we have
$\frac{\mathrm{FDP}}{\alpha'}\rightarrow 1$ in probability as $p, q\rightarrow\infty$.
\end{theorem}

 Theorem \ref{th3} shows that the FDP  of the proposed estimator $\widehat{\supp(\O)}\otimes \widehat{\supp(\Ga)}$ can be estimated consistently by $\alpha'$.   Note that by Theorems \ref{th1} and \ref{th2}, the sparsity conditions $\omega_{1}=o(\omega)$ and $\kappa_{1}=o(\kappa)$ imply that $\frac{\mathrm{FDP}_{1}}{\alpha} \rightarrow 1$ and $\frac{\mathrm{FDP}_{2}}{\alpha} \rightarrow 1$ in probability when $p, q \rightarrow \infty$. Therefore, one can replace $a_0$ and $b_0$ in FDP in \eqref{eq:FDP_all} by $\alpha a$ and $\alpha b$, respectively, and achieve the result in Theorem \ref{th3}. In fact, we can still obtain the guarantee of FDP of $\widehat{\supp(\O)}\otimes \widehat{\supp(\Ga)}$  even without the sparsity conditions $\omega_{1}=o(\omega)$ and $\kappa_{1}=o(\kappa)$. For any $\varepsilon>0$,  Theorems \ref{th1} and \ref{th2} show that  $\mathbb{P}(\mathrm{FDP}_1 \leq \alpha(1+\varepsilon) ) \rightarrow 1 $ and  $\mathbb{P}(\mathrm{FDP}_2 \leq \alpha(1+\varepsilon) ) \rightarrow 1 $ as $p, q \rightarrow \infty$ regardless of the sparsity conditions. This further implies the following guarantee on the FDP of $\widehat{\supp(\O)}\otimes \widehat{\supp(\Ga)}$ for any $\varepsilon>0$:
\[
\pr\Big{(}\frac{\text{FDP}}{\alpha \{(2ab+aq+bp)/\max(ab+aq+pb,1)\}}\leq 1+\varepsilon\Big{)}\rightarrow 1, \; \text{as} \; p, q \rightarrow \infty.
\]

\subsection{Power analysis}
\label{sec:power}
We next study the statistical power of the proposed method by considering the following class of alternatives.  We assume that for some $c>4$,
\begin{eqnarray}\label{a38}
|\rho^{\Ga}_{ij\cdot}|=c\sqrt{\frac{A_{p}\log q}{(n-1)p}}\quad\mbox{and}\quad |\rho^{\O}_{kl\cdot}|=c\sqrt{\frac{B_{q}\log p}{(n-1)q}},\quad (i,j)\in\mathcal{H}_{1},~
(k,l)\in\mathcal{H}^{'}_{1}.
\end{eqnarray}

We will show in the next theorem that the power of the support estimators will converge to 1 as $p, q \rightarrow \infty$.

\begin{theorem}\label{th4} Let the dimension $(p,q)$ satisfy $p\leq (nq)^{r}$ and $q\leq (np)^{r}$ for some $r>0$. Assume that (C1)-(C3), (\ref{cd2}) and (\ref{cd22}) hold. We have $\supp(\Ga)\subseteq\widehat{\supp(\Ga)}$, $\supp(\O)\subseteq\widehat{\supp(\O)}$ and
$\supp(\O\otimes \Ga)\subseteq\widehat{\supp(\O)}\otimes\widehat{\supp(\Ga)}$  with probability tending to one as $p, q\rightarrow\infty$.
\end{theorem}

Recall that the power is defined by the ratio between the number of true discoveries in $\widehat{\supp(\O)}\otimes\widehat{\supp(\Ga)}$ and the total number of non-zero off-diagonals in $\supp(\O\otimes \Ga)$. Thus, Theorem \ref{th4} shows that the power converges to 1 as $p, q \rightarrow \infty$. In addition, Theorem \ref{th4} shows that to detect the edge between $X_{ij}$ and $X_{kl}$, the corresponding partial correlation $\varrho_{ij,kl}= \rho^{\Ga}_{ij\cdot} \cdot \rho^{\O}_{kl\cdot}$ can be as small as $C\frac{1}{n-1}\sqrt{\frac{\log p \log q}{pq}}$ (note that $A_{p}$ and $B_{q}$ are bounded, see assumption (C1)). This is essentially different from the estimation of  vector-variate GGMs.
Actually, if we apply the method of estimation of  vector-variate GGMs to vec$(\textbf{X})$ directly (e.g., the method from \cite{RenSunZhangZhou2015}),  even for an individual test (detecting a single edge),  the magnitude of the  partial correlation $\varrho_{ij,kl}$ needs to be $C\frac{1}{\sqrt{n}}$.

\subsection{Convergence rate of the initial estimators of regression coefficients}

Finally, we present the next proposition, which shows that the convergence rate condition of $\hat{\be}_{j}$ in (\ref{c1}) and (\ref{cd2}) can be satisfied under some regular conditions. The convergence rate condition in \eqref{cd22} can be established similarly. This result establishes the consistency of Lasso for correlated samples, which in itself can yield a separate interesting result.

\begin{proposition}\label{pro3-3} Let $\delta$ in (\ref{a10}) be large enough. Suppose that (C1) holds and $\max_{1\leq j \leq q}|\be_{j}|_{0}=o\Big{(}\frac{\sqrt{np}}{(\log \max(q,np))^{3/2}}\Big{)}$. We have  $\hat{\be}_{j}(\delta)$ for $1\leq j \leq q$ are consistent in both $\ell_1$ and $\ell_2$ norms with the rate in (\ref{cd2}).
 \end{proposition}


\section{Numerical Results}
\label{sec:exp}

In this section, we present numerical results on both simulated and real data to investigate the performance of the proposed method on support recovery of matrix-variate data. In our experiment, we adopt the data-driven parameter-tuning approach from \cite{Liu2013} to tune the parameters (see Section \ref{sec:add_exp} in the supplement for details). Due to space constraints, some simulated experimental results and real data analysis are provided in Section \ref{sec:add_exp}  of the supplement.

\subsection{Simulated experiments}
\label{sec:simu}

In the simulated study, we construct $\O$ and $\Ga$ based on combinations of following graph structures used in \cite{Liu2013}.
\begin{enumerate}
\item Hub graph (``hub" for short). There are $p/10$ rows with sparsity 11. The rest of the rows have sparsity 2. In particular, we let $\O_1 = (\omega_{ij}), \omega_{ij} = \omega_{ji} = 0.5$ for $i = 10(k-1)+1$ and $10(k-1)+2 \leq j \leq 10(k-1)+10, 1 \leq k \leq p/10$. The diagonal $\omega_{ii}=1$ and other entries are zero. We also let $\O = \O_1+(|\min(\lambda_{\min})|+0.05) \mathbf{I}_p$ to make the matrix be positive definite.
\item Band graph (``band" for short). $\O = (\omega_{ij})$, where $\omega_{ii}=1$, $\omega_{i,i+1} = \omega_{i+1,i}=0.6$, $\omega_{i,i+2} = \omega_{i+2,i}=0.3$, $\omega_{ij} = 0$ for $|i-j|\geq 3$.
\item Erd\"os-R\'enyi random graph (``random" for short). There is an edge between each pair of nodes with probability $\min(0.05,5/p)$ independently. Let $\O_{1}=(\omega_{ij})$, $\omega_{ii}=1$ and $\omega_{ij}=u_{ij}\ast \delta_{ij}$ for $i\neq j$, where $u_{ij} \sim U(0.4,0.8)$ is the uniform random variable and $\delta_{ij}$ is the Bernoulli random variable with success probability $\min(0.05,5/p)$. $u_{ij}$ and $\delta_{ij}$ are independent. We also let $\O= \O_1 + (|\min(\lambda_{\min})|+0.05)\mathbf{I}_p$ so that the matrix is positive definite.
\end{enumerate}
The  matrix $\Ga$ is also constructed from one of the above three graph structures.

\begin{table}[!t]
\centering
\small
\caption{Averaged empirical FDP, the estimated FDP/FDR level $\alpha'$ in \eqref{eq:FDP_est} and power.}
\begin{tabular}{r r r r r r r r r r r r} \hline \hline
 $p$ & $q$ & $\O$ & $\Ga$  & \multicolumn{2}{c}{$n=20$} & \multicolumn{2}{c}{$n=100$} \\ \cmidrule(r){5-6} \cmidrule(r){7-8}
     &     &          &           & FDP ($\alpha'$)  & Power & FDP ($\alpha'$)  & Power
  \\ \hline
 100 & 100 & hub & hub & 0.192 (0.146) & 1.000 &0.155 (0.145) & 1.000 \\
    &   & hub & band & 0.158 (0.152) & 1.000 &0.146 (0.152) & 1.000 \\
    &   & hub & random & 0.188 (0.154) & 0.916 &0.156 (0.154) & 1.000 \\
    &   & band & band & 0.138 (0.161) & 1.000 &0.154 (0.162) & 1.000 \\
    &   & band & random & 0.152 (0.164) & 0.998 &0.127 (0.163) & 1.000 \\
    &   & random & random & 0.161 (0.164) & 0.834 &0.104 (0.165) & 0.999 \\
 200 & 200 & hub & hub & 0.183 (0.146) & 1.000 &0.145 (0.145) & 1.000 \\
    &   & hub & band & 0.149 (0.152) & 1.000 &0.144 (0.152) & 1.000 \\
    &   & hub & random & 0.167 (0.154) & 0.981 &0.153 (0.154) & 1.000 \\
    &   & band & band & 0.138 (0.162) & 1.000 &0.148 (0.162) & 1.000 \\
    &   & band & random & 0.158 (0.164) & 1.000 &0.134 (0.163) & 1.000 \\
    &   & random & random & 0.171 (0.166) & 0.980 &0.134 (0.166) & 1.000 \\
 200 & 50 & hub & hub & 0.166 (0.145) & 1.000 &0.154 (0.145) & 1.000 \\
    &   & hub & band & 0.159 (0.152) & 1.000 &0.151 (0.152) & 1.000 \\
    &   & hub & random & 0.138 (0.154) & 0.991 &0.134 (0.153) & 1.000 \\
    &   & band & band & 0.127 (0.161) & 1.000 &0.146 (0.161) & 1.000 \\
    &   & band & random & 0.200 (0.163) & 0.894 &0.120 (0.163) & 0.992 \\
    &   & random & random & 0.194 (0.162) & 0.714 &0.141 (0.165) & 0.980 \\
 400 & 400 & hub & hub & 0.160 (0.145) & 1.000 &0.141 (0.145) & 1.000 \\
    &   & hub & band & 0.146 (0.152) & 1.000 &0.144 (0.152) & 1.000 \\
    &   & hub & random & 0.172 (0.154) & 0.999 &0.151 (0.154) & 1.000 \\
    &   & band & band & 0.169 (0.162) & 1.000 &0.148 (0.162) & 1.000 \\
    &   & band & random & 0.159 (0.164) & 1.000 &0.142 (0.164) & 1.000 \\
    &   & random & random & 0.180 (0.166) & 1.000 &0.147 (0.166) & 1.000 \\ \hline
\end{tabular}
\label{tab:FDP_Power}
\normalsize
\end{table}

For each combination of $\O$ and $\Ga$, we generate $n$ ($n=20$ or $n=100$) samples $(\bX^{(k)})_{k=1}^n$, where each $\bX^{(k)} \sim N_{p,q}(\mathbf{0}, \S \otimes \ps)$ with $\S=\O^{-1}$ and $\ps=\Ga^{-1}$. We consider different settings of $p$ and $q$, i.e., $(p,q)=(100,100)$, $(p,q)=(200,50)$,  $(p,q)=(200,200)$ and $(p,q)=(400,400)$. The FDR level $\alpha$ for the support recovery of $\Ga$ and $\O$ are set to $0.1$ (the observations for other $\alpha$s are similar and thus omitted for space considerations). The parameters $\lambda$ and $\delta$ are tuned using the data-driven approach in \eqref{ats}. All the simulation results are based on 100 independent replications.

In Table \ref{tab:FDP_Power}, we report the averaged true FDP  in \eqref{eq:FDP_all}, the FDP estimator  $\alpha'$ in \eqref{eq:FDP_est} and the power for estimating $\supp(\O \otimes \Ga)$ over 100 replications. On the one hand, according to Theorem \ref{th3}, it will be desirable that the true FDP is close to $\alpha'$. On the other hand, we are aiming for a large power. In particular, the power of the estimator $\widehat{\supp(\O)}\otimes \widehat{\supp(\Ga)}$  can be calculated using the following simple formula. Let $A$ and $B$ be the number of nonzero off-diagonals in $\O$ and $\Ga$. Recall the definition of $a_0$, $a$, $b_0$, $b$ in \eqref{eq:FDP_all} and we have

\begin{equation}\label{eq:power}
\text{Power}=\frac{(pb+a(q+b))-(a_0(q+b)+(a-a_0)b_0+pb_0)}{pB+A(q+B)}=\frac{p(b-b_0)+(a-a_0)(q+b-b_0)}{pB+A(q+B)},
\end{equation}
\normalsize
where the numerator is the number of true discoveries in $\widehat{\supp(\O)}\otimes \widehat{\supp(\Ga)}$ and the denominator is the total number of nonzero off-diagonals in $\supp(\O \otimes \Ga)$. From Table \ref{tab:FDP_Power},  in all settings with $n=100$, the true FDPs are close to their estimates $\alpha'$ and the powers are all very close to 1. For $n=20$, which is a more challenging case due to the small sample size, the true FDPs are still controlled by their estimates $\alpha'$ for most graphs. When $\O$ and $\Ga$ are both hub or random graphs, the true FDPs are slightly larger than the corresponding estimates. In terms of power with $n=20$, when $\O$ and $\Ga$ are both generated from random graphs and either $p$ or $q$ is small (e.g., $p=q=100$ or $p=200, q=50$)  the powers could be away from 1 (but still above 0.7); while for all other cases, the powers are still close to 1. We examined the cases in which the power is much less than one and found that our FDP procedure generates overly sparse estimators, which leads to lower powers. In fact, a lower power for a small $n$ and $p$ (or $q$) is expected since we essentially use $np$ correlated samples to estimate $\text{supp}(\Ga)$ and $nq$ correlated samples to estimate $\text{supp}(\O)$.  

Due to space constraints, we relegate the following simulation studies to the supplement:
\begin{enumerate}
\item In Section \ref{sec:supp_box} of the supplement, we present the boxplots of FDPs over 100 replications. The plots show that FDPs are well concentrated, which suggests that the performance of the proposed estimator is quite stable.
\item In Section \ref{sec:supp_A_p} of the supplement, we provide experimental results on the estimation of $\hat{A}_p$, which empirically verify our theoretical result in Proposition \ref{prop2}.
\item  In Section \ref{sec:Leng} of the supplement, we compare our procedure with the penalized likelihood approach in \cite{LengTang2012}. We observe that when $p,q$ are small as compared to $n$, the penalized likelihood approach still achieves good support recovery performance (e.g., the case $n=100, p=q=20$ as reported in \citet{LengTang2012}). When $p, q$ are comparable to or lager than $n$, our testing based method achieves better support recovery performance.
\item In Section \ref{sec:decorrelation} of the supplement, we provide some empirical evidences to show that the de-correlation approach in Remark \ref{remark_decorr} cannot control FDP well.
\item In Section \ref{sec:perturbation} of the supplement,  we further present  simulation studies when the covariance matrix does not follow the form of a Kronecker product.
\end{enumerate}

\subsubsection{ROC curves}
\label{sec:ROC}

\begin{figure}[!t]
\centering
\subfigure[t][$f=1$]{
  \includegraphics[width=0.33\textwidth]{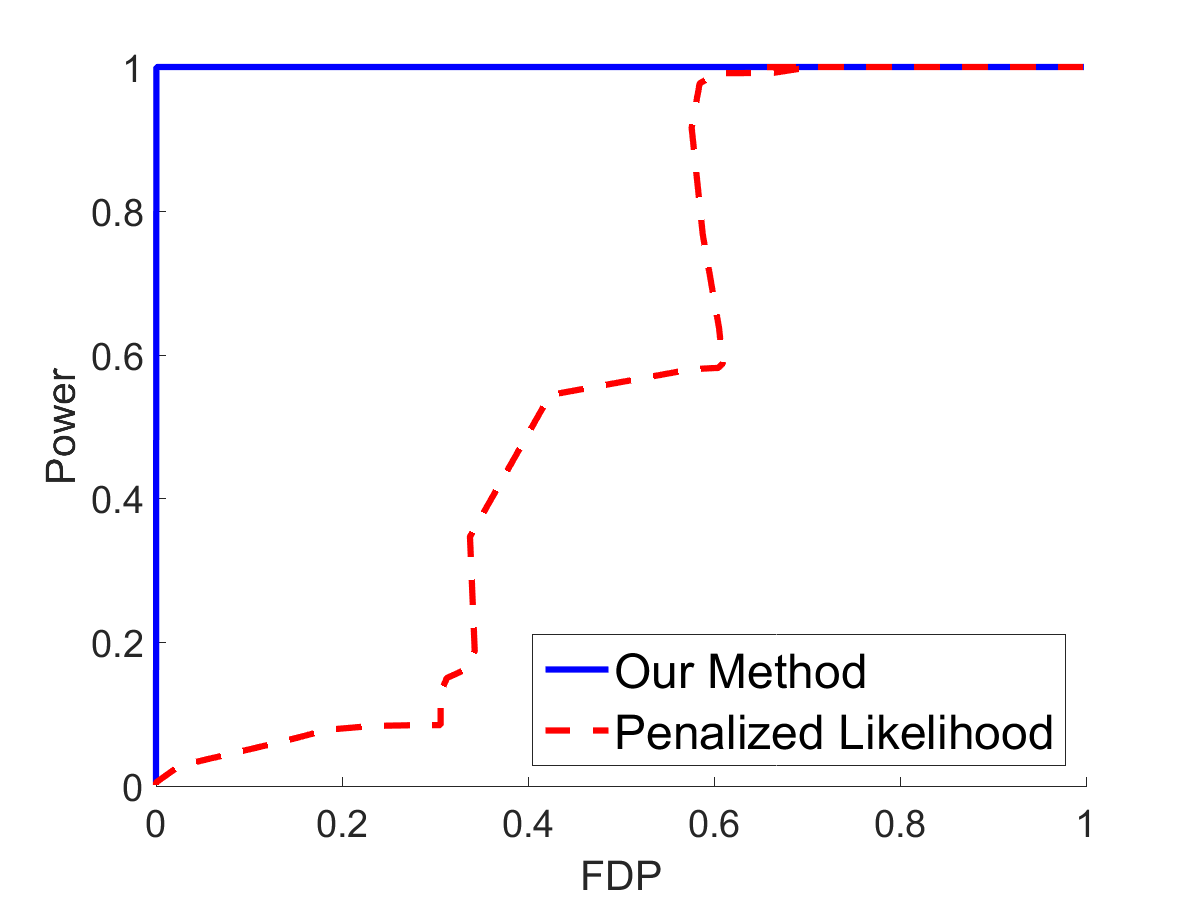}
    \label{fig:roc_factor_1}
}\hspace{-8mm}
\subfigure[t][$f=2$]{
  \includegraphics[width=0.33\textwidth]{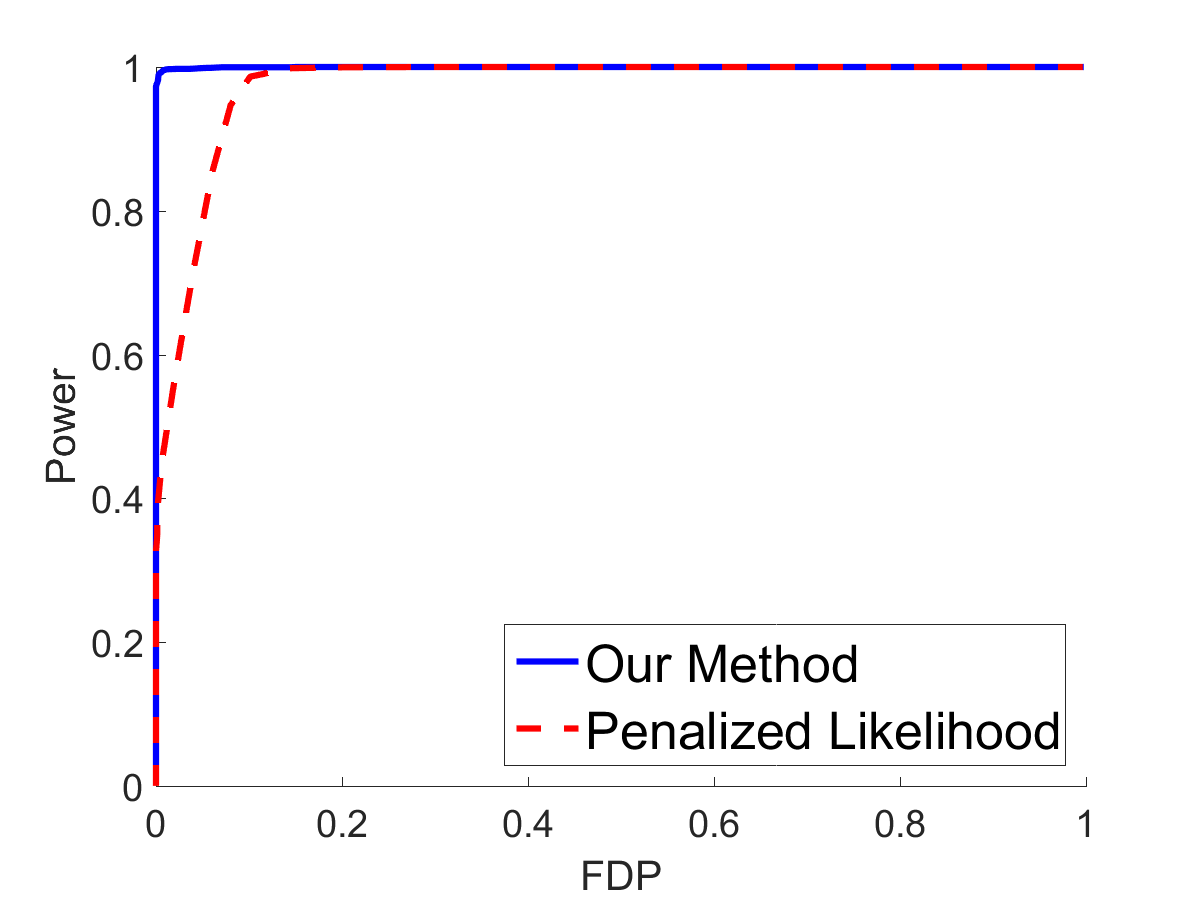}
    \label{fig:roc_factor_2}
}\hspace{-8mm}
\subfigure[t][$f=3$]{
  \includegraphics[width=0.33\textwidth]{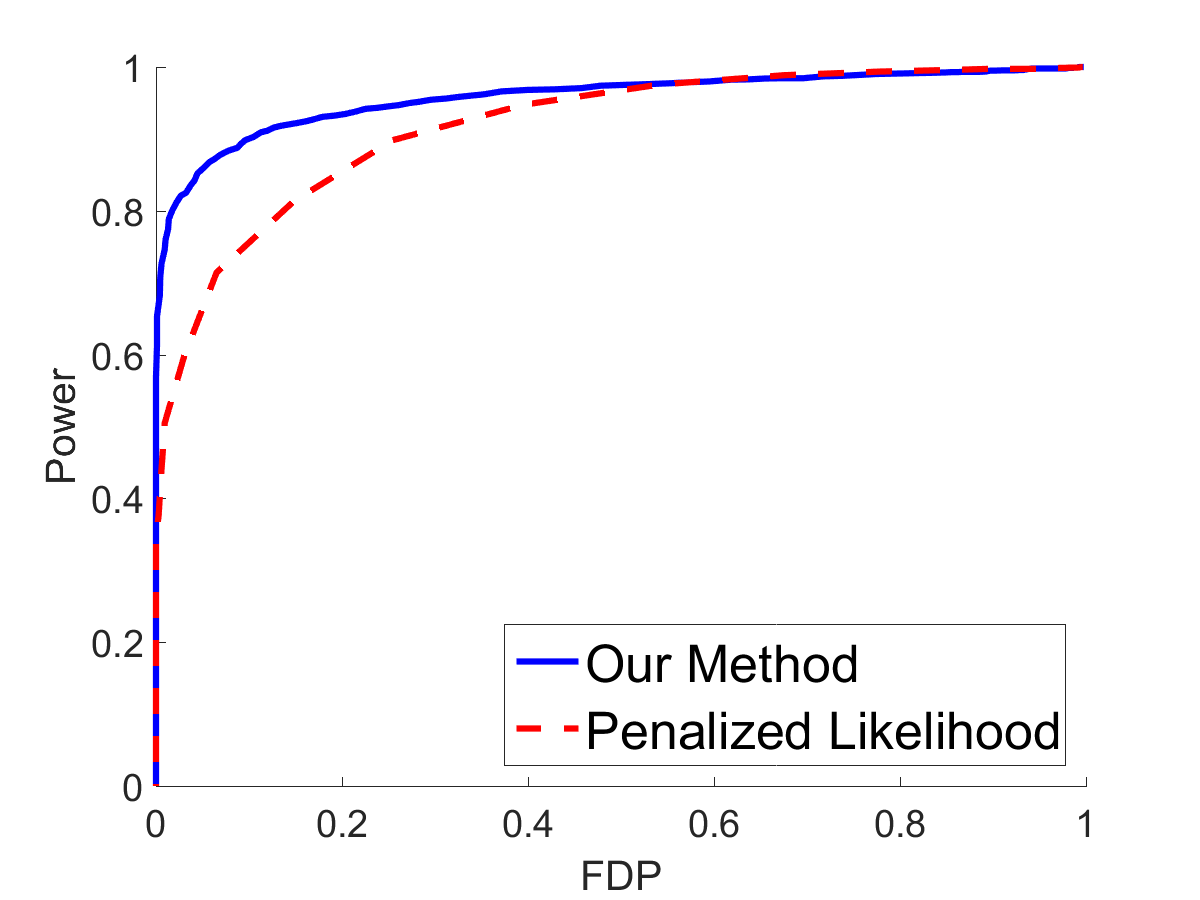}
    \label{fig:roc_factor_3}
}\hspace{-8mm}
\hspace{-8mm}
\caption{ROC curves for different signal strength  when $\O$ and $\Ga$ are bandable matrices.}
\label{fig:roc_factor}
\end{figure}

In this study, we present comparisons between our method and the penalized likelihood approach in \cite{LengTang2012} (with the SCAD penalty) in terms of the ROC curve. We construct the precision matrices $\O$ and $\Ga$ using the graph structures in Section \ref{sec:simu} but introducing an additional factor $f$ to tune the signal strength. In particular, for hub graph, we set $\O_1 = (\omega_{ij})$ with  $\omega_{ij} = \omega_{ji} = 0.5/f$; for band graph, we set $\omega_{i,i+1} = \omega_{i+1,i}=0.6/f$ and $\omega_{i,i+2} = \omega_{i+2,i}=0.3/f$; and for random graph, we choose $u_{ij} \sim U(0.4/f,0.8/f)$. When $f=1$, the construction of $\O$ and $\Ga$ is the same as that in Section \ref{sec:simu}. The higher the value of $f$ is, the weaker the signal strength is. Due to space constraints, we only report the comparison  when $n=20$, $p=q=100$ and observations are similar for other settings of $n, p,$ and $q$.

We first compare the support recovery performance for different signal strengths by varying $f=1,2,3$ and fix $\O$ and $\Ga$ to be bandable matrices.   We note that the usual ROC curve for binary classification is based on false positive rate (a.k.a. $1 -$ specificity) vs true positive rate (a.k.a.  sensitivity or power). However, our problem is essentially a highly unbalanced classification and thus the standard ROC curve is not suitable. In other words, in our high-dimensional setting, $\O \otimes \Ga$ is a highly sparse matrix and thus the false positive rate will  be extremely small for any reasonable choice of $\alpha$ or regularization parameter that give a small number of discoveries.  Therefore, we choose to report the ROC curve in terms of FDP vs power, from which, one can easily compare powers for different methods under the same level of FDP.
As one can see from Figure \ref{fig:roc_factor}, our method achieves better performance than the method in \cite{LengTang2012} for different signal strengths. When the factor $f=3$, the ROC curve of our method is still almost vertical. In Section \ref{sec:add_ROC} in the supplement, we fix the factor $f=3$ and consider different types of $\O$ and $\Ga$. For most cases, our method still achieves better performance.


\subsection{Real data analysis}
\label{sec:real}

For the real data analysis, we investigate the performance of the proposed method on two real datasets, the U.S. agricultural export data from \citet{LengTang2012} and the climatological data from \cite{Lozano:09}. Due to space constraints, the details of real data analysis are provided in Section \ref{sec:supp_real} in the supplement.

\section{Discussions and future work}
\label{sec:disc}

In this paper, we propose new test statistics with FDR control guarantees for graph estimation from matrix-variate Gaussian data. To handle the correlation structure among ``row samples" and ``column samples", we develop the \emph{variance correlation technique}. The proposed variance correlation technique can be directly extended to address the problem of learning high-dimensional GGMs with correlated samples, which has not been studied in the existing literature but finds many important applications in practice. We leave this extension as a future work.

To establish the FDR control result,  the correlation among ``row samples" makes the theoretical analysis significantly more challenging than the \emph{i.i.d.} case and all the analysis in the \emph{i.i.d.} case must be carefully tailored. For example, we need to establish the consistency for Lasso estimators from correlated samples. We also need a few new large deviation bounds on sample covariance matrices with correlated samples (see Section \ref{sec:tech_lemma} in the supplement). 

There are several future directions of this work. First, although our paper mainly focuses on the support recovery and graph estimation, it is also interesting to estimate the Kronecker product precision matrix based on the multiple testing framework.  Moreover, our work relies on the the Kronecker product structure, it is interesting to consider other forms of the covariance matrices, e.g.,  the true covariance matrix does not exactly follow Kronecker product structure but close to that structure.



\vskip 14pt
\noindent {\large\bf Supplementary Materials}

The supplementary material consists of several technical lemmas and the proofs of our propositions and theorems. Moreover, it includes additional simulation studies and real data analysis.
\par

\vskip 14pt
\noindent {\large\bf Acknowledgements}

The authors are very grateful to two anonymous referees and the associate editor for their detailed
and constructive comments that considerably improved the quality of this paper. We would also
like to thank Yichen Zhang for helpful discussions and Chenlei Leng and Cheng Yong Tang for sharing the code of \cite{LengTang2012} and U.S. export data. Weidong Liu's research is supported by NSFC, Grants  No. 11322107 and No. 11431006,  the Program for Professor of Special Appointment (Eastern Scholar) at Shanghai Institutions of Higher Learning,  Shanghai Shuguang Program, Shanghai Youth Talent Support Program and 973 Program (2015CB856004).
\par


\section*{Supplementary Material}
\setcounter{section}{0} 
\section{Technical Lemmas}
\label{sec:tech_lemma}

We introduce several technical lemmas, which will be used throughout the proofs of
our propositions and theorems. In particular, let $\ps=(\psi_{ij})_{q\times q}$ and recall the definition of $\hat{\ps}=(\hat{\psi}_{ij})_{q\times q}$ from Section \ref{sec:init}.   We first prove a maximal concentration inequality on $\hat{\psi}_{ij}$ in Lemma \ref{le1}. Note that this result differs from the standard concentration inequality on the sample covariance matrix with \emph{i.i.d.} samples since the ``row samples" for constructing $\hat{\ps}$ are correlated.

\begin{lemma}\label{le1} We have for any $M>0$, there exists a constant $C$ such that
	\begin{eqnarray*}
		\pr\Big{(}\max_{1\leq i\leq j\leq q}\Big{|}\hat{\psi}_{ij}-\frac{\tr(\S)}{p}\psi_{ij}\Big{|}\geq C\sqrt{\frac{\log\max(q,np)}{np}}\Big{)}=O((q+np)^{-M}).
	\end{eqnarray*}
\end{lemma}

\noindent{\bf Proof.} Recall that for any pair of $i \in [q]$ and $j \in [q]$
\begin{eqnarray}\label{eq:hat_psi}
\hat{\psi}_{ij}&=&\frac{1}{(n-1)p}\sum_{k=1}^{n}\sum_{l=1}^{p}(X^{(k)}_{li}-\bar{X}_{li})(X^{(k)}_{lj}-\bar{X}_{lj})\nonumber \\
&=&\frac{1}{(n-1)p}\sum_{l=1}^{p}\sum_{k=1}^{n}(X^{(k)}_{li}-\bar{X}_{li})(X^{(k)}_{lj}-\bar{X}_{lj}), \
\end{eqnarray}
where $\bar{X}_{li}=\frac{1}{n} \sum_{k=1}^n X^{(k)}_{li}$ and $\bar{X}_{lj}=\frac{1}{n} \sum_{k=1}^n X^{(k)}_{lj}$. Without loss of generality, we assume that $\m=0$.

Let $\textbf{A}\in \text{R}^{n\times n}$ be an orthogonal matrix with the last row $(\frac{1}{\sqrt{n}},\ldots,\frac{1}{\sqrt{n}})$. Let
$\Y_{li}=(Y^{(1)}_{li},\ldots,Y^{(n)}_{li})^{'}=\textbf{A}(X^{(1)}_{li},\ldots,X^{(n)}_{li})^{'} \in \R^{n \times 1}$. So we have $\sqrt{n}\bar{X}_{li}=Y^{(n)}_{li}$ and
\begin{eqnarray}\label{eq:X_Y}
\sum_{k=1}^{n}(X^{(k)}_{li}-\bar{X}_{li})(X^{(k)}_{lj}-\bar{X}_{lj})=\Y_{li}^{'}\Y_{lj}-Y^{(n)}_{li}Y^{(n)}_{lj}=\sum_{k=1}^{n-1}Y^{(k)}_{li}Y^{(k)}_{lj}.
\end{eqnarray}
Since $(X^{(1)}_{li},\ldots,X^{(n)}_{li})^{'} \sim N(\0, \sigma_{ll} \psi_{ii} \I_{n \times n})$, $(Y^{(1)}_{li},\ldots,Y^{(n-1)}_{li})^{'}\sim N(\0, \sigma_{ll}\psi_{ii}\I_{(n-1)\times(n-1)})$.
Let $\textbf{Y}_{k}=(Y^{(k)}_{li})_{1\leq l\leq p,1\leq i\leq q}$ for $1\leq k\leq n-1$.  We have $\textbf{Y}_{k}\sim N(\0,\S\otimes\ps)$ and $\textbf{Y}_{k}$, $1\leq k\leq n-1$, are independent. Let us define $\Z_k =\S^{-1/2}\Y_k \sim N(\0, \I_{p \times p} \otimes \ps)$. Let $\Z_{ki}$  be the $i$-th column of $\Y_k$. Then $(\textbf{Z}_{ki},\textbf{Z}_{kj})\sim N(0,\I_{p\times p}\otimes\ps_{[i,j]})$, where $\ps_{[i,j]}=\left(
\begin{array}{cc}
\varphi_{ii} & \varphi_{ij} \\
\varphi_{ji} &\varphi_{jj} \\
\end{array}
\right).$

Let the $\textbf{U}^{'}\textbf{D}\textbf{U}$ be the eigenvalue decomposition of $\S$, where $\textbf{U}$ is an orthogonal matrix and $\textbf{D}=\diag(\lambda^{(1)}_{1},\ldots,\lambda^{(1)}_{p})$. Define $(\textbf{W}_{ki},\textbf{W}_{kj}):=(\textbf{U}\textbf{Z}_{ki},\textbf{U}\textbf{Z}_{kj}) \in \R^{p \times 2}$ where $\textbf{W}_{ki}=(w_{ki,1},\ldots,w_{ki,p})^{'} \in \R^{p \times 1}$ and $\textbf{W}_{kj}=(w_{kj,1},\ldots,w_{kj,p})^{'} \in \R^{p \times 1}$. Since $\mathbf{U}$ is an orthogonal matrix, $(\textbf{W}_{ki},\textbf{W}_{kj})\sim N(\0,\I_{p\times p}\otimes\ps_{[i,j]})$, which also  implies that $(w_{ki,l},w_{kj,l})$ are independent for $1\leq l\leq p$.

Now combining \eqref{eq:hat_psi} and \eqref{eq:X_Y}, we have,
\begin{eqnarray}\label{a4}
\hat{\psi}_{ij}
&=& \frac{1}{(n-1)p} \sum_{l=1}^p \sum_{k=1}^{n-1} \Y_{ki}'\Y_{ki} \cr
&=&\frac{1}{(n-1)p}\sum_{k=1}^{n-1}(\textbf{U}\textbf{Z}_{ki})^{'}\textbf{D}\textbf{U}\textbf{Z}_{kj}\cr
&=&\frac{1}{(n-1)p}\sum_{k=1}^{n-1}\sum_{l=1}^{p}\lambda^{(1)}_{l}w_{ki,l}w_{kj,l}.
\end{eqnarray}
We further note that
\begin{eqnarray*}
	\E \hat{\psi}_{ij} = \frac{1}{(n-1)p}\sum_{k=1}^{n-1}\sum_{l=1}^{p}\lambda^{(1)}_{l}\psi_{ij} = \frac{\tr(\S)}{p} \psi_{ij}.
\end{eqnarray*}
Put $w_{kij,l}=w_{ki,l}w_{kj,l}-\ep w_{ki,l}w_{kj,l}$. We have for some $\eta>0$ such that $\ep e^{2\eta|\lambda^{(1)}_{l}w_{kij,l}|}\leq K$ for some $K>0$, uniformly in $i,j,l,k$. It implies that
\begin{eqnarray*}
	\sum_{k=1}^{n-1}\sum_{l=1}^{p}\ep(\lambda^{(1)}_{l}w_{kij,l})^{2}e^{\eta|\lambda^{(1)}_{l}w_{kij,l}|}&\leq& K\sum_{k=1}^{n-1}\sum_{l=1}^{p}(\ep(\lambda^{(1)}_{l}w_{kij,l})^{4})^{1/2}\cr
	&=&\sqrt{2}K(n-1)\|\S\|^{2}_{F}(\varphi_{ii}\varphi_{jj}+\varphi^{2}_{ij}).
\end{eqnarray*}
By the exponential inequality in Lemma 1 in Cai and Liu (2011) and $\|\S\|^{2}_{F}/p\leq \lambda^{(1)}_{p}$, for any $M>0$, there exists a constant $C>0$,
\begin{eqnarray*}
	\pr\Big{(}|\hat{\psi}_{ij}-\ep \hat{\psi}_{ij}|\geq C\sqrt{\frac{\log\max(q,np)}{np}}\Big{)}=O((q+np)^{-M}).
\end{eqnarray*}
This proves Lemma \ref{le1}.
\qed

The next concentration inequality involves the residuals. In particular, for $k \in [n]$, $l \in [p]$ and $j \in [q]$,  we define:
\begin{equation*}
\tilde{\varepsilon}^{(k)}_{lj}=\varepsilon^{(k)}_{lj}-\frac{1}{n}\sum_{k=1}^{n}\varepsilon^{(k)}_{lj}=:\varepsilon^{(k)}_{lj}-\bar{\varepsilon}_{lj}, \quad \hat{\sigma}_{jj,\varepsilon}=\frac{1}{(n-1)p}\sum_{k=1}^{n}\sum_{l=1}^{p} \left( \tilde{\varepsilon}^{(k)}_{lj}\right)^2.
\end{equation*}

\begin{lemma}\label{le2} For any $M>0$, there exists a constant $C$ such that
	\begin{eqnarray*}
		\pr\Big{(}\max_{1\leq i\leq q}\max_{1\leq h\leq q,h\neq i}\Big{|}\frac{1}{np}\sum_{k=1}^{n}\sum_{l=1}^{p}\tilde{\varepsilon}^{(k)}_{li}(X^{(k)}_{lh}-\bar{X}_{lh})\Big{|}\geq C\sqrt{\frac{\log\max(q,np)}{np}}\Big{)}=O((q+np)^{-M})
	\end{eqnarray*}
	and
	\begin{eqnarray*}
		\pr\Big{(}\max_{1\leq i\leq p}\Big{|}\frac{1}{np}\sum_{k=1}^{n}\sum_{l=1}^{p}\tilde{\varepsilon}^{(k)}_{li}(\textbf{X}^{(k)}_{l,-i}-\bar{\textbf{X}}_{l,-i})\be_{i}\Big{|}\geq C\sqrt{\frac{\log\max(q,np)}{np}}\Big{)}=O((q+np)^{-M}).
	\end{eqnarray*}
\end{lemma}

\noindent{\bf Proof.} Recall that
\begin{eqnarray*}
	\varepsilon^{(k)}_{li}=X^{(k)}_{li}-\alpha_{li}-\textbf{X}^{(k)}_{l,-i}\be_{i}.
\end{eqnarray*}
Set $\va^{(k)}_{i}=(\varepsilon^{(k)}_{1i},\ldots,\varepsilon^{(k)}_{pi})^{'}$ and $\Y^{(k)}_{i}=(\textbf{X}^{(k)}_{1,-i}\be_{i},\ldots,\textbf{X}^{(k)}_{p,-i}\be_{i})^{'}$.
It is easy to see that $\Cov(\va^{(k)}_{i})=\gamma_{ii}^{-1}\S$. Let $\textbf{X}^{(k)}_{h}$ be the $h$-th column of $\textbf{X}^{(k)}$. Since $\va^{(k)}_{i}$ and $\textbf{X}^{(k)}_{h}$ are independent for $h \neq i$, for the $p \times 2$ matrix $(\va^{(k)}_{i},\textbf{X}^{(k)}_{h})$, we have $\Cov((\va^{(k)}_{i},\textbf{X}^{(k)}_{h}))=\S\otimes \diag(\gamma_{ii}^{-1},\psi_{ii})$ for $h\neq i$.

In addition, $\X^{(k)}_{\cdot, -i} \sim N(\0, \S \otimes \ps_{-i, -i})$ and $\be_i = -\frac{1}{\gamma_{ii}}\Ga_{-i, i}$, we have
\[
\Cov(\Y^{(k)}_{i})= \frac{1}{\gamma_{ii}^2}\tr(\Ga_{-i,i}\Ga_{i,-i} \ps_{-i, -i} ) \S.
\]
Further, by the fact that,
\begin{align*}
\tr(\Ga_{-i,i}\Ga_{i,-i} \ps_{-i, -i} ) =   \Ga_{i,-i} \ps_{-i, -i}\Ga_{-i,i} = -\gamma_{ii} \ps_{i,-i}\Ga_{-i,i}=-\gamma_{ii}(1-\psi_{ii}\gamma_{ii}),
\end{align*}
which further implies that,
\[
\Cov(\Y^{(k)}_{i})=\frac{\psi_{ii}\gamma_{ii}-1}{\gamma_{ii}} \S.
\]
Since $\va^{(k)}_{i}$ and $\Y^{(k)}_{i}$ are independent,
$\Cov((\va^{(k)}_{i},\Y^{(k)}_{i}))=\S\otimes \diag(\gamma_{ii}^{-1},(\psi_{ii}\gamma_{ii}-1)/\gamma_{ii})$.
Following exactly the same proof of Lemma \ref{le1}, where we replace $(X^{(k)}_{li},X^{(k)}_{lj})$ by $(\varepsilon^{(k)}_{li},X^{(k)}_{lh})$ or $(\varepsilon^{(k)}_{li},\textbf{X}^{(k)}_{l,-i}\be_{i})$ and $\ps_{[i,j]}$ by $\diag(\gamma_{ii}^{-1},\psi_{ii})$ or $\diag(\gamma_{ii}^{-1},(\psi_{ii}\gamma_{ii}-1)/\gamma_{ii})$,
we can obtain Lemma \ref{le2} immediately. \qed

\begin{lemma}\label{le3} (i). We have, as $np\rightarrow\infty$,
	\begin{eqnarray*}
		\frac{\sum_{k=1}^{n}\sum_{l=1}^{p}(\tilde{\varepsilon}^{(k)}_{li}\tilde{\varepsilon}^{(k)}_{lj}-\ep \tilde{\varepsilon}^{(k)}_{li}\tilde{\varepsilon}^{(k)}_{lj})}{\sqrt{n_{1}\|\S\|^{2}_{F}}}\rightarrow N\Big{(}0,\frac{1}{\gamma_{ii}\gamma_{jj}}+\frac{\gamma^{2}_{ij}}{(\gamma_{ii}\gamma_{jj})^{2}}\Big{)}
	\end{eqnarray*}
	in distribution.
	
	(ii).  For any $M>0$, there exists a constant $C$ such that
	\begin{eqnarray*}
		\pr\Big{(}\max_{1\leq i\leq j\leq q}\Big{|}\hat{\sigma}_{ij,\varepsilon}-\frac{\tr(\S)}{p}\frac{\gamma_{ij}}{\gamma_{ii}\gamma_{jj}}\Big{|}\geq C\sqrt{\frac{\log\max(q,np)}{np}}\Big{)}=O((q+np)^{-M}).
	\end{eqnarray*}
\end{lemma}

\noindent{\bf Proof.} Note that $\Cov(\varepsilon^{(k)}_{li},\varepsilon^{(k)}_{li})=\frac{\gamma_{ij}}{\gamma_{ii}\gamma_{jj}}$. It is easy to show that $\Cov((\va^{(k)}_{i},\va^{(k)}_{j}))=\S\otimes \Delta_{[i,j]}$, where
$\Delta_{[i,j]}=\left(
\begin{array}{cc}
\frac{1}{\gamma_{ii}} & \frac{\gamma_{ij}}{\gamma_{ii}\gamma_{jj}} \\
\frac{\gamma_{ji}}{\gamma_{ii}\gamma_{jj}} &\frac{1}{\gamma_{jj}}  \\
\end{array}
\right).$
As in the proof of Lemma \ref{le1}, we can write
\begin{eqnarray}\label{a16}
\sum_{k=1}^{n}\sum_{l=1}^{p}\tilde{\varepsilon}^{(k)}_{li}\tilde{\varepsilon}^{(k)}_{lj}
&=&\sum_{k=1}^{n-1}\sum_{l=1}^{p}\lambda^{(1)}_{l}\eta_{ki,l}\eta_{kj,l},
\end{eqnarray}
where $(\eta_{ki,l},\eta_{kj,l})$, $1\leq l\leq p$, $1\leq k\leq n-1$, are i.i.d. $N(0,\Delta_{[i,j]})$ random vectors. Note that $\Var(\eta_{ki,l}\eta_{kj,l})=\frac{1}{\gamma_{ii}\gamma_{jj}}+\frac{\gamma^{2}_{ij}}{(\gamma_{ii}\gamma_{jj})^{2}}$ and
$\sum_{l=1}^{p}(\lambda^{(1)}_{l})^{2}=\|\S\|^{2}_{\text{F}}$.
(i) follows from Lindeberg-Feller central limit theorem. (ii) follows from the exponential inequality in Lemma 1 in Cai and Liu (2011).
\qed

\section{Proof of Proposition \ref{prop}--\ref{prop2} on the properties of the proposed test statistics}

\noindent{\bf Proof of Proposition \ref{prop}.} For notational simplicity, let $n_{1}=n-1$
and $\zeta^{(k)}_{lji}=\tilde{\varepsilon}^{(k)}_{lj}+(X_{li}-\bar{X}_{li})\beta_{i,j}$.
Note that for all $k \in [n]$ and $i\neq j$,
\[
\hat{\varepsilon}^{(k)}_{lji}=\tilde{\varepsilon}^{(k)}_{lj}+(X_{li}-\bar{X}_{li})\beta_{i,j}
-(\textbf{X}^{(k)}_{l,-j}-\bar{\textbf{X}}_{l,-j})(\hat{\be}_{j,\setminus i}-\be_{j,\setminus i}),
\]
which implies that
\begin{eqnarray}\label{p1}
\hat{\varepsilon}^{(k)}_{lij}\hat{\varepsilon}^{(k)}_{lji}&=&\zeta^{(k)}_{lij}\zeta^{(k)}_{lji}-
\zeta^{(k)}_{lij}(\textbf{X}^{(k)}_{l,-j}-\bar{\textbf{X}}_{l,-j})(\hat{\be}_{j,\setminus i}-\be_{j,\setminus i})\cr
& &-\zeta^{(k)}_{lji}(\textbf{X}^{(k)}_{l,-i}-\bar{\textbf{X}}_{l,-i})(\hat{\be}_{i,\setminus j}-\be_{i,\setminus j})\cr
& &+(\hat{\be}_{i,\setminus j}-\be_{i,\setminus j})^{'}(\textbf{X}^{(k)}_{l,-i}-\bar{\textbf{X}}_{l,-i})^{'}(\textbf{X}^{(k)}_{l,-j}-\bar{\textbf{X}}_{l,-j})(\hat{\be}_{j,\setminus i}-\be_{j,\setminus i}).
\end{eqnarray}
Let $\sigma=\tr(\S)/p$. By the assumption (C1), we have $c^{-1} \leq \sigma \leq c$.  For the last term in (\ref{p1}), by Cauchy-Schwarz inequality, we have
\begin{align*}
&\left|  \frac{1}{n_1 p} \sum_{k=1}^{n}\sum_{l=1}^{p}(\hat{\be}_{i,\setminus j}-\be_{i,\setminus j})^{'}(\textbf{X}^{(k)}_{l,-i}-\bar{\textbf{X}}_{l,-i})^{'}(\textbf{X}^{(k)}_{l,-j}-\bar{\textbf{X}}_{l,-j})(\hat{\be}_{j,\setminus i}-\be_{j,\setminus i}) \ \right|\\
\leq & \max_{1 \leq i,j \leq p} \left| (\hat{\be}_{i,\setminus j}-\be_{i,\setminus j})^{'} \hat{\ps}_{-i,-i}(\hat{\be}_{i,\setminus j}-\be_{i,\setminus j}) \right|.
\end{align*}
For any $i,j \in [q]$, we have
\begin{eqnarray*}
	|(\hat{\be}_{i,\setminus j}-\be_{i,\setminus j})^{'}\hat{\ps}_{-i,-i}(\hat{\be}_{i,\setminus j}-\be_{i,\setminus j})|&\leq& |(\hat{\be}_{i,\setminus j}-\be_{i,\setminus j})^{'}(\hat{\ps}_{-i,-i}-\sigma\ps_{-i,-i})(\hat{\be}_{i,\setminus j}-\be_{i,\setminus j})|\cr
	& &+\sigma|(\hat{\be}_{i,\setminus j}-\be_{i,\setminus j})^{'}\ps_{-i,-i}(\hat{\be}_{i,\setminus j}-\be_{i,\setminus j})|.
\end{eqnarray*}
By Lemma \ref{le1},
\begin{eqnarray*}
	\max_{1 \leq i,j \leq q}|(\hat{\be}_{i,\setminus j}-\be_{i,\setminus j})^{'}(\hat{\ps}_{-i,-i}-\sigma\ps_{-i,-i})(\hat{\be}_{i,\setminus j}-\be_{i,\setminus j})|=O_{\pr}\Big{(}a^{2}_{n1}\sqrt{\frac{\log\max(q,np)}{np}}\Big{)}.
\end{eqnarray*}
Moreover,
\begin{eqnarray*}
	|(\hat{\be}_{i,\setminus j}-\be_{i,\setminus j})^{'}\ps_{-i,-i}(\hat{\be}_{i,\setminus j}-\be_{i,\setminus j})|= O_{\pr}(\lambda_{\max}(\ps)|\hat{\be}_{i}-\be_{i}|_{2}^{2})
\end{eqnarray*}
uniformly in $i \in [q]$.
Combining the above arguments,
\begin{eqnarray}\label{a35}
&&\left|  \frac{1}{n_1 p} \sum_{k=1}^{n}\sum_{l=1}^{p}(\hat{\be}_{i,\setminus j}-\be_{i,\setminus j})^{'}(\textbf{X}^{(k)}_{l,-i}-\bar{\textbf{X}}_{l,-i})^{'}(\textbf{X}^{(k)}_{l,-j}-\bar{\textbf{X}}_{l,-j})(\hat{\be}_{j,\setminus i}-\be_{j,\setminus i}) \ \right|\cr
&&\quad=O_{\pr}\Big{(}a_{n2}^{2}+a^{2}_{n1}\sqrt{\frac{\log\max(q,np)}{np}}\Big{)}.
\end{eqnarray}
Under the null $H_{0ij}: \gamma_{ij}=0$, we have
$\zeta^{(k)}_{lji}=\tilde{\varepsilon}^{(k)}_{lj}$. Note that
\begin{eqnarray}\label{a40}
(\textbf{X}^{(k)}_{l,-j}-\bar{\textbf{X}}_{l,-j})(\hat{\be}_{j,\setminus i}-\be_{j,\setminus i})=(\textbf{X}^{(k)}_{l,-\{i,j\}}-\bar{\textbf{X}}_{l,-\{i,j\}})(\hat{\be}_{j,-i}-\be_{j,-i}).
\end{eqnarray}
So
\begin{eqnarray*}
	\zeta^{(k)}_{lij}(\textbf{X}^{(k)}_{l,-j}-\bar{\textbf{X}}_{l,-j})(\hat{\be}_{j,\setminus i}-\be_{j,\setminus i})&=&\sum_{h\neq i,j}\tilde{\varepsilon}^{(k)}_{li}(X^{(k)}_{lh}-\bar{X}_{lh})
	(\hat{\beta}_{h,j}-\beta_{h,j}),
\end{eqnarray*}
where $\hat{\be}_{j}=(\hat{\beta}_{1,j},\ldots,\hat{\beta}_{p-1,j})^{'}$ and we set $\hat{\beta}_{p,j}=0$.
By Lemma \ref{le2} (i),
\begin{eqnarray}\label{a8}
&& \max_{1\leq i\leq j\leq q}\Big{|}\sum_{h\neq i,j}\frac{1}{n_1p}\sum_{k=1}^{n}\sum_{l=1}^{p}\tilde{\varepsilon}^{(k)}_{li}(X^{(k)}_{lh}-\bar{X}_{lh})
(\hat{\beta}_{h,j}-\beta_{h,j})\Big{|}\cr
& \leq &  \max_{1\leq i \leq j \leq q}\max_{h\neq i,j}\left|\frac{1}{n_1p}\sum_{k=1}^{n}\sum_{l=1}^{p}\tilde{\varepsilon}^{(k)}_{li}(X^{(k)}_{lh}-\bar{X}_{lh})  \right| |\hat{\be}_j-\be_j|_1\cr
&= & O_{\pr}\Big{(}a_{n1}\sqrt{\frac{\log\max(q,np)}{np}}\Big{)}.
\end{eqnarray}
A similar inequality holds for the third term on the right hand side of (\ref{p1}). Therefore, for $i\neq j$, under $\gamma_{ij}=0$,
\begin{eqnarray}\label{p4}
\frac{1}{n_{1}p}\sum_{k=1}^{n}\sum_{l=1}^{p}\hat{\varepsilon}^{(k)}_{lij}\hat{\varepsilon}^{(k)}_{lji}
&=&\frac{1}{n_{1}p}\sum_{k=1}^{n}\sum_{l=1}^{p}\tilde{\varepsilon}^{(k)}_{li}\tilde{\varepsilon}^{(k)}_{lj}\cr
& &+O_{\pr}\Big{(}(a^{2}_{n1}+a_{n1})\sqrt{\frac{\log\max(q,np)}{np}}+a_{n2}^{2}\Big{)}
\end{eqnarray}
uniformly in $1\leq i\neq j\leq q$. By (\ref{a35}) and (\ref{a8}) with $i=j$, we obtain that
\begin{eqnarray}\label{a41}
\hat{r}_{ii}
=\frac{1}{n_{1}p}\sum_{k=1}^{n}\sum_{l=1}^{p}(\tilde{\varepsilon}^{(k)}_{li})^{2}+O_{\pr}\Big{(}(a^{2}_{n1}+a_{n1})\sqrt{\frac{\log\max(q,np)}{np}}+a_{n2}^{2}\Big{)}
\end{eqnarray}
uniformly in $1\leq i\leq q$.
The proof of  Proposition \ref{prop} is complete by Lemma \ref{le3}.
\qed

\noindent{\bf Proof of Proposition \ref{propv}.} By Lemma \ref{le1}, it is easy to show that
\begin{eqnarray}\label{a33}
\frac{1}{n_{1}p}\sum_{k=1}^{n}\sum_{l=1}^{p}(X_{li}-\bar{X}_{li})(\textbf{X}^{(k)}_{l,-i}-\bar{\textbf{X}}_{l,-i})(\hat{\be}_{i,\setminus j}-\be_{i,\setminus j})=O_{\pr}(a_{n1})
\end{eqnarray}
uniformly in $i,j$.
Also, by (\ref{a40}) and (\ref{a8}),
\begin{eqnarray}\label{a34}
\frac{1}{n_1 p} \sum_{k=1}^{n}\sum_{l=1}^{p}\tilde{\varepsilon}^{(k)}_{li}(\textbf{X}^{(k)}_{l,-j}-\bar{\textbf{X}}_{l,-j})(\hat{\be}_{j,\setminus i}-\be_{j,\setminus i})=O_{\pr}\Big{(}a_{n1}\sqrt{\frac{\log\max(q,np)}{np}}\Big{)}.
\end{eqnarray}
By (\ref{a35}), (\ref{a41}), (\ref{a33}) and (\ref{a34}),
it suffices to prove that
\begin{eqnarray}\label{a36}
\frac{p}{\tr(\S)}\frac{\sqrt{\gamma_{ii}\gamma_{jj}}}{n_{1}p}\sum_{k=1}^{n}\sum_{l=1}^{p}\zeta^{(k)}_{lij}\zeta^{(k)}_{lji}-(1-\gamma_{ij}\psi_{ij})\rho^{\Ga}_{ij\cdot}\rightarrow 0
\end{eqnarray}
in probability. We have
\begin{eqnarray*}
	\ep\Big{[}(\varepsilon^{(k)}_{lj}+X_{li}\beta_{i,j})(\varepsilon^{(k)}_{li}+X_{lj}\beta_{i,j})\Big{]}=\sigma_{ll}(-\frac{\gamma_{ij}}{\gamma_{ii}\gamma_{jj}}
	+\psi_{ij}\frac{\gamma^{2}_{ij}}{\gamma_{ii}\gamma_{jj}}).
\end{eqnarray*}
Now, as the proof of Lemma \ref{le1}, we can write
\begin{eqnarray}\label{a37}
\sum_{k=1}^{n}\sum_{l=1}^{p}\zeta^{(k)}_{lij}\zeta^{(k)}_{lji}=\sum_{k=1}^{n-1}\sum_{l=1}^{p}\lambda^{(1)}_{l}\xi_{ki,l}\xi_{kj,l},
\end{eqnarray}
where $(\xi_{ki,l},\xi_{kj,l})$, $1\leq l\leq p$, are i.i.d. with $\ep \xi_{ki,l}\xi_{kj,l}=(-\frac{\gamma_{ij}}{\gamma_{ii}\gamma_{jj}}
+\psi_{ij}\frac{\gamma^{2}_{ij}}{\gamma_{ii}\gamma_{jj}})$. This proves (\ref{a36}). \qed

\noindent{\bf Proof of Proposition \ref{prop2}.} Let $\phi_{ij}=\frac{\tr(\Ga)}{q}\sigma_{ij}$ and define $\widetilde{\lambda}= \lambda \sqrt{\frac{\log \max(p, nq)}{nq}}$. We have
\begin{eqnarray*}
	\sum_{j=1}^{p}\hat{\sigma}^{2}_{ij,\lambda}=\sum_{j=1}^{p}(\hat{\sigma}^{2}_{ij}-\phi^{2}_{ij})I\{|\hat{\sigma}_{ij}|\geq  \widetilde{\lambda} \}
	+\sum_{j=1}^{p}\phi^{2}_{ij}I\{|\hat{\sigma}_{ij}|\geq \widetilde{\lambda}\}.
\end{eqnarray*}
Also by Lemma \ref{le1}, with probability tending to one,
\begin{eqnarray*}
	\sum_{j=1}^{p}\phi^{2}_{ij}I\{|\hat{\sigma}_{ij}|<\widetilde{\lambda}\}\leq\sum_{j=1}^{p}\phi^{2}_{ij}I\{|\phi_{ij}|<2\widetilde{\lambda}\}
	=O(\tilde{\lambda}^{2-\tau}s(p))
\end{eqnarray*}
uniformly in $1\leq i\leq p$ by the assumption (C2).
So, for the last term,
\begin{eqnarray*}
	\sum_{j=1}^{p}\phi^{2}_{ij}I\{|\hat{\sigma}_{ij}|\geq  \widetilde{\lambda}\}=\sum_{j=1}^{p}\phi^{2}_{ij}-\sum_{j=1}^{p}\phi^{2}_{ij}I\{|\hat{\sigma}_{ij}|<\widetilde{\lambda}\}
	=(1+O_{\pr}(\tilde{\lambda}^{2-\tau}s(p)))\sum_{j=1}^{p}\phi^{2}_{ij}
\end{eqnarray*}
uniformly in $1\leq i\leq p$. Moreover, with probability tending to  one,
\begin{eqnarray*}
	\sum_{j=1}^{p}|\hat{\sigma}^{2}_{ij}-\phi^{2}_{ij}|I\{|\hat{\sigma}_{ij}|\geq  \widetilde{\lambda}\}&=&\sum_{j=1}^{p}|\hat{\sigma}_{ij}+\phi_{ij}||\hat{\sigma}_{ij}-\phi_{ij}|I\{|\hat{\sigma}_{ij}|\geq  \widetilde{\lambda}\}\cr
	&\leq& C\sum_{j=1}^{p}|\phi_{ij}||\hat{\sigma}_{ij}-\phi_{ij}|I\{|\phi_{ij}|\geq \widetilde{\lambda}/2\}\cr
	&\leq& C \widetilde{\lambda}^{-(\tau-1)\vee 0}\sum_{j=1}^{p}|\phi_{ij}|^{\tau}|\hat{\sigma}_{ij}-\phi_{ij}|\cr
	&\leq& C\widetilde{\lambda}^{(2-\tau)\wedge 1}s(p),
\end{eqnarray*}
where the last inequality following from  $\max_{1\leq i\leq j\leq q}|\hat{\sigma}_{ij}-\phi_{ij}|=O_{\pr}(\tilde{\lambda})$
by Lemma \ref{le1}.
It implies that $\max_{1\leq i\leq q}|\sum_{j=1}^{p}\hat{\sigma}^{2}_{ij,\lambda}-\sum_{j=1}^{p}\phi^{2}_{ij}|=O_{\pr}(\widetilde{\lambda}^{(2-\tau)\wedge 1}s(p))$
and hence
\begin{eqnarray*}
	\frac{\|\hat{\S}_{\lambda}\|^{2}_{F}}{\|\S\|^{2}_{F}}=O_{\pr}(\widetilde{\lambda}^{(2-\tau)\wedge 1}s(p)).
\end{eqnarray*}
By Lemma \ref{le1}, we have $\tr(\hat{\S}_{\lambda})/\tr(\S)=O_{\pr}(\tilde{\lambda})$. This implies this proposition holds.
\qed

\section{Proof of Theorems \ref{th1}--\ref{th4} on FDR control and power analysis}

It is easy to see that the Benjamini and Hochberg (BH) method is equivalent to reject $H_{0ij}$ if $|\hat{T}_{ij}|\geq \hat{t}$, where
\begin{eqnarray}\label{a3}
\hat{t}=\inf\Big{\{}t\geq 0: \frac{G(t)(q^{2}-q)/2}{\max\{\sum_{1\leq i < j\leq q}I\{|\hat{T}_{ij}|\geq t\},1\}}\leq \alpha\Big{\}},
\end{eqnarray}
where $G(t):= 2 -2 \Phi(t)$. We first give some key lemmas which are
the generalization of Lemmas 6,1 and 6.2 in \cite{Liu2013} from i.i.d. case to independent case (but not necessarily identically distributed).

Let $\xi_{1},\ldots,\xi_{n}$ be independent $d$-dimensional random vectors with
mean zero. Define $|\cdot|_{(d)}$ by $|\z|_{(d)}=\min\{|z_{i}|; 1\leq i\leq d\}$ for $\z=(z_{1},\ldots,z_{d})^{'}$.
Let $(p,n)$ be  a sequence of  positive integers and the constants $c,r,b,\gamma,K,C$ mentioned below do not depend on $(p,n)$.

\begin{lemma}\label{le0} Suppose that $p\leq cn^{r}$ and $\max_{1\leq k\leq n}\ep|\xi_{k}|_{2}^{bdr+2+\epsilon}\leq K$ for some fixed $c>0$, $r>0$, $b>0$, $K>0$ and $\epsilon>0$. Assume that
	$\|\frac{1}{n}\Cov(\sum_{k=1}^{n}\xi_{k})-\I_{d}\|\leq C(\log p)^{-2-\gamma}$ for some $\gamma>0$ and $C>0$. Then we have
	\begin{eqnarray*}
		\sup_{0\leq t\leq \sqrt{b\log p}}\Big{|}\frac{\pr(|\sum_{k=1}^{n}\xi_{k}|_{(d)}\geq t\sqrt{n})}{(G(t))^{d}}-1\Big{|}\leq C(\log p)^{-1-\gamma_{1}}
	\end{eqnarray*}
	for  $\gamma_{1}=\min\{\gamma,1/2\}$.
\end{lemma}

Let  $\boldsymbol{\eta}_{k}=(\eta_{k1},\eta_{k2})^{'}$, $1\leq k\leq n$, are independent  $2$-dimensional random vectors with
mean zero.

\begin{lemma}\label{le00} Suppose that $p\leq cn^{r}$ and $\max_{1\leq k\leq n}\ep|\boldsymbol{\eta}_{k}|_{2}^{2br+2+\epsilon}<\infty$ for some fixed $c>0$, $r>0$, $b>0$ and $\epsilon>0$. Assume that
	$\sum_{k=1}^{n}\Var(\eta_{k1})=\sum_{k=1}^{n}\Var(\eta_{k2})=n$ and $|\frac{1}{n}\sum_{k=1}^{n}\Cov(\eta_{k1},\eta_{k2})|\leq \delta$ for some $0\leq \delta<1$. Then we have
	\begin{eqnarray*}
		\pr\Big{(}|\sum_{k=1}^{n}\eta_{k1}|\geq t\sqrt{n},|\sum_{k=1}^{n}\eta_{k2}|\geq t\sqrt{n}\Big{)}\leq C(t+1)^{-2}\exp(-t^{2}/(1+\delta))
	\end{eqnarray*}
	uniformly for $0\leq t\leq \sqrt{b\log p}$, where $C$ only depends on $c,b,r,\epsilon,\delta$.
\end{lemma}

The proofs of Lemmas \ref{le0} and \ref{le00} are   the same as those of Lemma 6.1 and 6.2 in \cite{Liu2013}.

Recall $\eta_{ki,l}$  in (\ref{a16}) and $\xi_{ki,l}$ in (\ref{a37}). For $1 \leq i < j \leq q$, let
\begin{eqnarray}\label{a9}
&&U_{ij}=\frac{\sum_{k=1}^{n-1}\sum_{l=1}^{p}\lambda^{(1)}_{l}(\eta_{ki,l}\eta_{kj,l}-\ep \eta_{ki,l}\eta_{kj,l}) (\gamma_{ii}\gamma_{jj})^{1/2}}{\sqrt{(n-1)pE_{p}}},\cr
&&V_{ij}=\frac{\sum_{k=1}^{n-1}\sum_{l=1}^{p}\lambda^{(1)}_{l}(\xi_{ki,l}\xi_{kj,l}-\ep \xi_{ki,l}\xi_{kj,l}) (\gamma_{ii}\gamma_{jj})^{1/2}}{\sqrt{(n-1)pE_{p}}},
\end{eqnarray}
where $E_{p}=p^{-1}\sum_{l=1}^{p}(\lambda^{(1)}_{l})^{2}$. Note that $\lambda^{(1)}_{l}$ are bounded away from zero and infinity. Also,
$\Var(\eta_{ki,l}\eta_{kj,l})=(\gamma_{ii}\gamma_{jj})^{-1}(1+\gamma^{2}_{ij}(\gamma_{ii}\gamma_{jj})^{-1})$, $\Var(U_{ij})=1+\gamma^{2}_{ij}(\gamma_{ii}\gamma_{jj})^{-1}$ and
$Corr(U_{ij},U_{kl})=Corr(\eta_{1i,1}\eta_{1j,1},\eta_{1k,1}\eta_{1l,1})$. By Lemma \ref{le0} with $d=1$, we have
\begin{eqnarray*}
	&&\max_{1\leq i,j\leq q}\sup_{0\leq t\leq 4\sqrt{\log q}}\Big{|}\frac{\pr(|U_{ij}|\geq t\sqrt{1+\gamma^{2}_{ij}(\gamma_{ii}\gamma_{jj})^{-1}})}{G(t)}-1\Big{|}\leq C(\log q)^{-1-\epsilon},\cr
	&&\max_{1\leq i,j\leq q}\sup_{0\leq t\leq 4\sqrt{\log q}}\Big{|}\frac{\pr(|V_{ij}|\geq t\sqrt{\Var(\xi_{1i,1}\xi_{1j,1})\gamma_{ii}\gamma_{jj}})}{G(t)}-1\Big{|}\leq C(\log q)^{-1-\epsilon},
\end{eqnarray*}
for some $\epsilon>0$. Therefore, $\max_{1\leq i,j\leq q}|U_{ij}|=O_{\pr}(\sqrt{\log q})$ and $\max_{1\leq i,j\leq q}|V_{ij}|=O_{\pr}(\sqrt{\log q})$.
This, together with (\ref{a41}), Lemma \ref{le3}, Proposition \ref{prop2}, the proof of Proposition \ref{propv}, (\ref{cd2}) and (\ref{p4}), implies that
\begin{eqnarray}\label{a39}
\max_{1\leq i<j\leq q}\Big{|}\hat{T}_{ij}+b_{nij}\sqrt{\frac{(n-1)p}{A_{p}}}(1-\gamma_{ij}\psi_{ij})\rho^{\Ga}_{ij\cdot}-V_{ij}\Big{|}=o_{\pr}((\log q)^{-1/2})
\end{eqnarray}
as $np, q \rightarrow\infty$, where $b_{nij}$ satisfies $\max_{1\leq i<j\leq q}|b_{nij}-1|\rightarrow 0$ in probability. Note that under the null $\gamma_{ij}=0$, $U_{ij}=V_{ij}$.
Now  Theorem \ref{th1}
follows from the proof of Theorem 3.1 in \cite{Liu2013} step by step, by using Lemmas \ref{le0} and \ref{le00} and replacing
$U_{ij}$ in \cite{Liu2013} by $U_{ij}$ in (\ref{a9}) and the sample size in \cite{Liu2013} by $(n-1)p$. The proof of Theorem \ref{th2} is similar.
Theorem \ref{th3} follows from the formula of FDP and Theorems \ref{th1} and \ref{th2}.

Under (\ref{a38}), we have $\Var(\xi_{1i,1}\xi_{1j,1})\gamma_{ii}\gamma_{jj}=1+o(1)$ uniformly in $i,j$. Hence $\max_{1\leq i<j\leq q}|V_{ij}|\leq (2+\delta)\sqrt{\log q}$ for some $\delta>0$ with probability tending to one. This shows that $$\pr\Big{(}\min_{(i,j)\in\mathcal{H}_{1}}|\hat{T}_{ij}|\geq (2+\delta^{'})\sqrt{\log q}\Big{)}\rightarrow 1.$$
for some $\delta^{'}>0$.  By the definition of $\hat{t}$ in \eqref{a3}, we have $\hat{t}\leq 2\sqrt{\log q}$ as $q\rightarrow\infty$. Thus,
$\pr(\mathcal{H}_{1}\subseteq\widehat{\text{supp}(\Ga)})\rightarrow 1$. Similarly, we can show that $\pr(\mathcal{H}^{'}_{1}\subseteq\widehat{\text{supp}(\O)})\rightarrow 1$. This finishes the proof of Theorem \ref{th4}.\qed

\section{Proof of Proposition \ref{pro3-3} on the convergence rate of $\hat{\be}_j$}
\label{sec:proof_init}
\noindent{\bf Proof of Proposition \ref{pro3-3}.} Define
\begin{eqnarray*}
	\hat{\a}_{j}=\frac{1}{(n-1)p}\sum_{k=1}^{n}\sum_{l=1}^{p}(\textbf{X}^{(k)}_{l,-j}-\bar{\textbf{X}}_{l,-j})^{'}(X^{(k)}_{lj}-\bar{X}_{lj}).
\end{eqnarray*}
We let $\theta_{nj}$ and $\bal_{j}$ denote $\theta_{nj}(\delta)=\delta\sqrt{\frac{\hat{\psi}_{jj} \log q}{np}}$ and $\bal_{j}(\delta)$ (defined in \eqref{eq:lasso_alpha}), respectively.
By the Karush-Kuhn-Tucker (KKT) condition, we have
\begin{eqnarray}\label{cs0}
\Big{|}\textbf{D}^{-1/2}_{j}\hat{\ps}_{-j,-j}\hat{\be}_{j}-\textbf{D}^{-1/2}_{j}\hat{\a}_{j}\Big{|}_{\infty}\leq \theta_{nj}.
\end{eqnarray}
By Lemma \ref{le1}, we have $c^{-1}\leq \min_{1\leq j\leq q-1}\D_{j}\leq \max_{1\leq j\leq q-1}\D_{j}\leq c$ for some $c>0$ with probability tending to one. This, together with
Lemma \ref{le2}, implies that, for sufficiently large $\delta$,
\begin{eqnarray}\label{de}
\Big{|}\frac{1}{(n-1)p}\textbf{D}^{-1/2}_{j}\sum_{k=1}^{n}\sum_{l=1}^{p}(\textbf{X}^{(k)}_{l,-j}-\bar{\textbf{X}}_{l,-j})^{'}\tilde{\varepsilon}^{(k)}_{lj}\Big{|}_{\infty}\leq
\frac{1}{2}\theta_{nj}
\end{eqnarray}
uniformly in $1\leq j\leq q$, with probability tending to one. Note that
$\tilde{\varepsilon}^{(k)}_{lj}=X^{(k)}_{lj}-\bar{X}_{lj}-(\textbf{X}^{(k)}_{l,-j}-\bar{\textbf{X}}_{l,-j})\be_{j}.
$
Therefore,
\begin{eqnarray}\label{cs}
\Big{|}\textbf{D}^{-1/2}_{j}\hat{\ps}_{-j,-j}\be_{j}-\textbf{D}^{-1/2}_{j}\hat{\a}_{j}\Big{|}_{\infty}\leq \frac{1}{2}\theta_{nj}
\end{eqnarray}
uniformly in $1\leq j\leq q$.
Note that inequalities (\ref{cs0}) and (\ref{cs}) imply that
\begin{eqnarray}\label{a12}
\Big{|}\textbf{D}^{-1/2}_{j}\hat{\ps}_{-j,-j}(\hat{\be}_{j}-\be_{j})\Big{|}_{\infty}\leq \frac{3}{2}\theta_{nj}.
\end{eqnarray}
Define $\boldsymbol{\Lambda}=\diag(\ps)^{-1/2}\ps \diag(\ps)^{-1/2}$.
For any subset $T\subset\{1,2,\cdots,q-1\}$ and $\nuu\in \R^{q-1}$ with $|T|=o\Big{(}\sqrt{\frac{np}{\log\max(q,np)}}\Big{)}$ and $|\nuu_{T^{c}}|_{1}\leq c|\nuu_{T}|_{1}$ for some $c>0$, by Lemma \ref{le1} and the conditions in Proposition \ref{pro3-3}, we have
\begin{eqnarray}\label{a11}
\nuu^{'}\textbf{D}^{-1/2}_{j}\hat{\ps}_{-j,-j}\textbf{D}^{-1/2}_{j}\nuu\geq \lambda_{\min}(\boldsymbol{\Lambda}_{-j,-j})|\nuu|_2^{2}-O_{\pr}\Big{(}\sqrt{\frac{\log\max(q,np)}{np}}\Big{)}|\nuu|^{2}_{1}\geq
|\nuu|_2^{2}/C,
\end{eqnarray}
for some constant $C>0$, where the first inequality follows from the fact
\begin{eqnarray*}
	|\nuu^{'}(\textbf{D}^{-1/2}_{j}\hat{\ps}_{-j,-j}\textbf{D}^{-1/2}_{j}-\boldsymbol{\Lambda}_{-j,-j})\nuu|\leq |\textbf{D}^{-1/2}_{j}\hat{\ps}_{-j,-j}\textbf{D}^{-1/2}_{j}-\boldsymbol{\Lambda}_{-j,-j}|_{\infty}|\nuu|^{2}_{1}
\end{eqnarray*}
and the second inequality follows from the fact $|\nuu|^{2}_{1}\leq (1+c)^{2}|\nuu_{T}|^{2}_{1}\leq (1+c)^{2}|T||\nuu|^{2}_{2}$.

Now let $T$ be the support of $\be_{j}$, $\bal_{j}=\D_{j}^{1/2}\be_{j}$ and $\nuu=\D^{1/2}_{j}(\hat{\be}_{j}-\be_{j})=\hat{\bal}_{j}-\bal_{j}$. We first show that $|\nuu_{T^{c}}|_{1}\leq 3|\nuu_{T}|_{1}$ uniformly in $1\leq j\leq q$ with probability tending to one. Define
\begin{eqnarray*}
	Q(\bal_j)&=&\frac{1}{2(n-1)p}\sum_{k=1}^{n}\sum_{l=1}^{p}(X^{(k)}_{lj}-\bar{X}_{lj}-(\textbf{X}^{(k)}_{l,-j}-\bar{\textbf{X}}_{l,-j})\D_{j}^{-1/2}\bal_j)^{2},\cr
	S(\bal_{j})&=&\textbf{D}^{-1/2}_{j}\hat{\a}_{j}-\textbf{D}^{-1/2}_{j}\hat{\ps}_{-j,-j}\be_{j}.
\end{eqnarray*}
Note that $S(\bal_{j})$ is the gradient of $Q(\bal_j)$. By the definition of $\hat{\bal}_{j}$,
we have
\begin{eqnarray*}
	Q(\hat{\bal}_{j})-Q(\bal_{j})\leq \theta_{nj}(\delta)|\bal_{j}|_{1}-\theta_{nj}|\hat{\bal}_{j}|_{1}\leq \theta_{nj}(|\nuu_{T}|_{1}-|\nuu_{T^{c}}|_{1}),
\end{eqnarray*}
and by (\ref{cs}), with probability tending to one,
\begin{eqnarray*}
	Q(\hat{\bal}_{j})-Q(\bal_{j})\geq S^{'}(\bal_{j})\nuu\geq -\frac{1}{2}\theta_{nj}|\nuu|_{1}=-\frac{1}{2}\theta_{nj}(|\nuu_{T}|_{1}+|\nuu_{T^{c}}|_{1})
\end{eqnarray*}
uniformly in $1\leq j\leq q$.
It follows from the above two inequalities that $|\nuu_{T^{c}}|_{1}\leq 3|\nuu_{T}|_{1}$. So by (\ref{a12}) and (\ref{a11}) we have
\begin{eqnarray*}
	|\nuu|_2^{2}&\leq& C\nuu^{'}\textbf{D}^{-1/2}_{j}\hat{\ps}_{-j,-j}\textbf{D}^{-1/2}_{j}\nuu\cr
	&\leq& C |\textbf{D}^{-1/2}_{j}\hat{\ps}_{-j,-j}\textbf{D}^{-1/2}_{j}\nuu|_{\infty} |\nuu|_{1}\cr
	&\leq& \frac{3}{2} C\theta_{nj}(|\nuu_{T}|_{1}+|\nuu_{T^{c}}|_{1})\cr
	& \leq & 6C \theta_{nj} |\nuu_{T}|_{1} \\
	&\leq& 6 C \theta_{nj}\sqrt{|\be_{j}|_{0}}|\nuu_{T}|_2
\end{eqnarray*}
uniformly in $1\leq j\leq q$ with probability tending to one.  By noting that $c^{-1}\leq \min_{1\leq j\leq q-1}\D_{j}\leq \max_{1\leq j\leq q-1}\D_{j}\leq c$ with probability tending to one, we have $|\hat{\be}_{j}-\be_{j}|_{2}\leq c|\nuu|_{2}$. Hence, by the conditions in Proposition \ref{pro3-3}, we have $a_{n2}=o_{\pr}\Big{(}(np\log q)^{-1/4}\Big{)}$. Note that $|\nuu|_{1}\leq 4 |\nuu_{T}|_{1}\leq 4\sqrt{|\be_{j}|_{0}}|\nuu_{T}|_{2}=o((\log \max(q,np))^{-1})$ uniformly in $1\leq j\leq q-1$ with probability tending to one. This proves Proposition \ref{pro3-3} holds.
\qed

\section{Additional Experiments}
\label{sec:add_exp}
In this section, we present some additional simulation studies and real data analysis. We first note that for the choice of tuning parameters,  our theoretical results will hold for any large enough constants $\lambda$ in \eqref{eq:thresh_sigma} for estimating $\hat{A}_p$ (see Proposition \ref{prop2}) and $\delta>0$ in \eqref{a10} for $\hat{\be}_{j}(\delta)$ (see Proposition \ref{pro3-3}). In our experiment, we will adopt a data-driven parameter-tuning strategy from \cite{Liu2013}.
In particular, $\lambda$ and $\delta$ are selected by
\begin{eqnarray}\label{ats}
(\hat{\lambda},\hat{\delta})=\argmin_{\lambda,\delta}\sum_{k=3}^{9}\left(\frac{\sum_{1\leq i\neq j\leq q}I\{|\hat{T}_{ij}(\lambda,\delta)|\geq \Phi^{-1}(1-\frac{k}{20})\}}{k(q^{2}-q)/10}-1\right)^{2},
\end{eqnarray}
where $\hat{T}_{ij}(\lambda,\delta)$ is the test statistic in (\ref{t1}) with an initial estimator $\hat{\be}_{j}(\delta)$ and $\hat{A}_{p}$ (depending on the threshold $\lambda$). The choice of $(\lambda,\delta)$ in (\ref{ats}) makes the distributions of  $\hat{T}_{ij}$, on average,  close to the standard normal distribution. We note that although the parameter searching is conducted on a two-dimensional grid on $\lambda$ and $\delta$, the main computational cost is the construction of $\hat{\be}_{j}(\delta)$, which is irrelevant  of $\lambda$. Therefore, the computational cost of the parameter searching is moderate.

\subsection{Boxplots of FDPs}
\label{sec:supp_box}

\begin{figure}[!t]
	\centering
	\subfigure[b][$p=100$, $q=100$]{
		\includegraphics[width=0.4\textwidth]{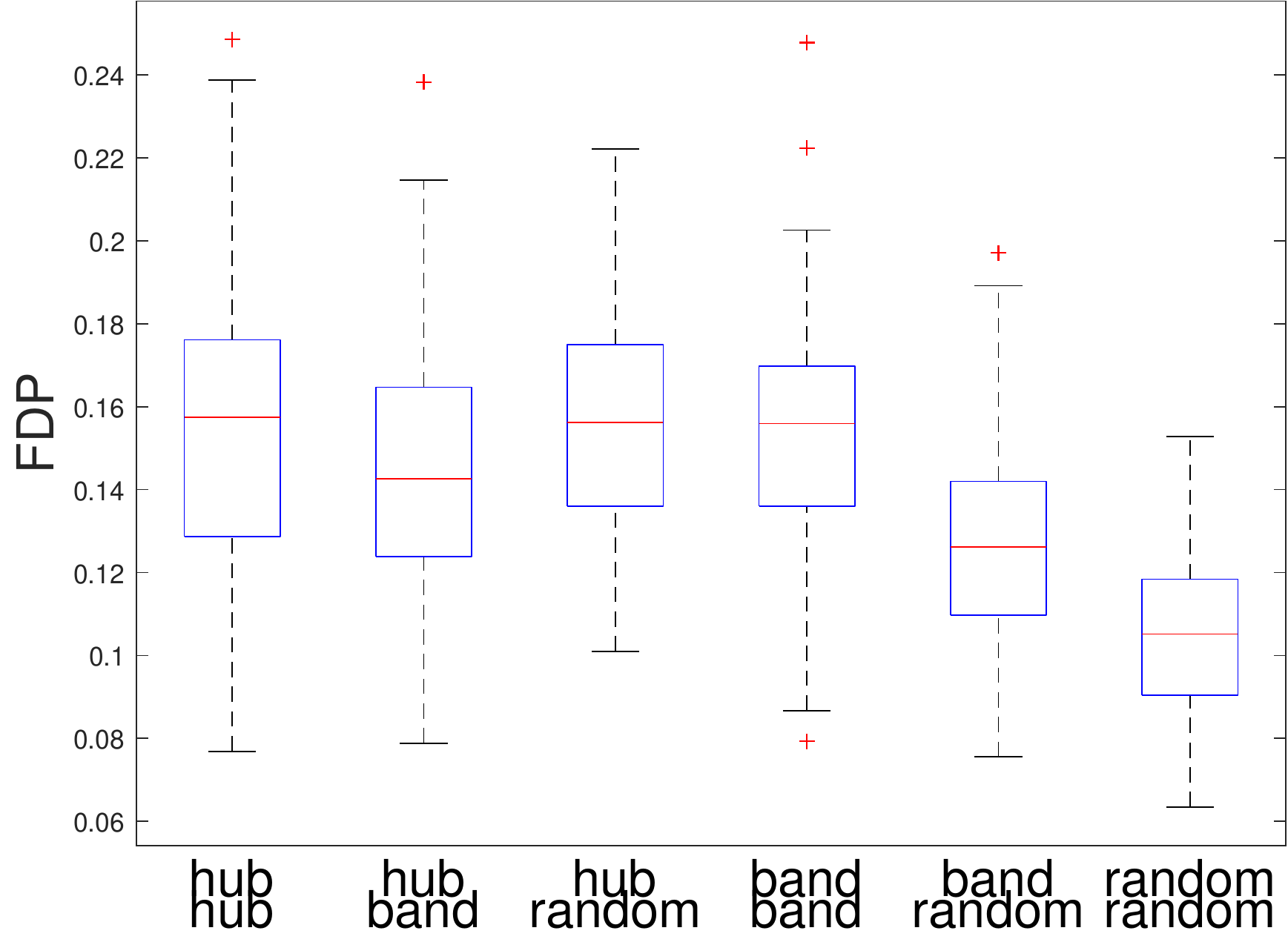}
		\label{fig:n100_p100_q100}
	}
	\subfigure[b][$p=200$, $q=50$]{
		\includegraphics[width=0.4\textwidth]{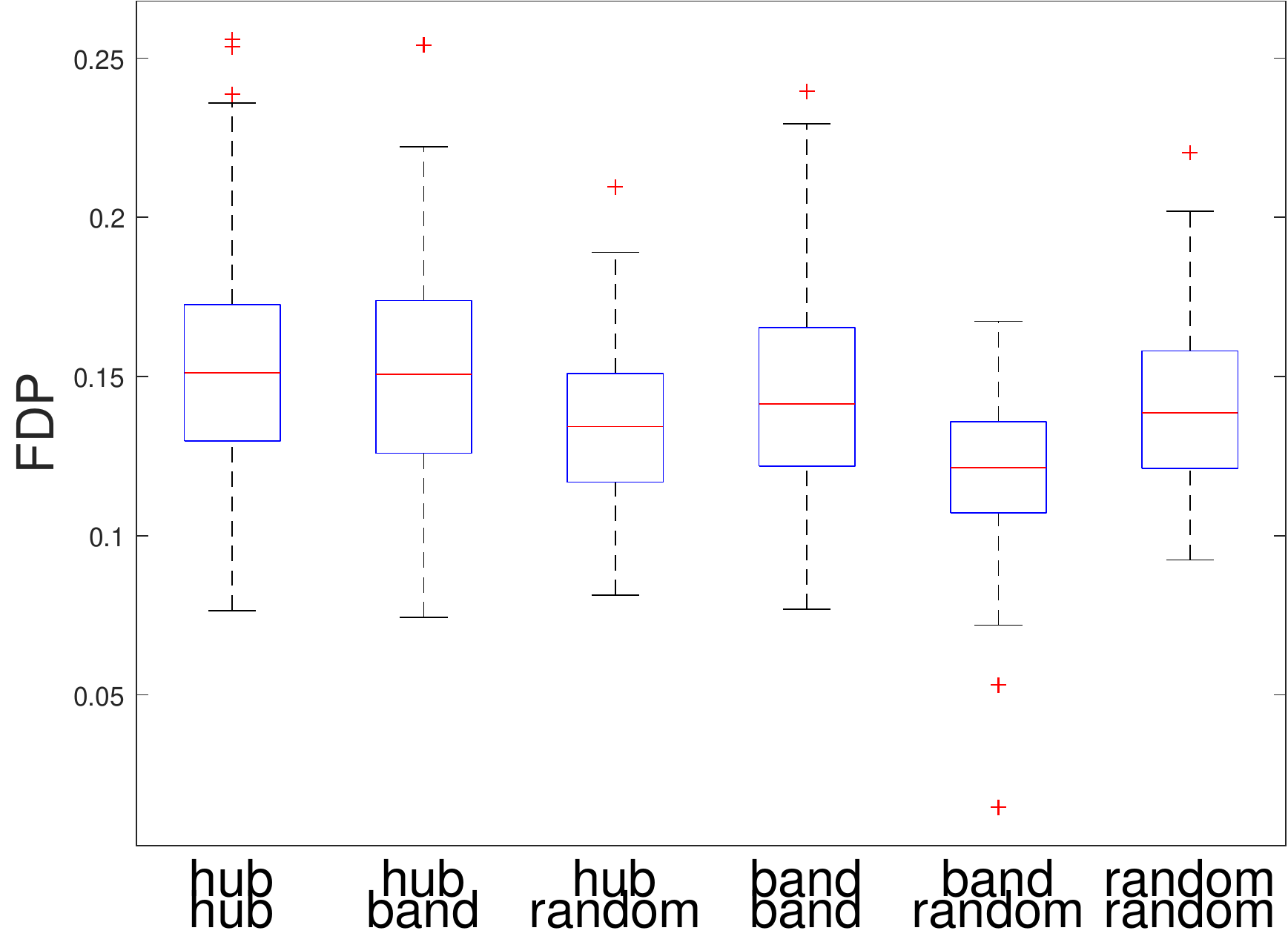}
		\label{fig:n100_p200_q50}
	}\\%
	\subfigure[b][$p=200$, $q=200$]{
		\includegraphics[width=0.4\textwidth]{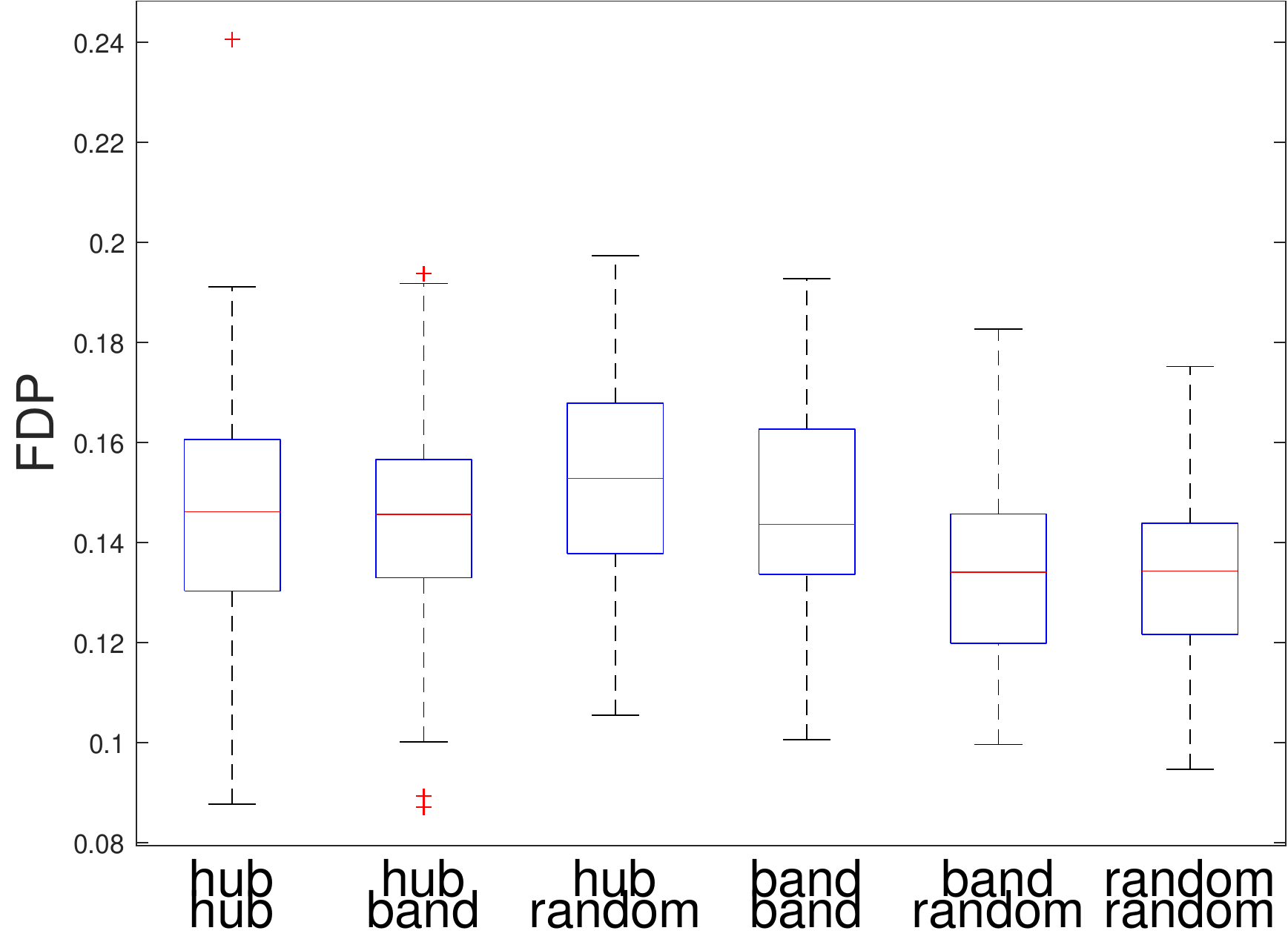}
		\label{fig:n100_p200_q50}
	}
	\subfigure[b][$p=400$, $q=400$]{
		\includegraphics[width=0.4\textwidth]{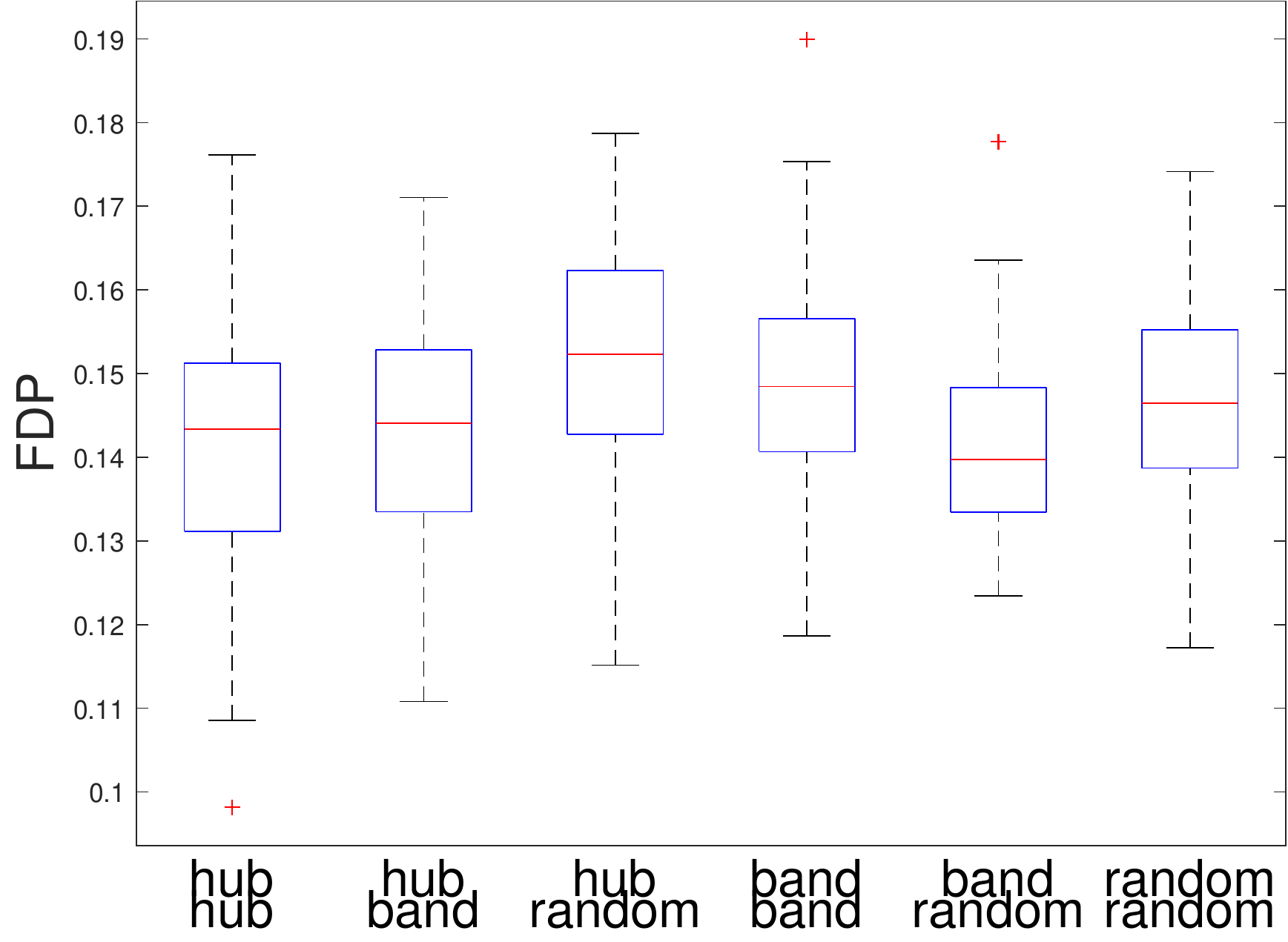}
		\label{fig:n100_p200_q50}
	}
	\caption{Boxplots for FDP when $n=100$ and $\alpha=0.1$. }
	\label{fig:box_n100}
\end{figure}

We present the boxplots of FDPs when $n=100$ over 100 replications in Figure \ref{fig:box_n100} for different $p$, $q$, and precision matrix structures. As we can see from Figure \ref{fig:box_n100}, FDPs are well concentrated, which suggests that the performance of the proposed estimator is quite stable.

\subsection{Estimation of $\widehat{A}_p$}
\label{sec:supp_A_p}

\begin{figure}[!t]
	\centering
	\subfigure[b][$n=20$ ]{
		\includegraphics[width=0.4\textwidth]{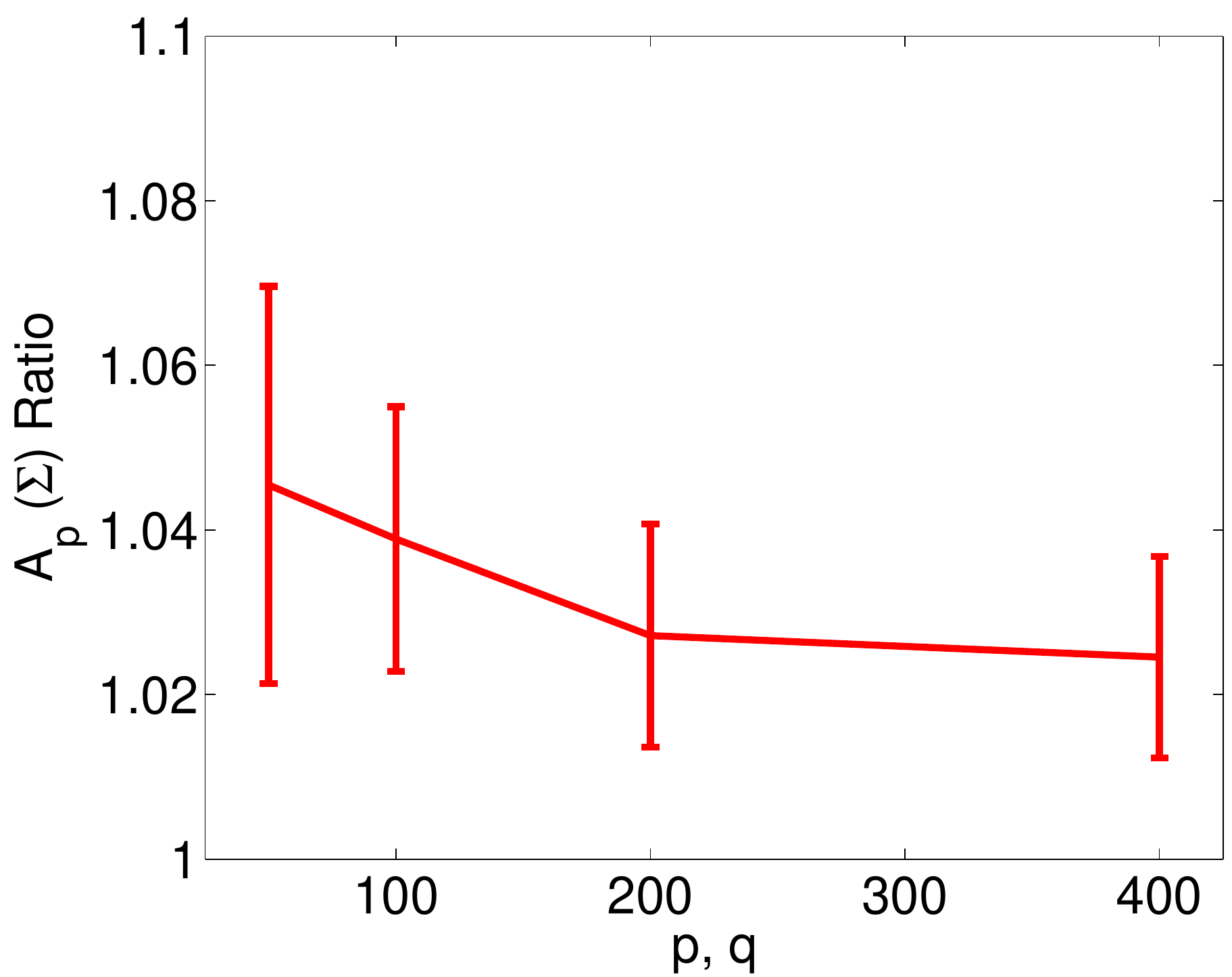}
		\label{fig:A_ratio_n_20}
	}
	\subfigure[b][$n=100$]{
		\includegraphics[width=0.4\textwidth]{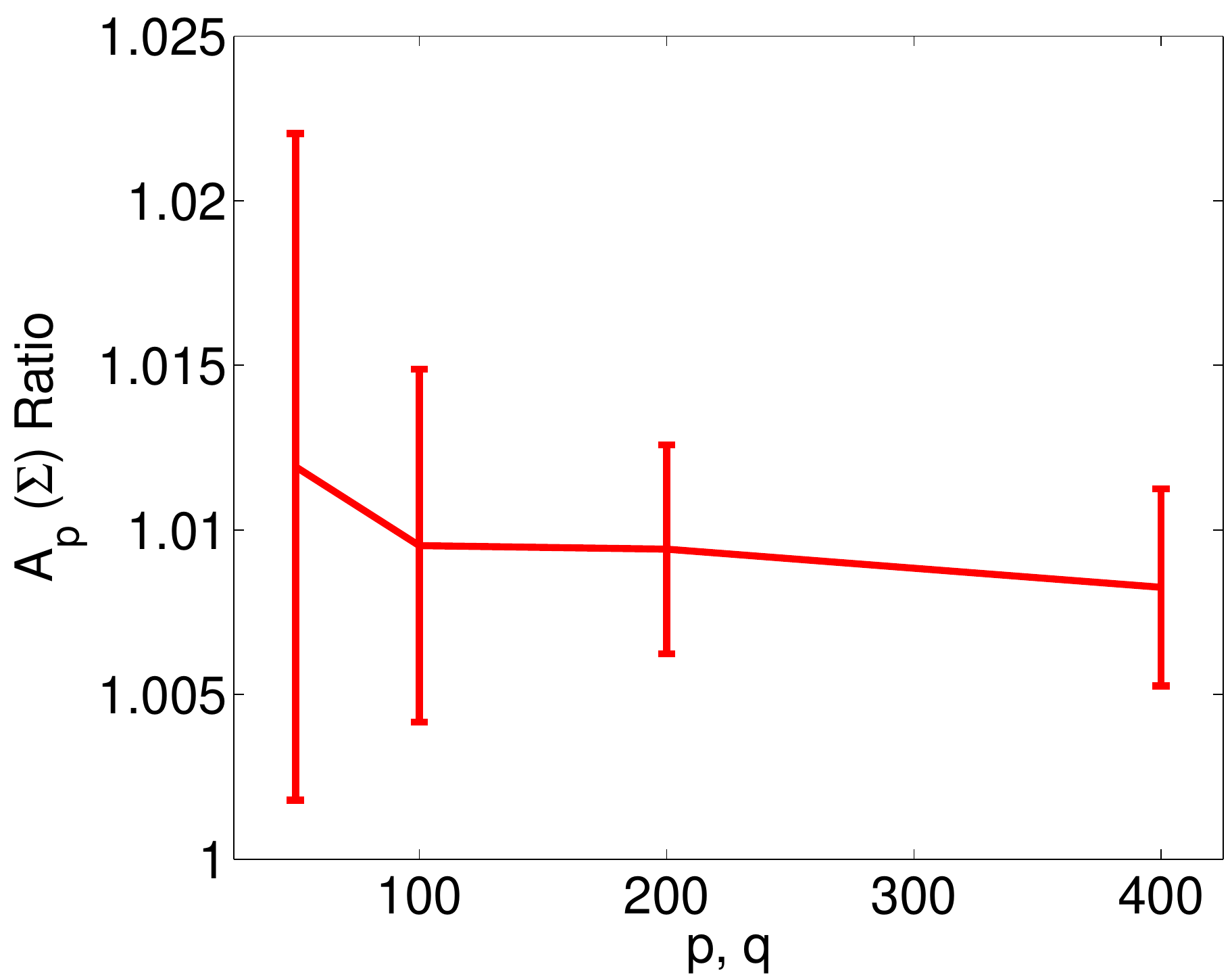}
		\label{fig:A_ratio_n_40}
	}
	\caption{The ratio $\widehat{A}_p/A_p$ for different $n$ and $p=q$ when $\O$ and $\Ga$ are hub graphs.}
	\label{fig:A_ratio_hub_hub}
\end{figure}

In Figure \ref{fig:A_ratio_hub_hub}, we plot the ratio $\widehat{A}_p/A_p$ for $n=20$ (left figure) and $n=100$ (right figure) as $p=q$ increases from $50$ to $400$. Due to space constraints, we only show the case when both $\O$ and $\Ga$ are generated from hub graphs (the plots when $\O$ and $\Ga$ are generated from other graphs structures are similar). As one can see from Figure \ref{fig:A_ratio_hub_hub}, when either $n$ is fixed and $p=q$ increases or $p=q$ is fixed and $n$ increases from 20 to 100, the mean ratio becomes closer to one and the standard deviation of the ratios decreases. This study empirically verifies Proposition \ref{prop2}, which claims that the ratio $\widehat{A}_p/A_p$  converges to 1 in probability as $nq \rightarrow \infty$.

\subsection{Comparison to the penalized likelihood approach}
\label{sec:Leng}

\begin{table}[!t]
	\centering
	\small
	\caption{Averaged empirical FDP and power using the penalized likelihood method with the SCAD penalty.}
	\begin{tabular}{r r r r r r r r r r r r} \hline \hline
		$p$ & $q$ & $\O$ & $\Ga$  & \multicolumn{2}{c}{$n=20$} & \multicolumn{2}{c}{$n=100$} \\ \cmidrule(r){5-6} \cmidrule(r){7-8}
		&     &          &           & FDP & Power & FDP   & Power
		\\ \hline
		100 & 100 & hub & hub & 0.037 & 0.347 &0.038 & 0.346 \\
		&   & hub & band & 0.455 & 0.410 &0.348 & 0.381 \\
		&   & hub & random & 0.722 & 0.926 &0.738 & 0.910 \\
		&   & band & band & 0.341 & 0.206 &0.332 & 0.167 \\
		&   & band & random & 0.339 & 0.964 &0.246 & 0.971 \\
		&   & random & random & 0.703 & 0.962 &0.622 & 0.943 \\
		200 & 200 & hub & hub & 0.027 & 0.322 &0.059 & 0.370 \\
		&   & hub & band & 0.357 & 0.314 &0.333 & 0.306 \\
		&   & hub & random & 0.629 & 0.903 &0.654 & 0.888 \\
		&   & band & band & 0.333 & 0.187 &0.333 & 0.167 \\
		&   & band & random & 0.246 & 0.966 &0.151 & 0.982 \\
		&   & random & random & 0.481 & 0.828 &0.326 & 0.863 \\
		200 & 50 & hub & hub & 0.060 & 0.381 &0.054 & 0.385 \\
		&   & hub & band & 0.362 & 0.330 &0.371 & 0.517 \\
		&   & hub & random & 0.754 & 0.927 &0.704 & 0.909 \\
		&   & band & band & 0.340 & 0.205 &0.334 & 0.174 \\
		&   & band & random & 0.588 & 0.833 &0.551 & 0.855 \\
		&   & random & random & 0.784 & 0.961 &0.607 & 0.947 \\
		400 & 400 & hub & hub & 0.946 & 0.978 &0.952 & 1.000 \\
		&   & hub & band & 0.982 & 1.000 &0.962 & 0.992 \\
		&   & hub & random & 0.949 & 0.999 &0.910 & 1.000 \\
		&   & band & band & 0.338 & 0.179 &0.336 & 0.175 \\
		&   & band & random & 0.233 & 0.848 &0.237 & 0.864 \\
		&   & random & random & 0.141 & 0.610 &0.099 & 0.644 \\
		\hline
	\end{tabular}
	\normalsize
	\label{tab:smgm}
	\normalfont
\end{table}

We compare our procedure with the penalized likelihood approach in \cite{LengTang2012}. We adopt the same  (regularization) parameter-tuning procedure as \citet{LengTang2012}, i.e., we generate  an extra random test dataset with the sample size equal to the training set and choose the parameter that maximizes the log-likelihood on the test dataset. Due to space constraints, we only report the result using the SCAD penalty \citep{Fan01} rather than the L1 penalty since the SCAD penalty leads to slightly better performance (also observed in \cite{LengTang2012}). The averaged empirical FDPs and powers for different settings of $n,p,q, \O, \Ga$ are shown in Table \ref{tab:smgm}. As one can see from Table \ref{tab:smgm},  each setting has either a large FDP or a small power. In fact, for those settings with small averaged FDPs (e.g., $n=100, p=200, q=50$ and $\O$ and $\Ga$ generated from hub graphs with the averaged FDP 0.054), the corresponding powers are also small (e.g., 0.385 for the aforementioned case), which indicates that the estimated $\widehat{\O}$ or $\hat{\Ga}$ is too sparse.  On the other hand, for those settings with large averaged powers (e.g., $n=100, p=q=400$ and $\O$ from hub and $\Ga$ from random with the averaged power equal to 1), the corresponding FDPs are also large (e.g., 0.910 for the aforementioned case), which indicates that the estimated $\widehat{\O}$ or $\hat{\Ga}$ is too dense.  We also note that when $p,q$ are small as compared to $n$, the penalized likelihood approach still achieves good support recovery performance (e.g., the case $n=100, p=q=20$ as reported in \citet{LengTang2012}). When $p, q$ are comparable to or lager than $n$, our testing based method achieves better support recovery performance.

\subsection{De-correlation method}
\label{sec:decorrelation}
\begin{figure}[!t]
	\centering
	\subfigure[t][$\O=$ hub, $\Ga=$ hub]{
		\includegraphics[width=0.31\textwidth]{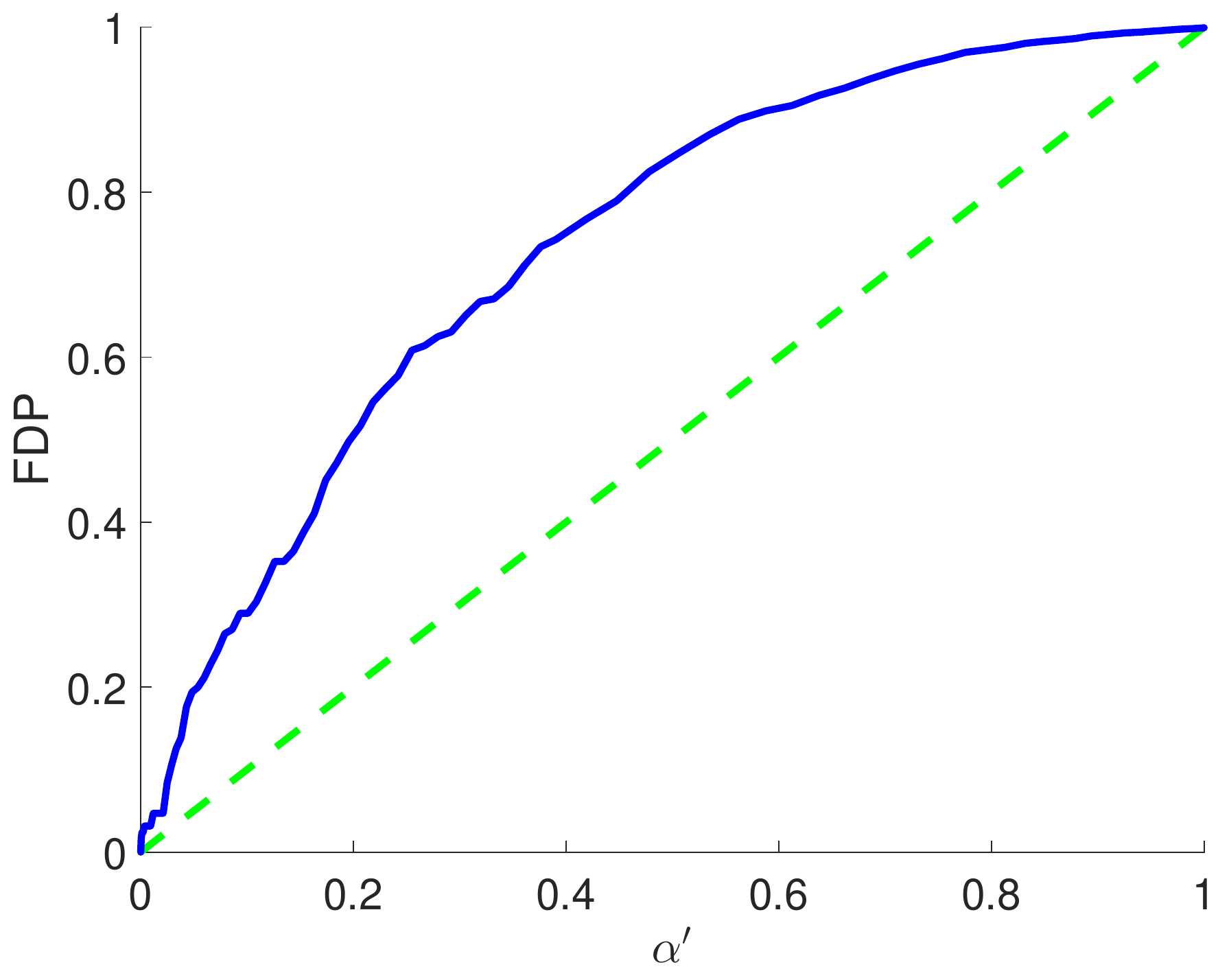}
		\label{fig:clime_4}
	}
	\subfigure[t][$\O=$hub, $\Ga=$ band]{
		\includegraphics[width=0.31\textwidth]{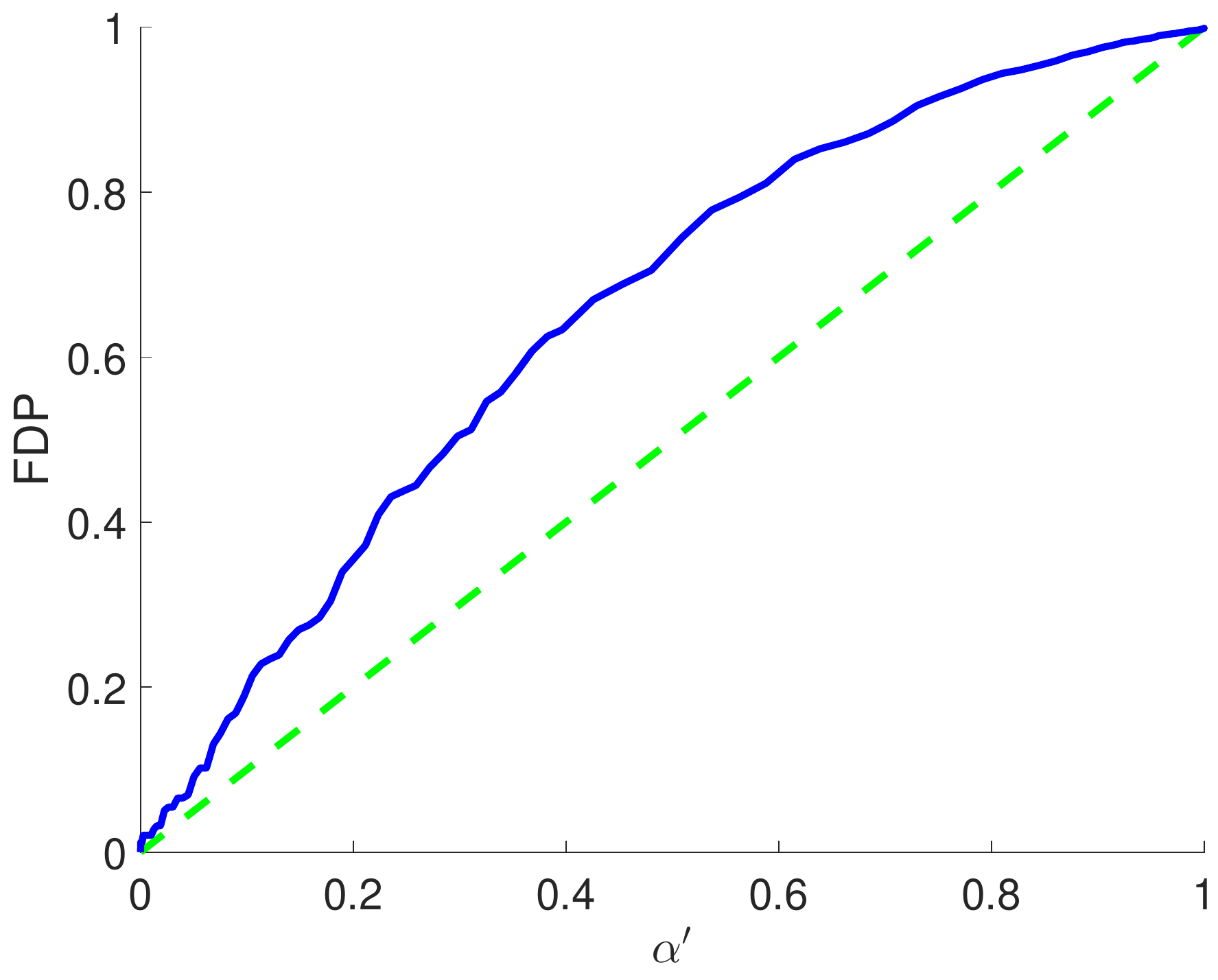}
		\label{fig:clime_3}
	}
	\subfigure[t][$\O=$ hub, $\Ga=$ random]{
		\includegraphics[width=0.31\textwidth]{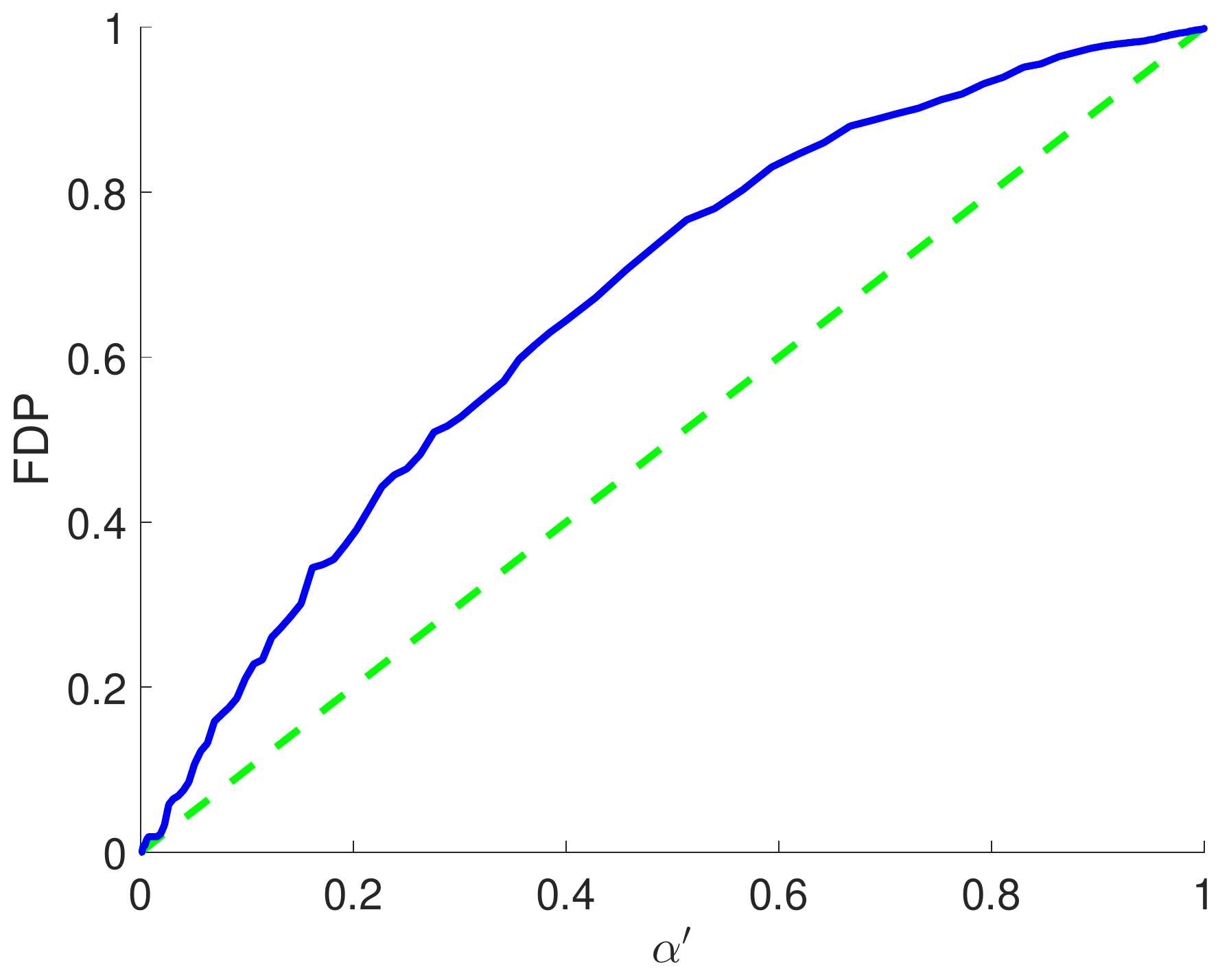}
		\label{fig:clime_5}
	}
	\\
	\subfigure[t][$\O=$ band, $\Ga=$ band]{
		\includegraphics[width=0.31\textwidth]{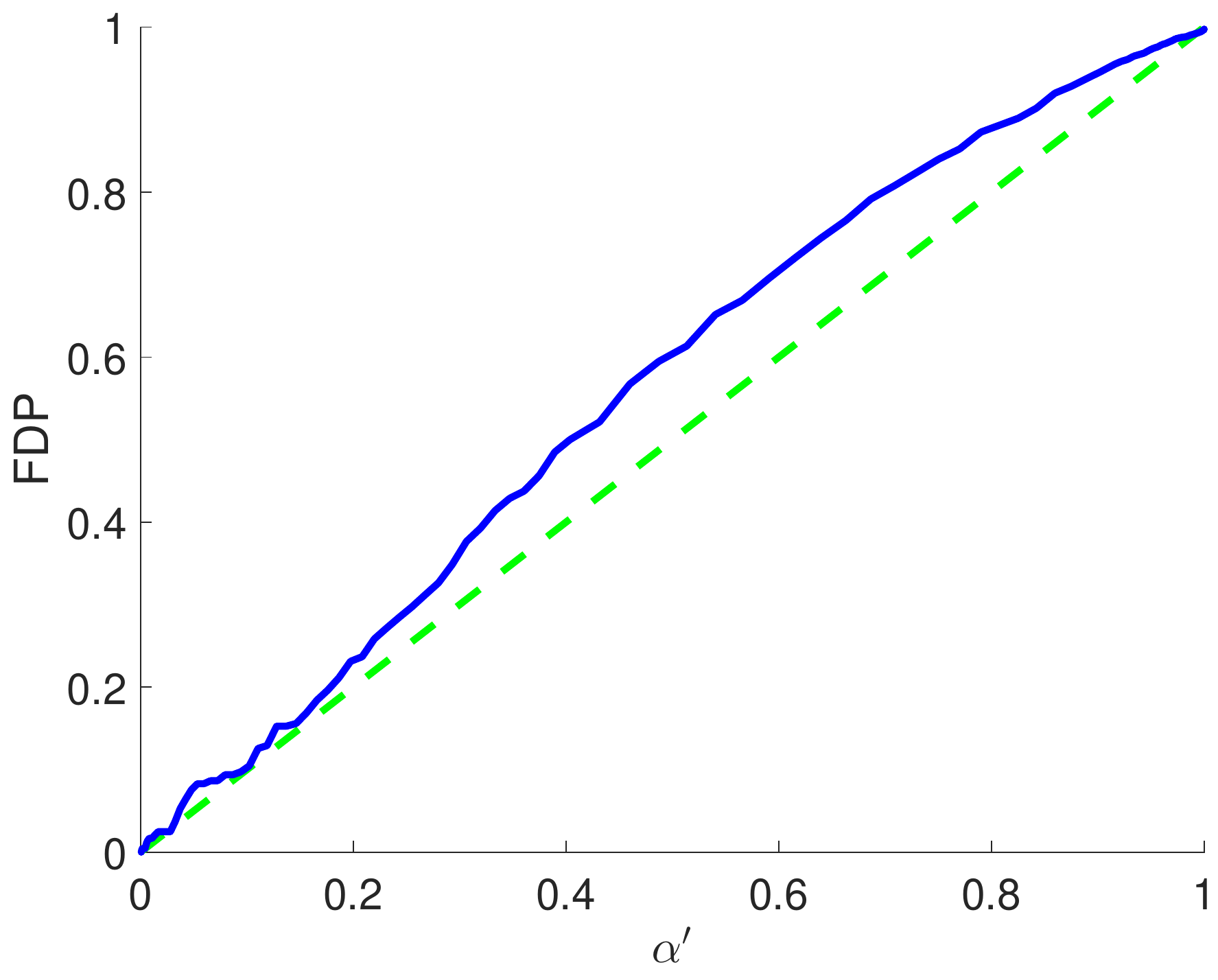}
		\label{fig:clime_1}
	}
	\subfigure[t][$\O=$ band, $\Ga=$ random]{
		\includegraphics[width=0.31\textwidth]{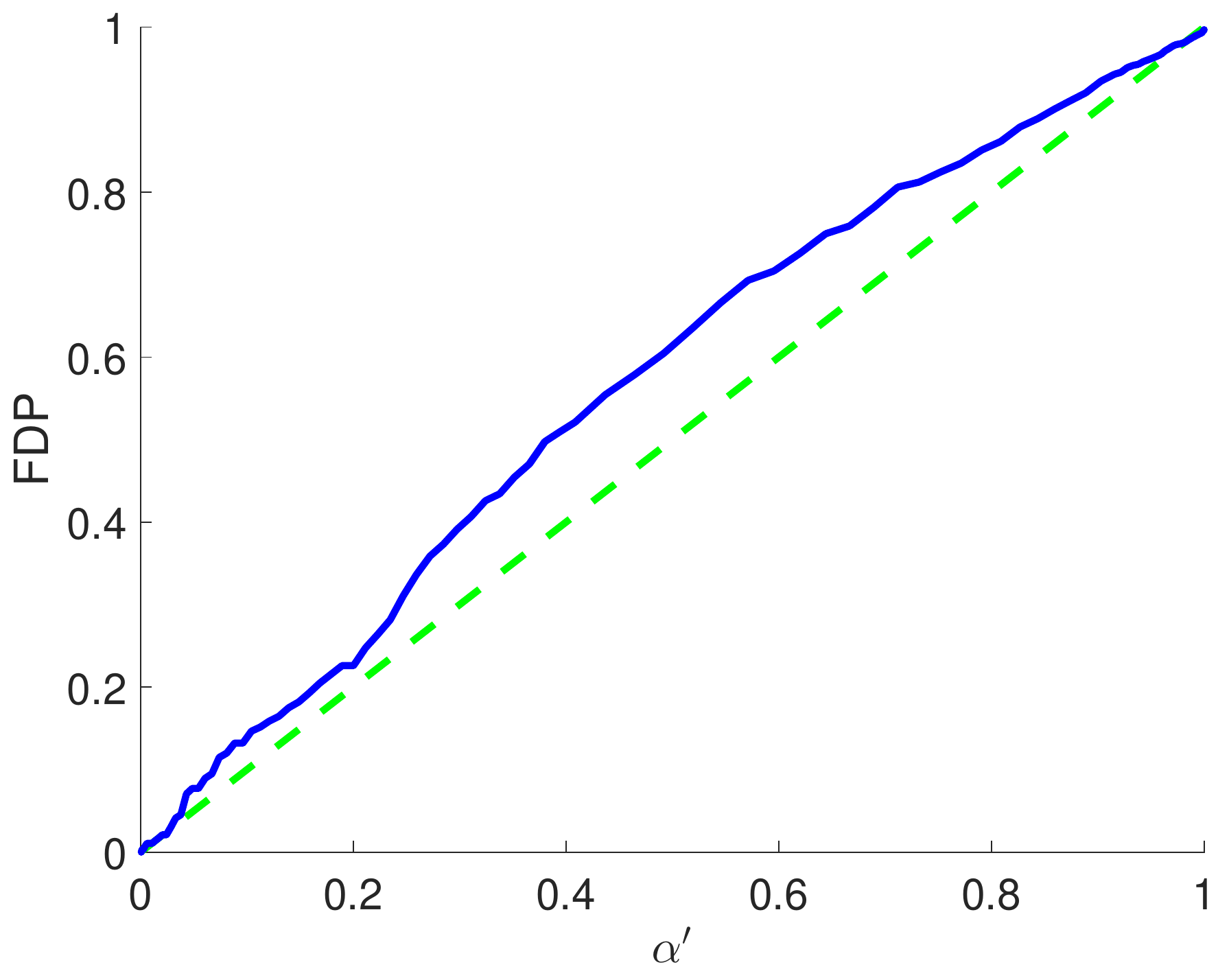}
		\label{fig:clime_2}
	}
	\subfigure[t][$\O=$ random, $\Ga=$ random]{
		\includegraphics[width=0.31\textwidth]{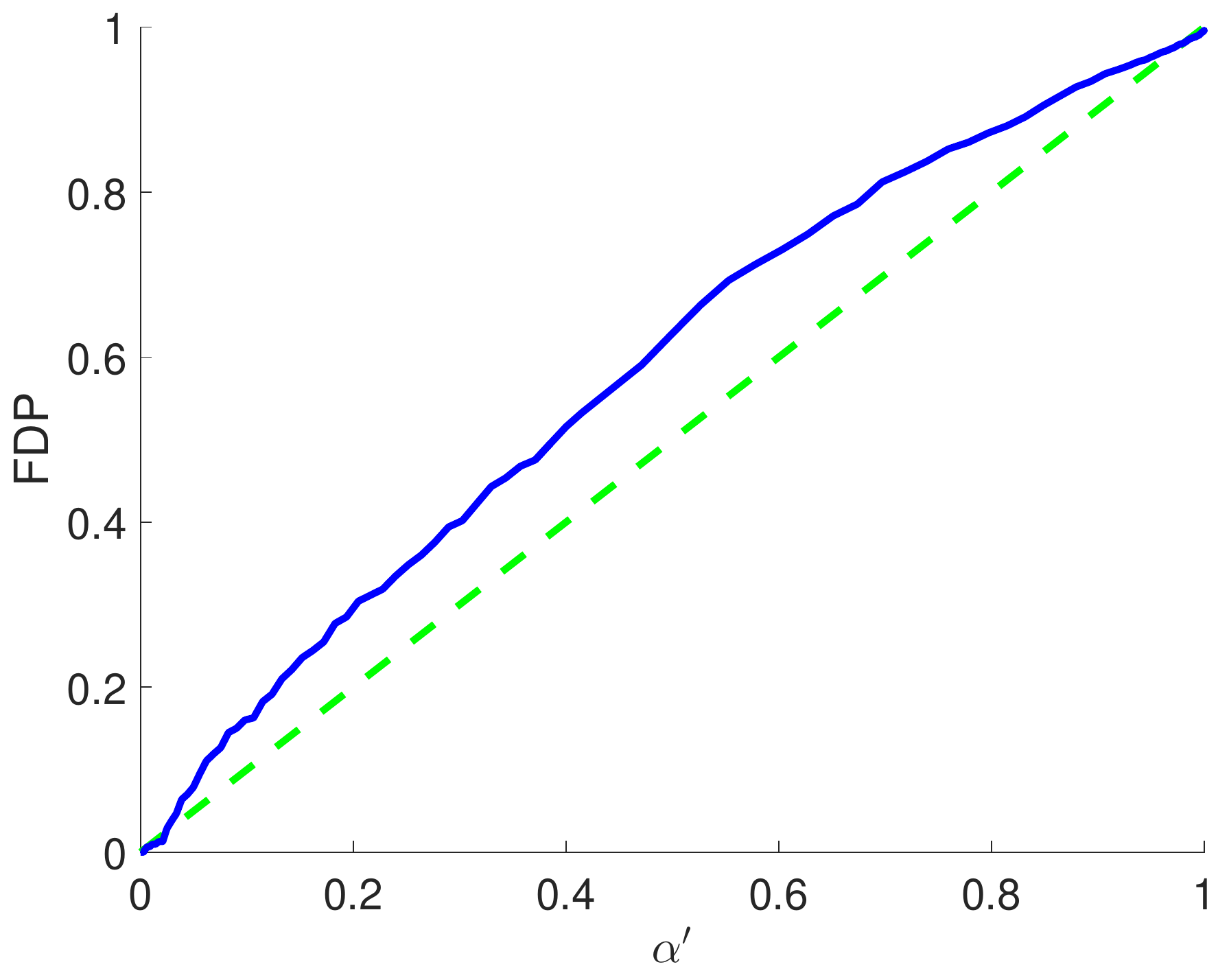}
		\label{fig:clime_6}
	}
	\caption{Averaged empirical FDPs (y-axis) against different estimated FDP $\alpha'$s (x-axis) for the de-correlation approach. For the purpose of controlling FDP, the blue line should be close to or below the dashed green line, which represents FDP=$\alpha'$. }
	\label{fig:clime}
\end{figure}

In Remark \ref{remark_decorr}, we illustrate why the de-correlation approach is not applicable in our problem setup from a theoretical perspective. Here, we provide some empirical evidences. 
Due to space constraints, we only report the comparison  when $n=20$, $p=q=100$, and, in fact, the performance becomes even worse when $p,q$ gets larger. Ideally, the empirical FDP should be close to (or below) the FDP estimate  $\alpha'$ in \eqref{eq:alpha_prime}. However, as one can see from Figure \ref{fig:clime},  the empirical FDP is much larger than the corresponding $\alpha'$ in many cases. Moreover, by setting the FDR level for individual $\Ga$ and $\O$ to be $\alpha=0.1$, we present the corresponding empirical FDP and $\alpha'$ in Table \ref{tab:FDP_clime}, where the FDP can be twice as large as $\alpha'$ in some cases. The  experimental results from Table \ref{tab:FDP_clime} and Figure \ref{fig:clime} empirically verify that the de-correlation approach does not control FDP well.


\begin{table}[!t]
	\centering
	\small
	\caption{Averaged empirical FDP and the estimated FDP $\alpha'$ for the de-correlation approach.}
	\begin{tabular}{r r c c} \hline
		$\O$ & $\Ga$  & FDP &($\alpha'$)
		\\ \hline
		hub & hub & 0.376 &(0.146) \\
		hub & band & 0.272 &(0.152) \\
		hub & random & 0.323 &(0.154) \\
		band & band & 0.176 &(0.161) \\
		band & random & 0.199 &(0.164) \\
		random & random & 0.250&(0.164) \\\hline
	\end{tabular}
	\label{tab:FDP_clime}
	\normalsize
\end{table}

\subsection{Simulation study when the covariance is not a Kronecker product}
\label{sec:perturbation}
\begin{figure}[!t]
	\centering
	\subfigure[t][$\O=$ band, $\Ga=$random]{
		\includegraphics[width=0.4\textwidth]{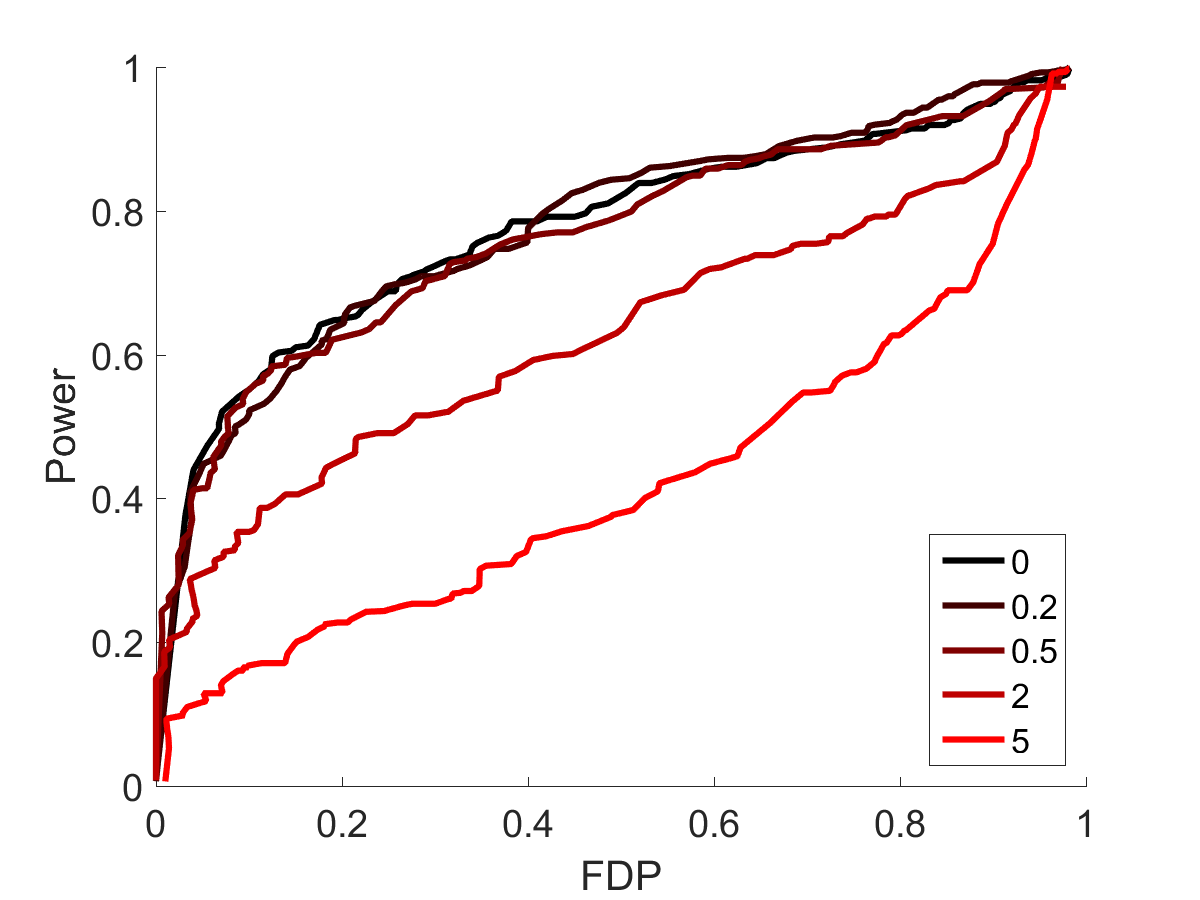}
		\label{fig:perturb_1}
	}\hspace{-8mm}
	\subfigure[t][$\O=$ hub,  $\Ga=$ random]{
		\includegraphics[width=0.4\textwidth]{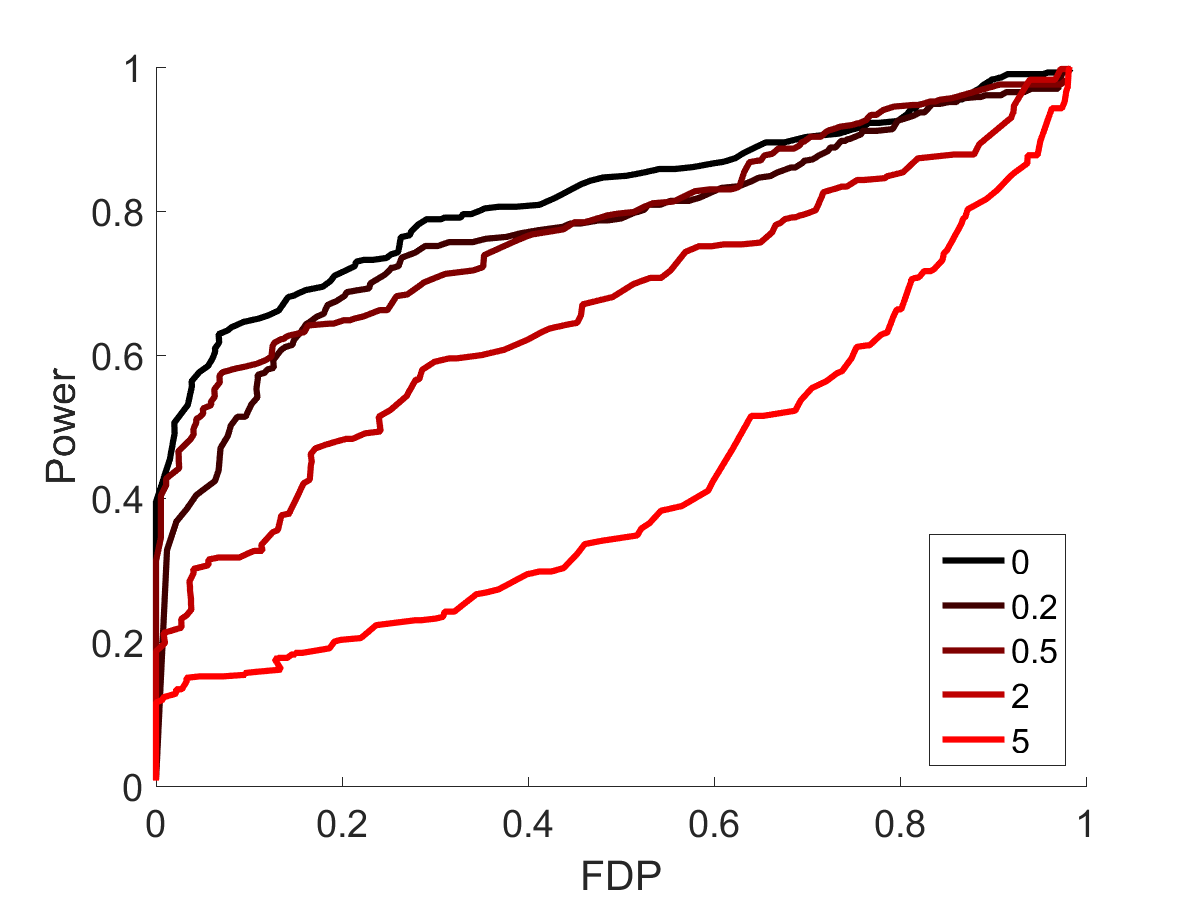}
		\label{fig:perturb_2}
	}\hspace{-8mm}
	\caption{ROC curves for different perturbation levels of  $\nu=0, 0.2, 0.5, 2 $ and $5$. The larger the $\nu$ is, the more ``red'' color of the ROC curve is (from black to red).}
	\label{fig:perturb}
\end{figure}

In this section, we present  simulation study when the covariance matrix does not follow the form of a Kronecker product. More precisely, we generate the covariance matrix  in the form of $\S \otimes \ps + \nu \mathbf{I}$, where $\mathbf{I}$ is the $pq \times pq$ identity matrix and $\nu$ is the level of perturbation. 
Due to space constraints, we only present the case when $n=20$, $p=q=30$, $\O$ is either a band or a hub graph, $\Ga$ is a random graph. The observation is similar for other settings.  Figure \ref{fig:perturb} plots the ROC curves for different perturbation parameters $\nu=0, 0.2, 0.5, 2 $, and $5$. As one can see, when the perturbation level $\nu$ is small, the ROC curve is almost identical to the case when the covariance is a Kronecker  product (i.e., $\nu=0$). However, when $\nu$ becomes larger, the support recovery performance becomes inferior.
\label{sec:perturbation}

\subsection{Additional ROC curve comparisons}
\label{sec:add_ROC}

\begin{figure}[!t]
	\centering
	\subfigure[t][$\O=$ hub, $\Ga=$ hub]{
		\includegraphics[width=0.31\textwidth]{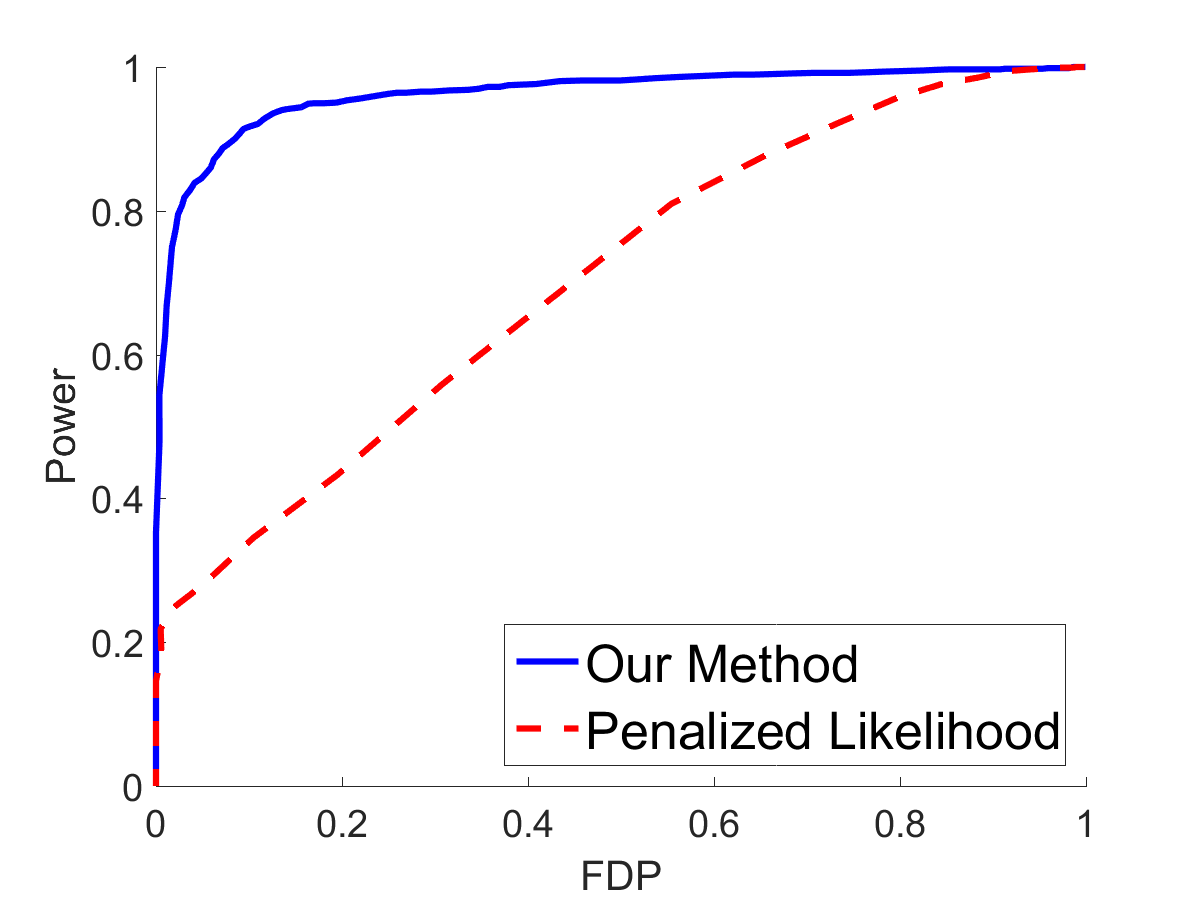}
		\label{fig:roc_type_4}
	}
	\subfigure[t][$\O=$hub, $\Ga=$ band]{
		\includegraphics[width=0.31\textwidth]{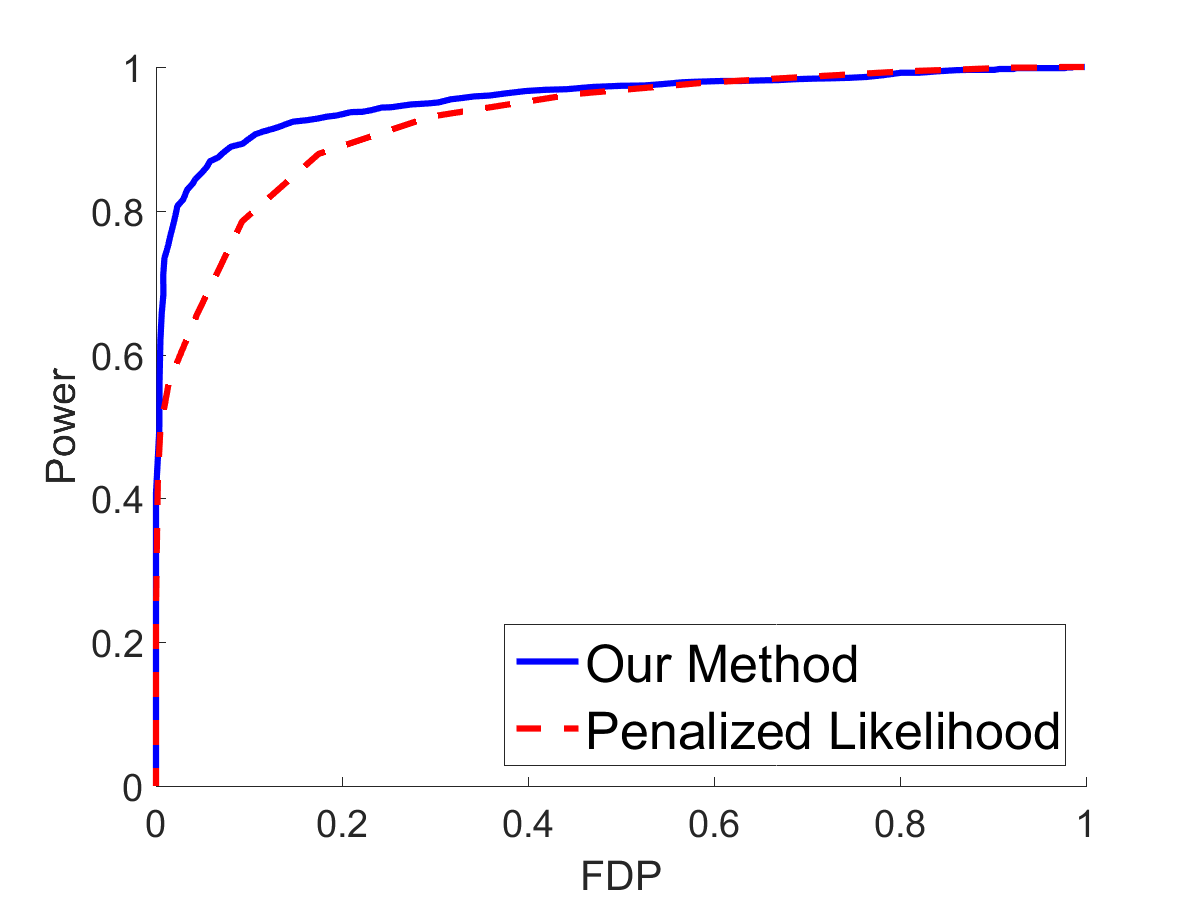}
		\label{fig:roc_type_3}
	}
	\subfigure[t][$\O=$ hub, $\Ga=$ random]{
		\includegraphics[width=0.31\textwidth]{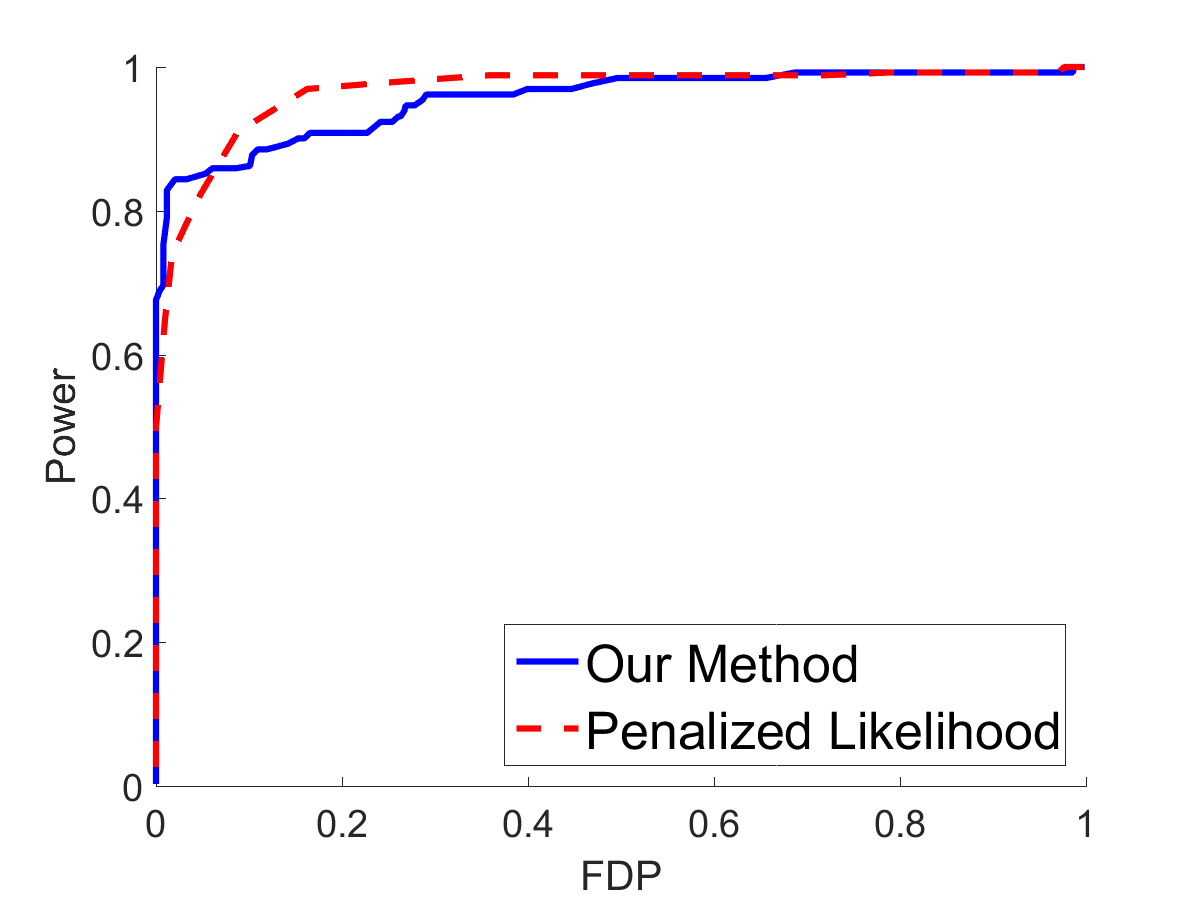}
		\label{fig:roc_type_5}
	}\\
	\subfigure[t][$\O=$ band, $\Ga=$ band]{
		\includegraphics[width=0.31\textwidth]{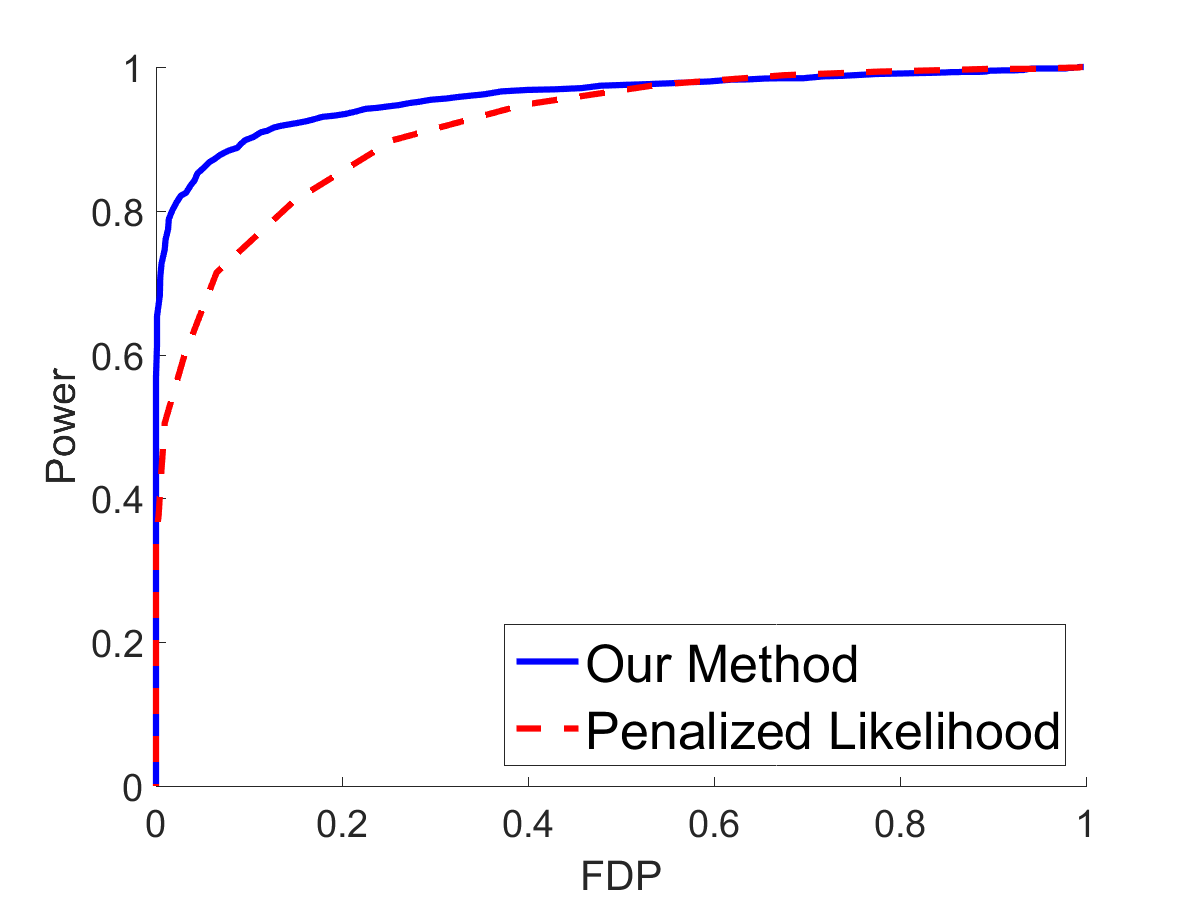}
		\label{fig:roc_type_1}
	}
	\subfigure[t][$\O=$ band, $\Ga=$ random]{
		\includegraphics[width=0.31\textwidth]{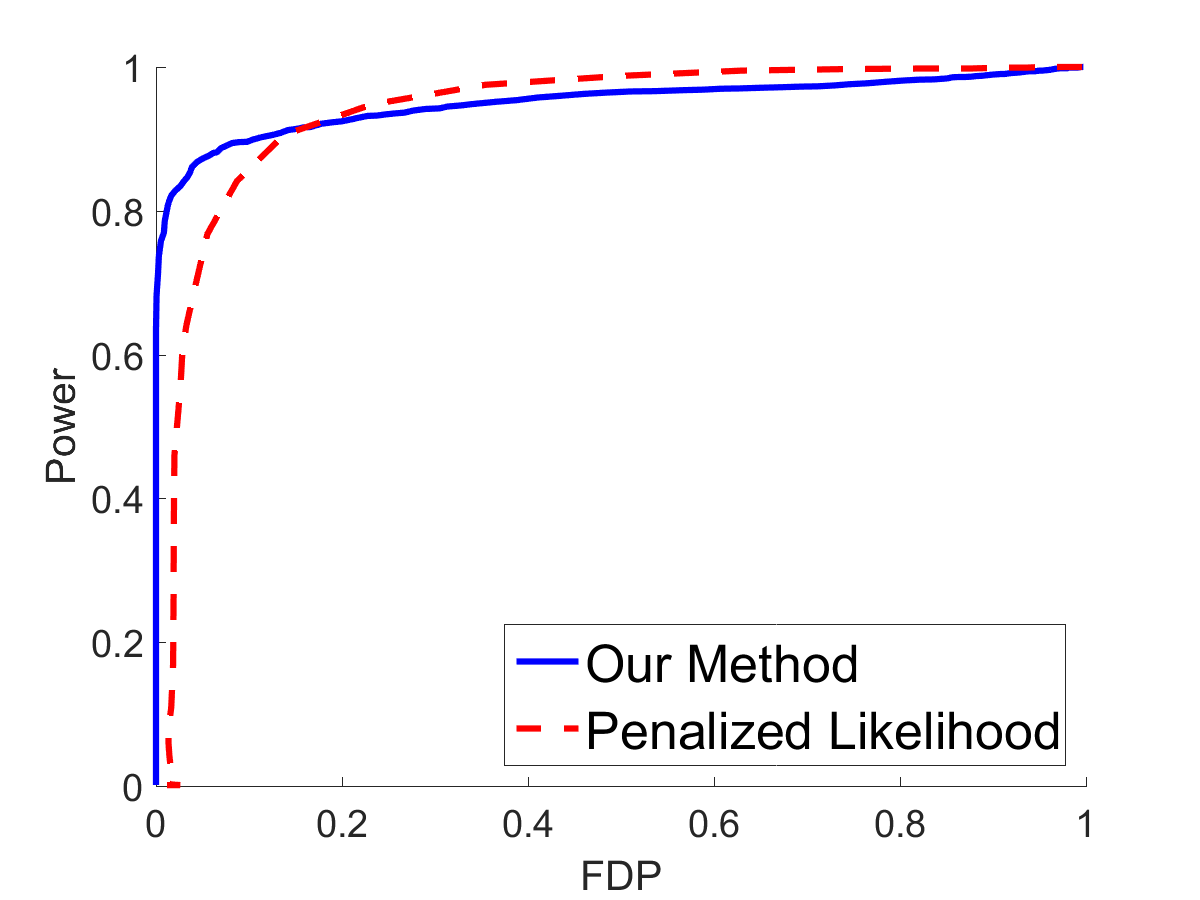}
		\label{fig:roc_type_2}
	}
	\subfigure[t][$\O=$ random, $\Ga=$ random]{
		\includegraphics[width=0.31\textwidth]{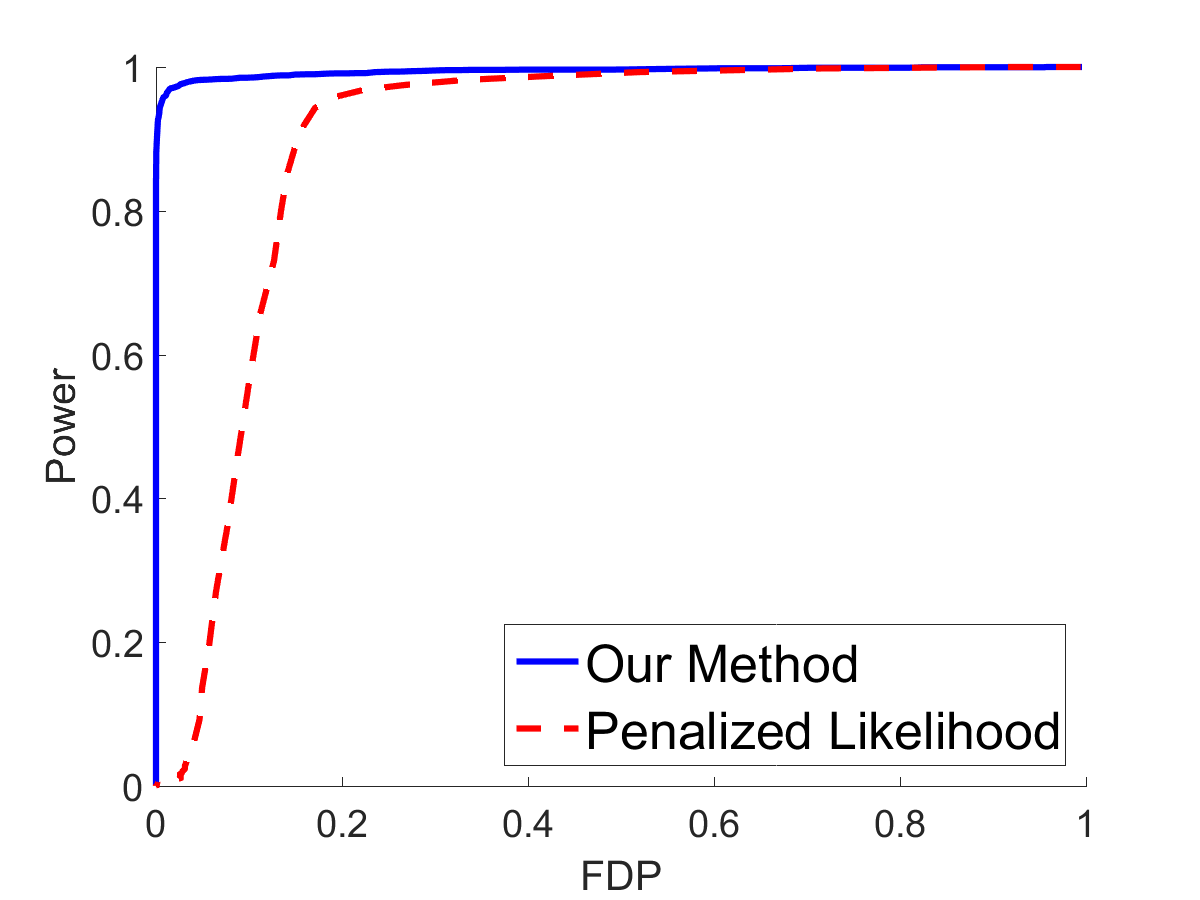}
		\label{fig:roc_type_6}
	}
	\caption{ROC curves for different types of the construction of $\O$ and $\Ga$ when $f=3$.}
	\label{fig_roc_type}
\end{figure}

In Figure \ref{fig_roc_type}, we fix the factor $f=3$ and consider different types of $\O$ and $\Ga$. For most cases, our method achieves better performance. The only exception is that, for hub/random and band/random graphs, the power of the penalized likelihood approach outperforms our method when FDP is large. However, for support recovery in high-dimensional settings, one is more interested in the scenario when FDP is very small. In such a case, our method consistently leads to a larger power than the penalized likelihood approach.

\subsection{Real data analysis}
\label{sec:supp_real}

In this section, we investigate the performance of the proposed method on two real datasets, the U.S. agricultural export data from \citet{LengTang2012} and the climatological data from \cite{Lozano:09}. 

\subsubsection{U.S. agricultural export data}
\label{sec:export}

\begin{table}[!t]
	\centering
	\caption{No. of edges  for the export data. For 13 regions, there are 78  possible edges in total. For 36 products, there are 630 possible edges in total.}
	\begin{tabular}{r c c c c c c c c c } \hline \hline
		&  \multicolumn{3}{c}{Region}  &        \multicolumn{3}{c}{Product}  \\ \cmidrule(l){2-4} \cmidrule(l){5-7}
		& $\alpha=0.1$ & $\alpha=0.2$ & $\alpha=0.3$ &   $\alpha=0.1$ & $\alpha=0.2$ & $\alpha=0.3$  \\
		No. of Edges     &  2 & 23 & 31 & 19 & 30  & 37 \\
		Density of the Graph         & 2.56\% &  29.49\% & 39.74\% & 3.01\% & 4.76\% & 5.87\%                    \\ \hline
	\end{tabular}
	\label{tab:export}
\end{table}

\begin{figure}[!t]
	\centering
	\subfigure[t][Graph for Regions ($\alpha=0.2$)]{
		\includegraphics[width=0.4\textwidth, height=0.4\textwidth, angle=-90]{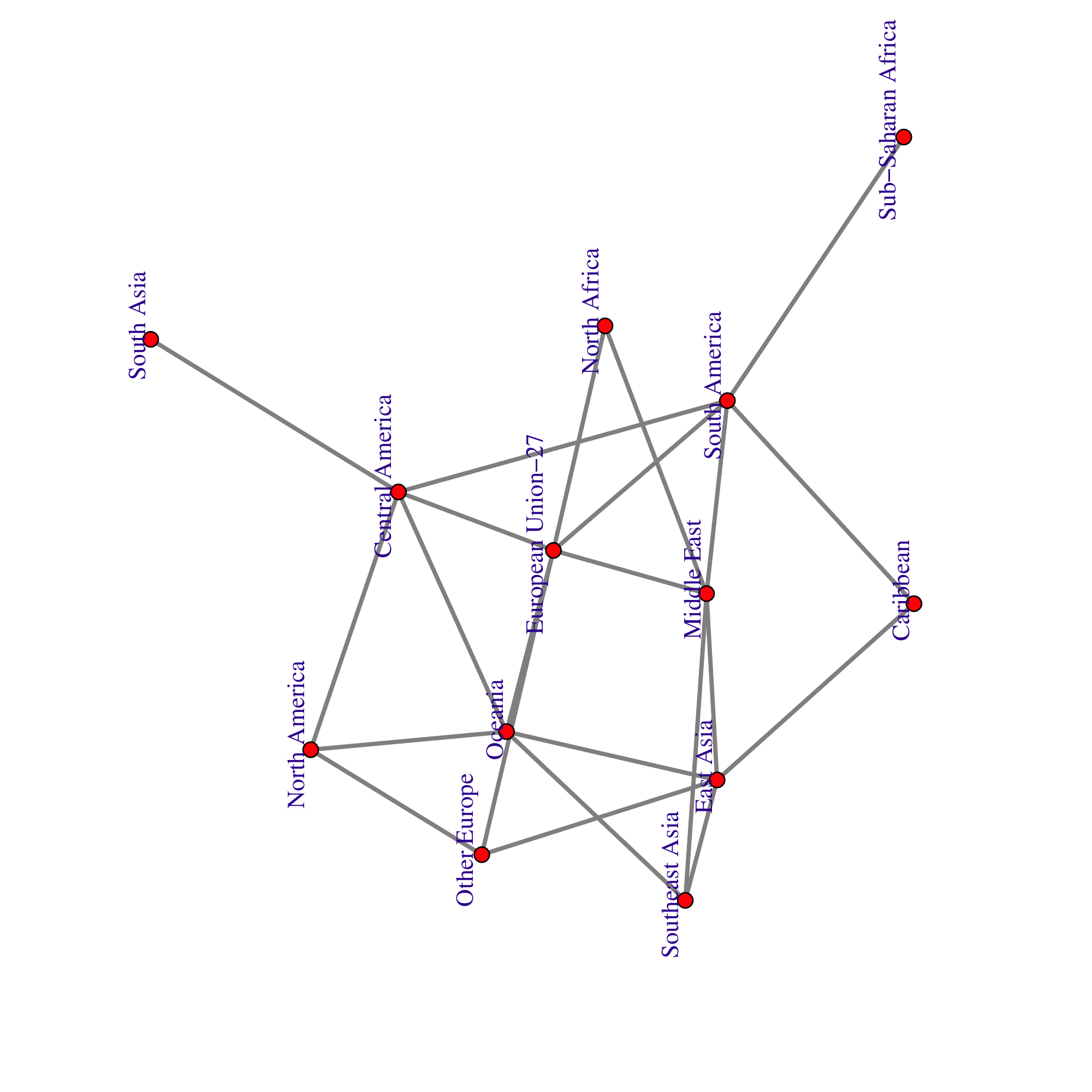}
		\label{fig:region_a_2}
	}\hspace{-8mm}
	\subfigure[t][Graph for Products ($\alpha=0.2$)]{
		\includegraphics[width=0.4\textwidth,  height=0.4\textwidth,angle=-90]{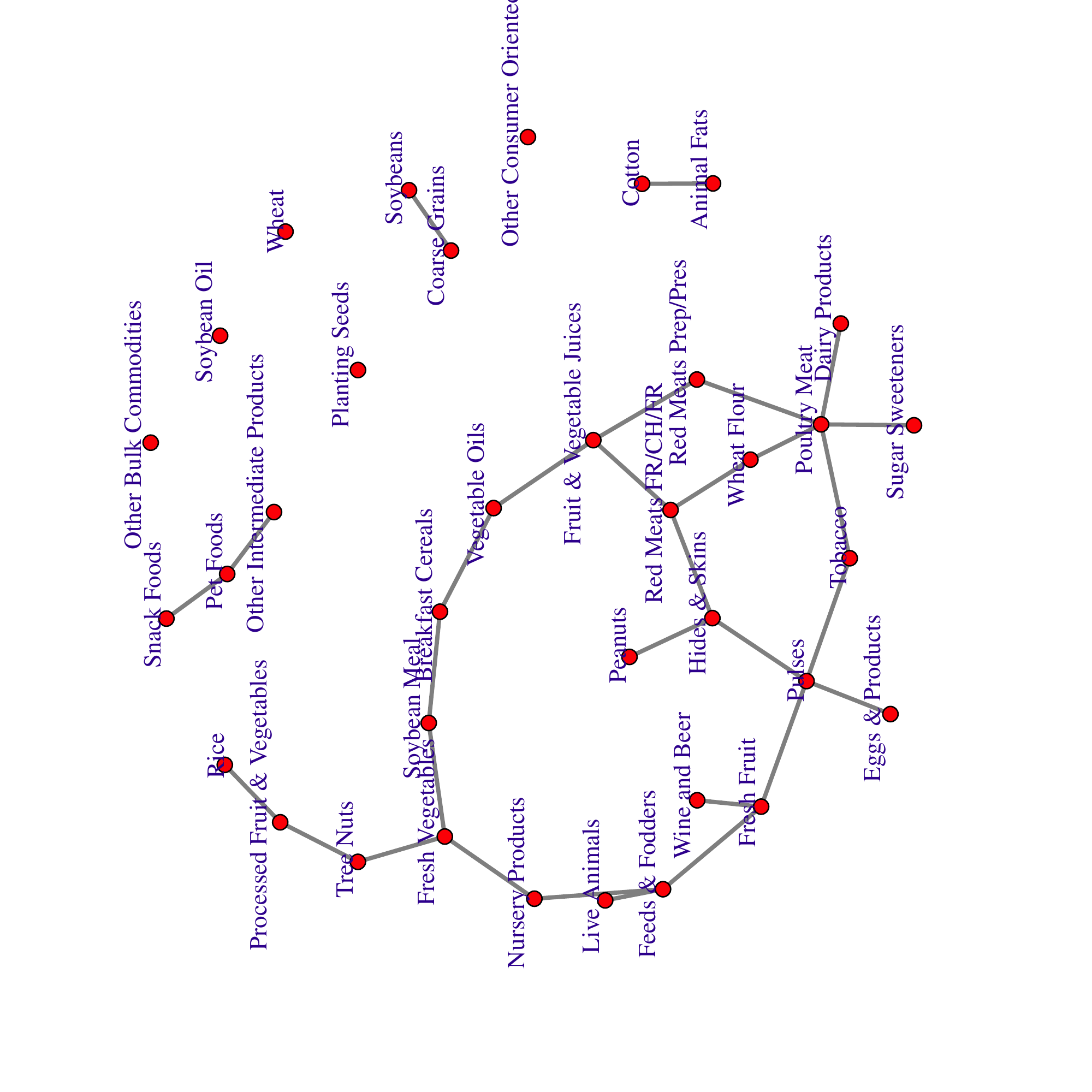}
		\label{fig:item_a_2}
	}\hspace{-8mm}\\
	\subfigure[t][Graph for Regions ($\alpha=0.3$)]{
		\includegraphics[width=0.4\textwidth, height=0.4\textwidth,angle=-90]{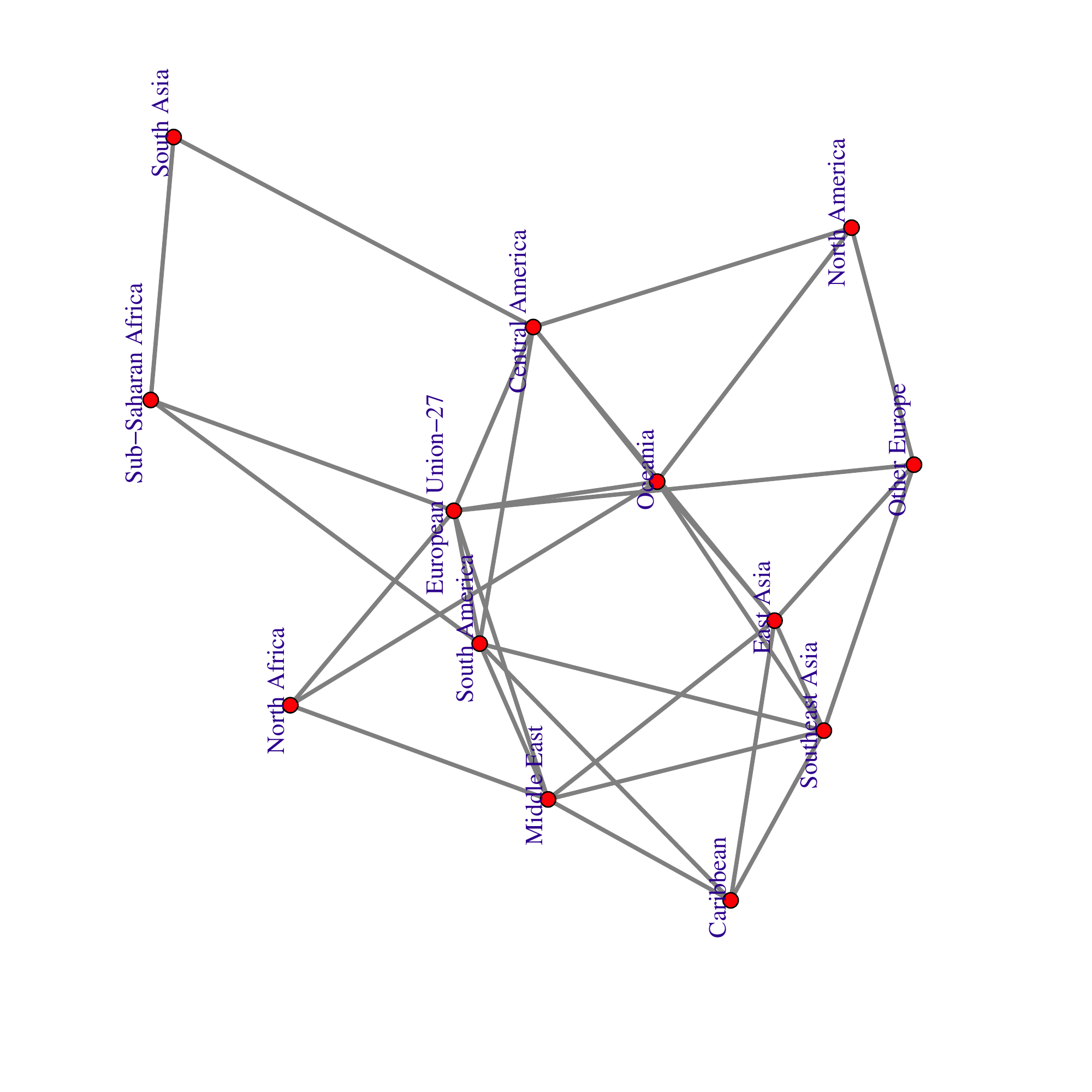}
		\label{fig:region_a_3}
	}\hspace{-8mm}
	\subfigure[t][Graph for Products ($\alpha=0.3$)]{
		\includegraphics[width=0.4\textwidth,  height=0.4\textwidth,angle=-90]{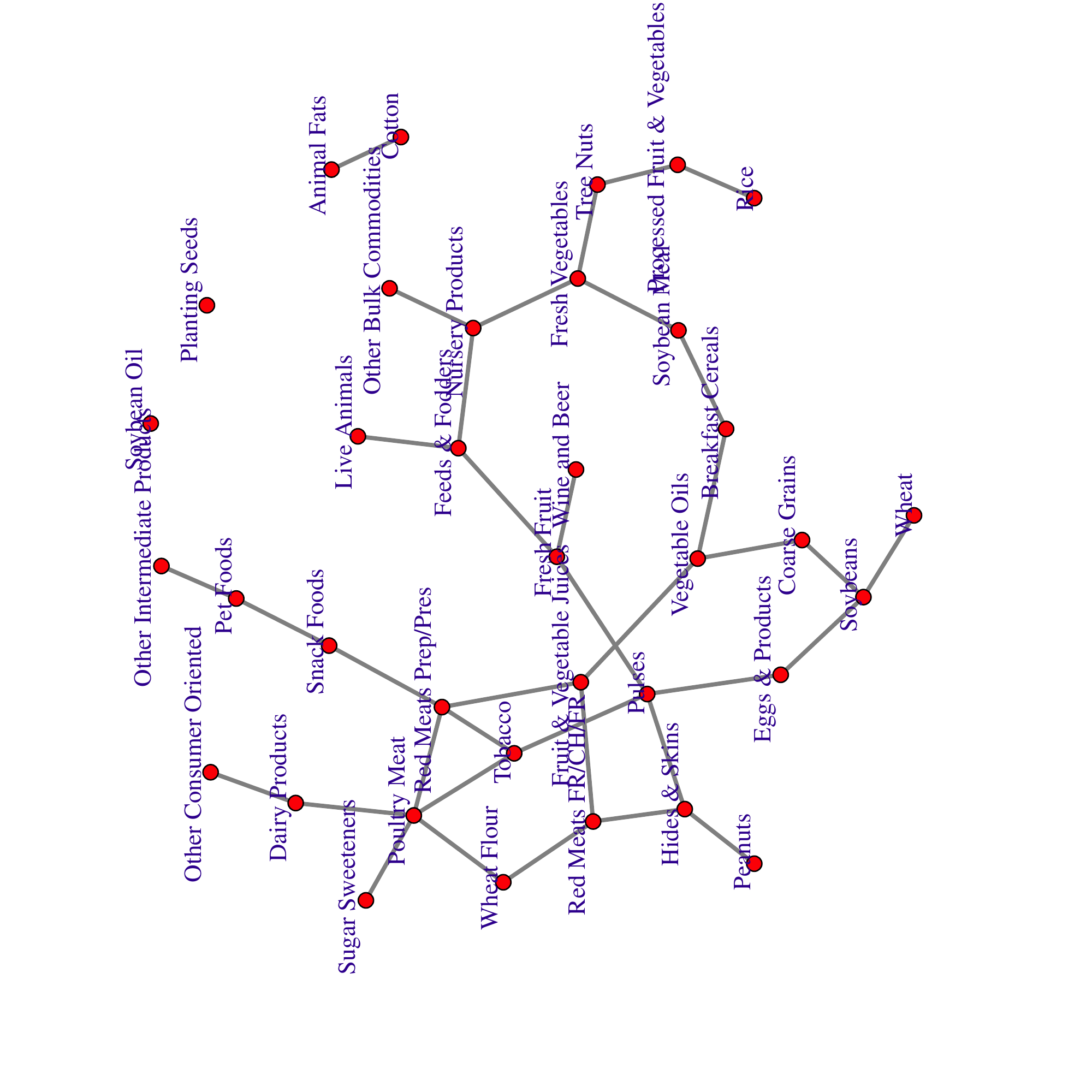}
		\label{fig:item_a_3}
	}\hspace{-8mm}
	\caption{Estimated graphs for the export data}
	\label{fig:export}
\end{figure}

We first apply our method to the U.S. agricultural export data studied in \citet{LengTang2012}. The dataset contains annual U.S. agriculture export data for 40 years, from 1970 to 2009. Each annual dataset contains the amount (in thousands U.S. dollars) of exports for $36$ products (e.g., pet foods, snack foods, breakfast cereals, soybean meal, meats, eggs, dairy products, etc.) in 13 different regions (e.g., North America, Central America, South America, South Asia, etc.). Thus, the dataset can be organized into $40$ matrix-variate observations, where each observation is a $(p=13) \times (q=36)$ matrix. We adopt the method proposed in \cite{LengTang2012} to remove the dependence in this matrix-variate time series data.  In particular, we take the logarithm of the original data plus one and then take the lag-one difference for each matrix observation so that the number of observations becomes $n=39$. Please refer to \cite{LengTang2012} for more details on the pre-processing of the data.

We apply the proposed FDR control procedure to estimate the support of the precision matrices for regions and products under different $\alpha \in \{0.1, 0.2, 0.3\}$. In Table \ref{tab:export}, we report the number of edges/discoveries for different $\alpha$'s. We observe that for the product graphs, the number of discoveries  is relatively small as compared to the number of hypotheses, which indicates that many pairs of products are conditionally independent. In Figure \ref{fig:export}, we plot the graphs corresponding to the estimated supports of $\O$ (corresponding to Regions) and $\Ga$ (corresponding to Products) for $\alpha=0.2$ and $\alpha=0.3$. Figures \ref{fig:region_a_2} and \ref{fig:region_a_3} show the estimated graphs for $p=13$ regions. As we can see, the regions in the following sets, $\{\text{East Asia}, \; \text{Southeast Asia}\}$, $\{\text{European Union}, \; \text{Other Europe}, \; \text{Oceania} \}$ and $\{\text{Central America}, \; \text{North America}, \; \text{South America}\}$, are always connected. Such an observation should be expected since regions in the aforementioned sets are close geographically. This observation is consistent with the result obtained by penalized likelihood approach in \citet{LengTang2012}, which claims that ``the magnitude between Europe Union and Other Europe, and that between East Asia and Southeast Asia are the strongest."  The regions South Asia, Sub-Saharan Africa, and North Africa connect to fewer regions. This observation is also consistent with the result in \citet{LengTang2012}, noting that ``interestingly, none of the 11 largest edges corresponds to either North Africa or Sub-Saharan Africa." The estimated graphs for products shown in Figures \ref{fig:item_a_2} and \ref{fig:item_a_3} are quite sparse, which indicates many pairs of products are conditionally independent given the information of the rest of the products. The product graphs also lead to many interesting observations.  For example,  the products in the following sets, $\{\text{Pet foods}, \; \text{Snack foods}, \; \text{Other Intermediate Products}\}$,

\noindent $\{\text{Dairy Products}, \text{Red Meats FR/CH/FR}, \text{Red Meats Prep/Pres}, \text{Poultry Meat}, \text{Wheat Flour}\}$, are always connected (not necessarily directly). Such observations also make sense since different kinds of meats and dairy products are closely related products and thus should be highly correlated.

\subsubsection{Climate data analysis}
\label{sec:climate}

In this section, we study the climatological data from \cite{Lozano:09}, which contains monthly data of $p=17$ different
meteorological factors during 144 months, from 1990 to 2002.  The observations span $q=125$ locations in the U.S. The  17 meteorological factors measured for each month include \textsf{\small CO$_2$}, \textsf{\small CH$_4$}, \textsf{\small
	H$_2$}, \textsf{\small CO}, average temperature (\textsf{\small
	TMP}), diurnal temperature range (\textsf{\small DTR}), minimum
temperate (\textsf{\small TMN}), maximum temperature (\textsf{\small
	TMX}), precipitation (\textsf{\small PRE}), vapor (\textsf{\small
	VAP}), cloud cover (\textsf{\small CLD}), wet days (\textsf{\small
	WET}), frost days (\textsf{\small FRS}), global solar radiation
(\textsf{\small GLO}), direct solar radiation (\textsf{\small DIR}),
extraterrestrial radiation (\textsf{\small ETR}) and extraterrestrial
normal radiation (\textsf{\small ETRN}). We note that we ignore the UV aerosol index factor in \cite{Lozano:09} since most measurements of this factor are missing. We adopt the same procedure as described in Section \ref{sec:export} to reduce the level of dependence in this matrix-variate time series data.

\begin{table}[!t]
	\centering
	\caption{No. of edges for the climate data. For $p=17$ meteorological  factors, there are 136 edges in total. For $q=125$ locations, there are 7,750 possible edges in total.}
	\begin{tabular}{r c c c c c c c c c} \hline \hline
		&  \multicolumn{3}{c}{Meteorological factors}  &        \multicolumn{3}{c}{Locations}  \\ \cmidrule(l){2-4} \cmidrule(l){5-7}
		& $\alpha=0.1$ & $\alpha=0.2$ & $\alpha=0.3$ &   $\alpha=0.1$ & $\alpha=0.2$ & $\alpha=0.3$  \\
		No. of Edges     &  30 & 40 & 42 & 1059 & 1539  & 2065 \\
		Density of the Graph & 22.05\% & 29.41\% & 30.88\% &  13.66\% & 19.85\% & 26.65\% \\\hline
	\end{tabular}
	\label{tab:climate}
\end{table}

\begin{figure}[!t]
	\centering
	\subfigure[t][Graph for meteorological factors ($\alpha=0.2$)]{
		\includegraphics[width=0.45\textwidth, height=0.45\textwidth, angle=-90]{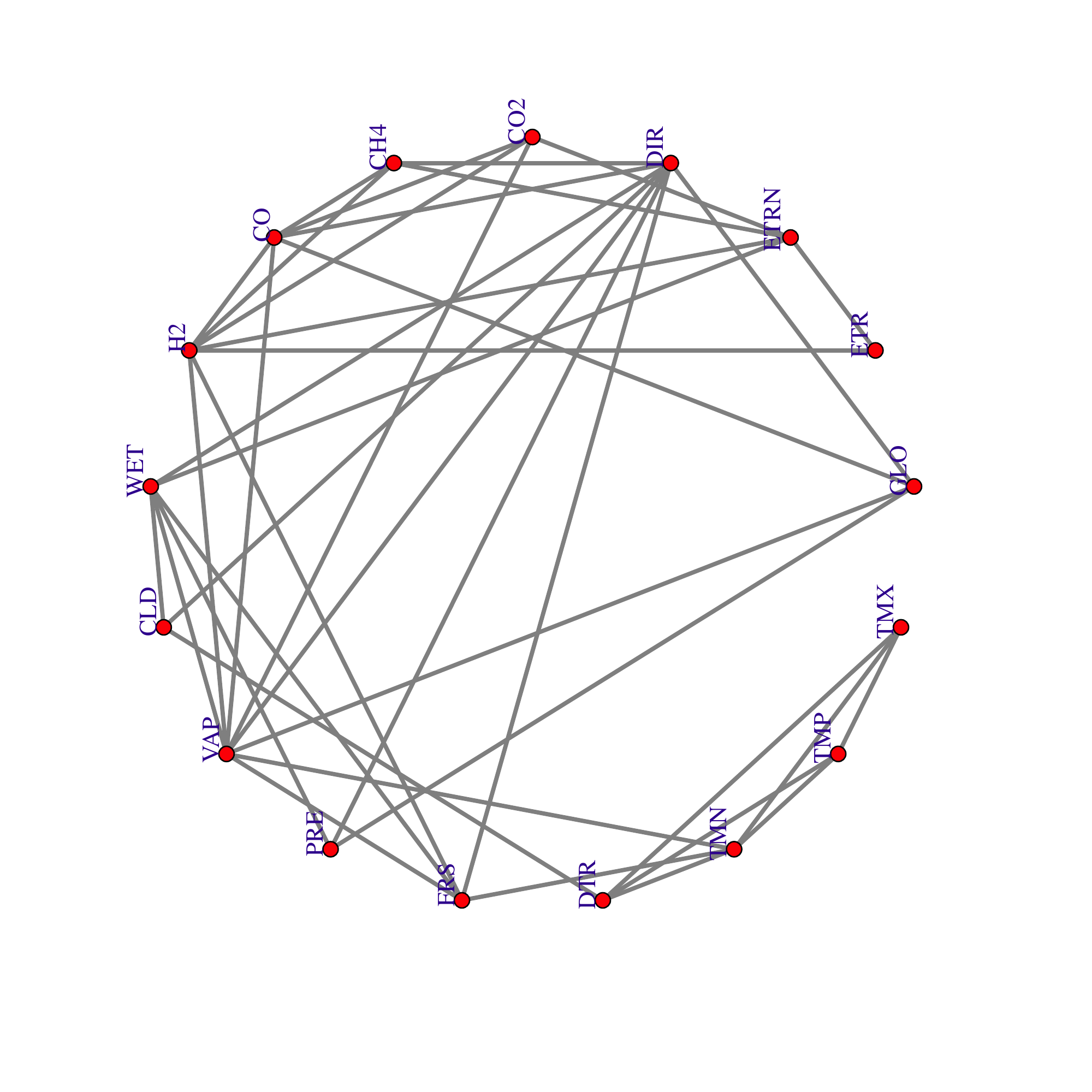}
		\label{fig:climate_a_2}
	}\hspace{-8mm}
	\subfigure[t][Graph for meteorological factors ($\alpha=0.3$)]{
		\includegraphics[width=0.45\textwidth,  height=0.45\textwidth,angle=-90]{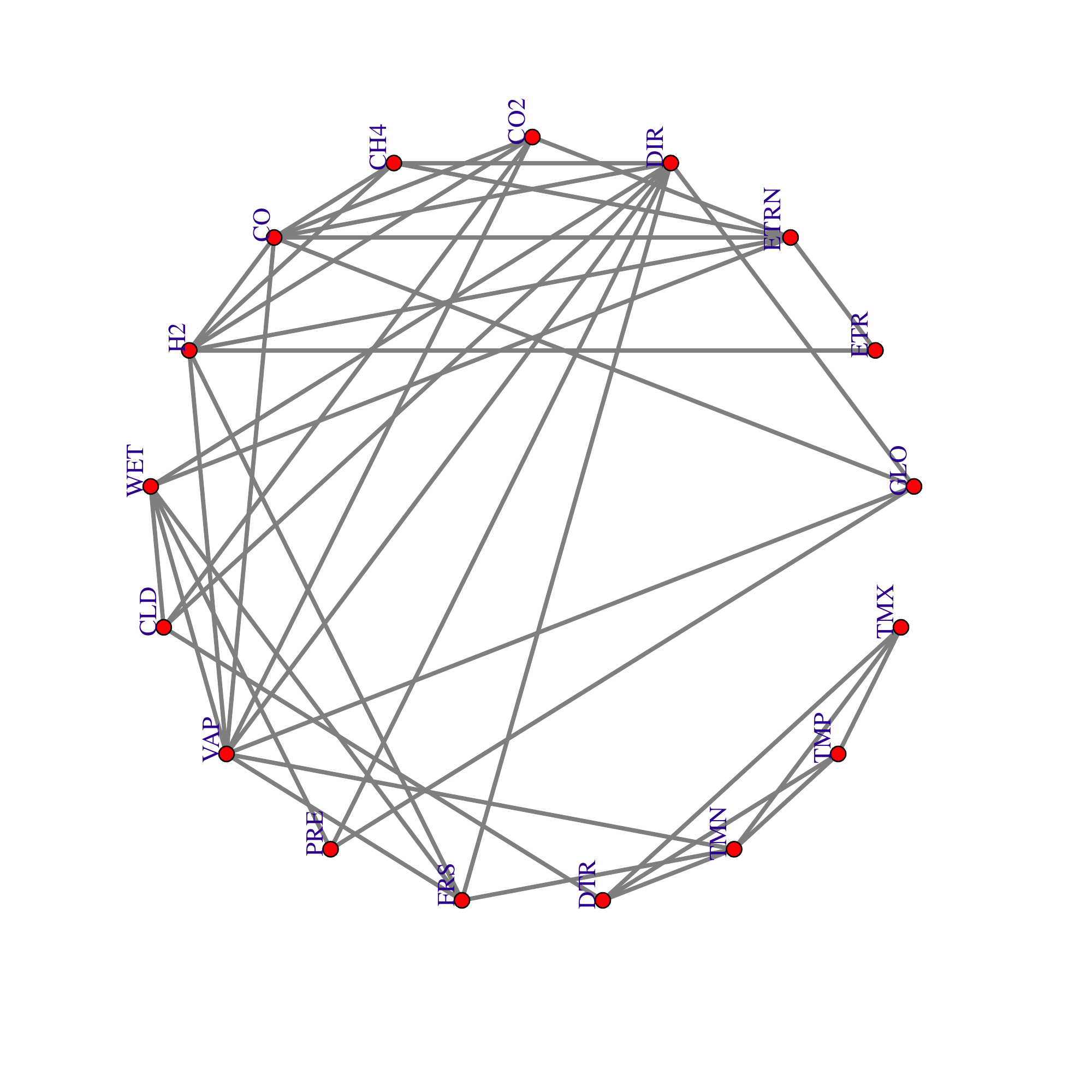}
		\label{fig:climate_a_3}
	}
	\caption{Estimated graphs for the climate data}
	\label{fig:climate}
\end{figure}

We apply the proposed FDR control procedure to estimate the support of the precision matrices for meteorological factors and locations under different $\alpha \in \{0.1, 0.2, 0.3\}$. In Table \ref{tab:climate}, we report the number of edges/discoveries for different $\alpha$'s. From Table \ref{tab:climate}, the number of discoveries for meteorological factors  is quite stable as $\alpha$ increases from 0.1 to 0.3. Moreover, the number of discoveries for locations is relatively large, which indicates many strong correlations among pairs of locations. We plot the graphs corresponding to the estimated supports of the precision matrices for meteorological factors in Figure \ref{fig:climate} (the plots for locations are omitted since they are too dense to visualize). An interesting observation is that the factors \textsf{TMX}, \textsf{TMP}, \textsf{TMN} and \textsf{DTR} form a clique.  This pattern is reasonable since the factors \textsf{TMX}, \textsf{TMP}, \textsf{TMN} and \textsf{DTR} are all related to temperature and thus should be highly correlated. Other sparsity patterns might also provide insight for understanding dependency relationships among meteorological factors.

\bibhang=1.7pc
\bibsep=2pt
\fontsize{9}{14pt plus.8pt minus .6pt}\selectfont
\renewcommand\bibname{\large \bf References}

\bibliographystyle{rss}
\bibliography{GGM}

\begin{thebibliography}{42}
\expandafter\ifx\csname natexlab\endcsname\relax\def\natexlab#1{#1}\fi
\expandafter\ifx\csname url\endcsname\relax
  \def\url#1{\texttt{#1}}\fi
\expandafter\ifx\csname urlprefix\endcsname\relax\def\urlprefix{URL: }\fi

\bibitem[{Allen and Tibshirani(2010)}]{AllenTibshirani2010}
Allen, G.~I. and Tibshirani, R. (2010) Transposable regularized covariance
  models with an application to missing data imputation.
\newblock \textit{Annals of Applied Statistics}, \textbf{4}, 764--790.

\bibitem[{Benjamini and Hochberg(1995)}]{Benjamini95}
Benjamini, Y. and Hochberg, Y. (1995) Controlling the false discovery rate: {a}
  practical and powerful approach to multiple testing.
\newblock \textit{Journal of the Royal Statistical Society. Series B
  (Statistical Methodology)}, \textbf{57}, 389--300.

\bibitem[{Bickle and Levina(2008)}]{bickel08regularized}
Bickle, P. and Levina, E. (2008) Regularized estimation of large covariance
  matrices.
\newblock \textit{Annals of Statistics}, \textbf{36}, 199--227.

\bibitem[{Bijma et~al.(2005)Bijma, De~Munck and Heethaar}]{Bijma05}
Bijma, F., De~Munck, J. and Heethaar, R. (2005) The spatiotemporal meg
  covariance matrix modeled as a sum of kronecker products.
\newblock \textit{NeuroImage}, \textbf{27}, 402--415.

\bibitem[{B\"{u}hlmann and van~de Geer(2011)}]{HighDimBook:11}
B\"{u}hlmann, P. and van~de Geer, S. (2011) \textit{Statistics for
  High-Dimensional Data --- Methods, Theory and Applications}.
\newblock Springer.

\bibitem[{Cai and Liu(2011)}]{Cai2011Adaptive}
Cai, T. and Liu, W. (2011) Adaptive thresholding for sparse covariance matrix
  estimation.
\newblock \textit{Journal of the American Statistical Association},
  \textbf{106}, 672--684.

\bibitem[{Cai et~al.(2016)Cai, Ren and Zhou}]{CaiRenZhou:16}
Cai, T., Ren, Z. and Zhou, H. (2016) Estimating structured high-dimensional
  covariance and precision matrices: Optimal rates and adaptive estimation.
\newblock \textit{Electronic Journal of Statistics}, \textbf{10}, 1--89.

\bibitem[{Cai et~al.(2011)Cai, Liu and Luo}]{CaiLiuLuo2011}
Cai, T.~T., Liu, W. and Luo, X. (2011) A constrained $\ell_1$ minimization
  approach to sparse precision matrix estimation.
\newblock \textit{Journal of the American Statistical Association},
  \textbf{106}, 594--607.

\bibitem[{Cand\`{e}s and Tao(2007)}]{CandesETao2007}
Cand\`{e}s, E. and Tao, T. (2007) The dantzig selector: statistical estimation
  when $p$ is much larger than $n$.
\newblock \textit{Annals of Statistics}, \textbf{35}, 2313--2351.

\bibitem[{Chen and Qin(2010)}]{Chen10Two}
Chen, S.~X. and Qin, Y.~L. (2010) A two-sample test for high-dimensional data
  with applications to gene-set testing.
\newblock \textit{Annals of Statistics}, \textbf{38}, 808--835.

\bibitem[{d'Aspremont et~al.(2008)d'Aspremont, Banerjee and
  El~Ghaoui}]{dAspremont08}
d'Aspremont, A., Banerjee, O. and El~Ghaoui, L. (2008) First-order methods for
  sparse covariance selection.
\newblock \textit{SIAM Journal on Matrix Analysis and its Applications},
  \textbf{30}, 56--66.

\bibitem[{Dawid(1981)}]{Dawid1981}
Dawid, A.~P. (1981) Some matrix-variate distribution theory: Notational
  considerations and a bayesian application.
\newblock \textit{Biometrika}, \textbf{68}, 265--274.

\bibitem[{Efron(2009)}]{Efron09}
Efron, B. (2009) Are a set of microarrays independent of each other?
\newblock \textit{Annals of Applied Statistics}, \textbf{3}, 922--942.

\bibitem[{Fan and Li(2001)}]{Fan01}
Fan, J. and Li, R. (2001) Variable selection via nonconcave penalized
  likelihood and its oracle properties.
\newblock \textit{Journal of the American Statistical Association},
  \textbf{96}, 1348--1360.

\bibitem[{Fan and Lv(2016)}]{FanY:16}
Fan, Y. and Lv, J. (2016) Innovated scalable efficient estimation in
  ultra-large gaussian graphical models.
\newblock \textit{Annals of Statistics}, \textbf{44}.

\bibitem[{Friedman et~al.(2008)Friedman, Hastie and
  Tibshirani}]{FriedmanHastieTibshirani2008}
Friedman, J., Hastie, T. and Tibshirani, R. (2008) Sparse inverse covariance
  estimation with the graphical lasso.
\newblock \textit{Biostatistics}, \textbf{9}, 432--441.

\bibitem[{van~de Geer et~al.(2014)van~de Geer, B¨¹hlmann, Ritov and
  Dezeure}]{vandegeer2014}
van~de Geer, S., B¨¹hlmann, P., Ritov, Y. and Dezeure, R. (2014) On
  asymptotically optimal confidence regions and tests for high-dimensional
  models.
\newblock \textit{Annals of Statistics}, \textbf{42}, 1166--1202.

\bibitem[{Gupta and Nagar(1999)}]{GuptaNagar1999}
Gupta, A.~K. and Nagar, D.~K. (1999) \textit{Matrix Variate Distributions}.
\newblock Chapman Hall.

\bibitem[{Huang and Chen(2015)}]{HuangChen2015}
Huang, F. and Chen, S. (2015) Joint learning of multiple sparse matrix
  {G}aussian graphical models.
\newblock \textit{IEEE Transactions on Neural Networks and Learning Systems},
  \textbf{26}, 2606 -- 2620.

\bibitem[{Kalaitzis et~al.(2013)Kalaitzis, Lafferty, Lawrence and
  Zhou}]{Kalaitzis13Bi}
Kalaitzis, A., Lafferty, J., Lawrence, N.~D. and Zhou, S. (2013) The
  bigraphical lasso.
\newblock In \textit{Proceedings of the 30th International Conference on
  Machine Learning}.

\bibitem[{Lam and Fan(2009)}]{Lam2009Sparsistency}
Lam, C. and Fan, J. (2009) Sparsistency and rates of convergence in large
  covariance matrix estimation.
\newblock \textit{Annals of Statistics}, \textbf{37}, 4254--4278.

\bibitem[{Leng and Tang(2012)}]{LengTang2012}
Leng, C. and Tang, C.~Y. (2012) Sparse matrix graphical models.
\newblock \textit{Journal of the American Statistical Association},
  \textbf{107}, 1187--1200.

\bibitem[{Liu et~al.(2012)Liu, Han, Yuan, Lafferty and
  Wasserman}]{LiuHanYuanLaffertyWasserman2012}
Liu, H., Han, F., Yuan, M., Lafferty, J. and Wasserman, L. (2012) High
  dimensional semiparametric {G}aussian copula graphical models.
\newblock \textit{Annals of Statistics}, \textbf{40}, 2293--2326.

\bibitem[{Liu(2013)}]{Liu2013}
Liu, W. (2013) {G}aussian graphical model estimation with false discovery rate
  control.
\newblock \textit{Annals of Statistics}, \textbf{41}, 2948--2978.

\bibitem[{Liu and Shao(2014)}]{liu2014}
Liu, W. and Shao, Q.~M. (2014) Phase transition and regularized bootstrap in
  large-scale $t$-tests with false discovery rate control.
\newblock \textit{Annals of Statistics}, \textbf{42}, 2003--2025.

\bibitem[{Lozano et~al.(2009)Lozano, Li, Niculescu-Mizil, Liu, Perlich, Hosking
  and Abe}]{Lozano:09}
Lozano, A.~C., Li, H., Niculescu-Mizil, A., Liu, Y., Perlich, C., Hosking, J.
  and Abe, N. (2009) Spatial-temporal causal modeling for climate change
  attribution.
\newblock In \textit{Proceedings of the 15th ACM SIGKDD International
  Conference on Knowledge Discovery and Data Mining}.

\bibitem[{Ma et~al.(2007)Ma, Gong and Bohnert}]{MaGongBohnert2007}
Ma, S., Gong, Q. and Bohnert, H.~J. (2007) An arabidopsis gene network based on
  the graphical {G}aussian model.
\newblock \textit{Genome Research}, \textbf{17}, 1614--1625.

\bibitem[{Meinshausen and B\"{u}hlmann(2006)}]{MeinshausenBuhlmannP2006}
Meinshausen, N. and B\"{u}hlmann, P. (2006) High-dimensional graphs and
  variable selection with the lasso.
\newblock \textit{Annals of Statistics}, \textbf{34}, 1436--1462.

\bibitem[{Ravikumar et~al.(2011)Ravikumar, Wainwright, Raskutti and
  Yu}]{RavikumarWainwrightRaskuttiYu2011}
Ravikumar, P., Wainwright, M., Raskutti, G. and Yu, B. (2011) High-dimensional
  covariance estimation by minimizing $l_{1}$-penalized log-determinant
  divergence.
\newblock \textit{Electronic Journal of Statistics}, \textbf{5}, 935--980.

\bibitem[{Ren et~al.(2016)Ren, Kang, Fan and Lv}]{Ren:16}
Ren, Z., Kang, Y., Fan, Y. and Lv, J. (2016) Tuning-free heterogeneity pursuit
  in massive networks.
\newblock ArXiv preprint arXiv:1606.03803.

\bibitem[{Ren et~al.(2015)Ren, Sun, Zhang and Zhou}]{RenSunZhangZhou2015}
Ren, Z., Sun, T., Zhang, C.~H. and Zhou, H.~H. (2015) Asymptotic normality and
  optimalities in estimation of large {G}aussian graphical model.
\newblock \textit{Annals of Statistics}, \textbf{43}, 991--1026.

\bibitem[{Rothman et~al.(2008)Rothman, Bickel, Levina and
  Zhu}]{RothmanBickelLevinaZhu2008}
Rothman, A., Bickel, P., Levina, E. and Zhu, J. (2008) Sparse permutation
  invariant covariance estimation.
\newblock \textit{Electronic Journal of Statistics}, \textbf{2}, 494--515.

\bibitem[{Schafer and Strimmer(2005)}]{SchaferStrimmer2005}
Schafer, J. and Strimmer, K. (2005) An empirical {B}ayes approach to inferring
  large-scale gene association networks.
\newblock \textit{Bioinformatics}, \textbf{21}, 754--764.

\bibitem[{Tibshirani(1996)}]{Tibshirani:96}
Tibshirani, R. (1996) Regression shrinkage and selection via the lasso.
\newblock \textit{Journal of the Royal Statistical Society. Series B
  (Statistical Methodology)}, \textbf{58}, 267--288.

\bibitem[{Tsiligkaridis et~al.(2013)Tsiligkaridis, Hero and
  Zhou}]{Tsiligkaridis13}
Tsiligkaridis, T., Hero, A.~O. and Zhou, S. (2013) Convergence properties of
  kronecker graphical lasso algorithms.
\newblock \textit{IEEE Transactions on Signal Processing}, \textbf{61},
  1743--1755.

\bibitem[{Xue and Zou(2012)}]{XueZou2012}
Xue, L. and Zou, H. (2012) Regularized rank-based estimation of
  high-dimensional nonparanormal graphical models.
\newblock \textit{Annals of Statistics}, \textbf{40}, 2541--2571.

\bibitem[{Yin and Li(2012)}]{YinLi12}
Yin, J. and Li, H. (2012) Model selection and estimation in matrix normal
  graphical model.
\newblock \textit{Journal of Multivariate Analysis}, \textbf{107}, 119--140.

\bibitem[{Ying and Liu(2013)}]{Ying13}
Ying, Y. and Liu, H. (2013) High-dimensional semiparametric bigraphical models.
\newblock \textit{Biometrika}, \textbf{100}, 655--670.

\bibitem[{Yuan(2010)}]{Yuan2010}
Yuan, M. (2010) Sparse inverse covariance matrix estimation via linear
  programming.
\newblock \textit{Journal of Machine Learning Research}, \textbf{11},
  2261--2286.

\bibitem[{Yuan and Lin(2007)}]{YuanLin2007}
Yuan, M. and Lin, Y. (2007) Model selection and estimation in the {G}aussian
  graphical model.
\newblock \textit{Biometrika}, \textbf{94}, 19--35.

\bibitem[{Zhou(2014)}]{Zhou2014}
Zhou, S. (2014) Gemini: Graph estimation with matrix variate normal instances.
\newblock \textit{Annals of Statistics}, \textbf{42}, 532--562.

\bibitem[{Zhu et~al.(2014)Zhu, Shen and Pan}]{Zhu:14}
Zhu, Y., Shen, X.~T. and Pan, W. (2014) Structural pursuit over multiple
  undirected graphs.
\newblock \textit{Journal of the American Statistical Association},
  \textbf{109}, 1683--1696.

\end{thebibliography}

\vskip .65cm
\noindent
Department of Information, Operations \& Management Sciences, Stern School of Business, New York University
\vskip 2pt
\noindent
Email: xchen3@stern.nyu.edu
\vskip 2pt

\noindent
Department of Mathematics, Institute of Natural Sciences and MOE-LSC, Shanghai Jiao Tong University.
\vskip 2pt
\noindent
Email:  weidongl@sjtu.edu.cn.

\end{document}